\documentclass[12pt]{article}

\usepackage{scicite}
\usepackage{graphicx,color}
\usepackage{times}

\usepackage{rotating}
\usepackage{amsmath, amssymb, amsfonts}
\usepackage{upgreek}
\usepackage{bbm,dsfont}
\usepackage{bm}
\usepackage{breqn}
\usepackage[unicode=true, pdfusetitle,
	colorlinks=false, breaklinks=true, backref=false,
	bookmarks=true, bookmarksnumbered=false, bookmarksopen=false,
	pdfborder={0 0 0}, pdfborderstyle={}]{hyperref}

\renewcommand{\vec}[1]{{\bf #1}}

\newcommand{\be}{\begin{equation}}
\newcommand{\ee}{\end{equation}}

\usepackage{paralist, tabularx}

\graphicspath{ {./_Figures/} }

\newenvironment{sciabstract}{%
\begin{quote} \bf}
{\end{quote}}

\renewcommand\refname{References and Notes}

\newcounter{lastnote}

\usepackage{bm}

\newcommand{\cso}{Cu$_{2}$OSeO$_{3}$}

\newcommand{\fcs}{Fe$_{1-x}$Co$_{x}$Si}

\title{Topological magnon band structure of \\
emergent Landau levels in a skyrmion lattice}

\author
{T. Weber,$^{1,\ast}$ 
D. M. Fobes,$^{2}$
J. Waizner,$^{3}$ 
P. Steffens,$^{1}$ 
G. S. Tucker,$^{4,5}$\\
M. B\"ohm,$^{1}$
L. Beddrich,$^{6,7}$
C. Franz,$^{6,7}$
H. Gabold,$^{6,7}$
R. Bewley,$^{8}$\\
D. Voneshen,$^{8,9}$
M. Skoulatos,$^{6,7}$
R. Georgii,$^{6,7}$
G. Ehlers,$^{10}$
A. Bauer,$^{6,11}$\\
C. Pfleiderer,$^{6,11,12}$
P. B\"oni,$^{6}$
M. Janoschek,$^{2,13,14}$
M. Garst,$^{3,15,16,17}$
\\
\\
\normalsize{$^{1}$Institut Laue-Langevin, 71 avenue des Martyrs, CS 20156, 38042 Grenoble cedex 9, France}\\
\normalsize{$^{2}$Los Alamos National Laboratory (LANL), Los Alamos, NM, USA}\\
\normalsize{$^{3}$Institut f\"ur Theoretische Physik, Universit\"at zu K\"oln, 50937 K\"oln, Germany}\\
\normalsize{$^{4}$Laboratory for Neutron Scattering and Imaging (LNS), Paul Scherrer Institut (PSI), }\\
\normalsize{CH-5232 Villigen, Switzerland}\\
\normalsize{$^{5}$Laboratory for Quantum Magnetism, EPF Lausanne, }
\normalsize{CH-1015 Lausanne, Switzerland}\\
\normalsize{$^{6}$ Physik-Department, Technical University of Munich, 85748 Garching, Germany}\\
\normalsize{$^{7}$ MLZ, Technical University of Munich, 85748 Garching, Germany}\\
\normalsize{$^{8}$ ISIS Facility, Rutherford Appleton Laboratory, Chilton, Didcot, OX11 0QX, Oxfordshire, UK}\\
\normalsize{$^{9}$ Department of Physics, Royal Holloway University of London, Egham, TW20 0EX, UK}\\
\normalsize{$^{10}$Neutron Technologies Division, Oak Ridge National Laboratory, Oak Ridge, USA}\\
\normalsize{$^{11}$ ZQE, Technical University of Munich, 85748 Garching, Germany}\\
\normalsize{$^{12}$ MCQST, Technical University of Munich, 85748 Garching, Germany}\\
\normalsize{$^{13}$Laboratory for Neutron and Muon Instrumentation (LIN), Paul Scherrer Institut (PSI), }\\
\normalsize{CH-5232 Villigen, Switzerland}\\
\normalsize{$^{14}$ Physik-Institut, Universit\"{a}t Z\"{u}rich, CH-8057 Z\"{u}rich, Switzerland}\\
\normalsize{$^{15}$ Institut f\"ur Theoretische Physik, Technical University of Dresden, 01062 Dresden, Germany}\\
\normalsize{$^{16}$ Institut f\"ur Theoretische Festk\"orperphysik, Karlsruhe Institute of Technology,}\\
\normalsize{76131 Karlsruhe, Germany}\\
\normalsize{$^{17}$ Institute for Quantum Materials and Technology, Karlsruhe Institute of Technology,}\\
\normalsize{76131 Karlsruhe, Germany}\\
\normalsize{$^\ast$Corresponding author: E-mail: tweber@ill.fr}
}

\date{}

\begin{document} 


\baselineskip24pt


\maketitle 

\clearpage
\begin{sciabstract}
The motion of a spin excitation across topologically non-trivial magnetic order exhibits a deflection that is analogous to the effect of the Lorentz force on an electrically charged particle in an orbital magnetic field. We used polarized inelastic neutron scattering to investigate the propagation of magnons (i.e., bosonic collective spin excitations) in a lattice of skyrmion tubes in manganese silicide. For wave vectors perpendicular to the skyrmion tubes, the magnon spectra are consistent with the formation of finely spaced emergent Landau levels that are characteristic of the fictitious magnetic field used to account for the nontrivial topological winding of the skyrmion lattice. This provides evidence of a topological magnon band structure in reciprocal space, which is borne out of the nontrivial real-space topology of a magnetic order.
\end{sciabstract}

\clearpage

In quantum mechanics, the movement of an electrically charged particle perpendicular to a magnetic field results in a Lorentz force that causes an orbital motion at discrete energy values known as Landau levels. The formation of Landau levels is ubiquitous in a wide range of condensed matter systems, causing, for instance, quantum oscillations in metals and quantum Hall phenomena in two-dimensional electron gases, where the latter reflects the formation of topological electronic bands with a finite Chern number. When the spin of a moving particle adiabatically adjusts to the local magnetization, the geometrical properties of a smooth magnetization texture give rise to an emergent magnetic field, $B_{\rm em}$, and an emergent Lorentz force \cite{volovik_linear_1987,bruno_topological_2004}. This raises the question of whether a magnetization texture may generate an analogous cyclotron motion of collective spin excitations, causing the formation of Landau levels and topological magnon bands.

Studies of thermal and magnon Hall effects in frustrated magnets \cite{2005_Strohm_PRL,2010_Katsura_PRL,2010_Onose_Science}, as well as certain tailored systems \cite{2013_Shindou_PRB}, strongly suggest the existence of topological magnon bands. Microscopic evidence includes spectroscopic surveys \cite{2015_Chisnell_PRL, 2020_Bo_PRX} and selected band crossings at very high energies far above the low-lying excitations \cite{2018_Yao_NatPhys, 2020_Zhang_PRR}. Skyrmion lattices in chiral magnets offer a particularly simple setting to explore these questions \cite{binz_chirality_2008,van_hoogdalem_magnetic_2013,iwasaki_theory_2014,gobel_family_2018,2016_Molina_NJP,2017_Mook_PRB}. Owing to the non-trivial topology in these systems, the average emergent magnetic field is a multiple of a flux quantum $|\langle B_{\rm em} \rangle|  = s \frac{4\pi \hbar}{e}/\mathcal{A}_{\rm UC}$ per area $\mathcal{A}_{\rm UC}$ of the skyrmion, where $e$ represents a coupling constant and $\hbar$ is the reduced Planck constant \cite{SOM}. Here, $s$ is the spin of the particle (i.e., $s = 1/2$ for an electron- or hole-like excitation and $s = 1$ for a magnon). Overwhelming evidence of this fictitious magnetic field has been observed for electron- and hole-like excitations (i.e., fermions) in terms of a topological Hall signal and spin transfer torques \cite{neubauer_topological_2009,2010_Jonietz_Science}. In contrast, only indirect evidence has been inferred for a fictitious magnetic field acting on magnons (i.e., bosons) from the rotational motion of skyrmion lattice domains \cite{mochizuki_thermally_2014}.

Here, we used polarized inelastic neutron scattering to determine the predicted dispersion and orbital motion of magnons in the skyrmion lattice of MnSi. Neutron and microwave studies of spin excitations in MnSi have been reported for the topologically trivial magnetic phases, namely the helical, conical, and spin-polarized states \cite{Jano10,Grigoriev15, Kugler15,Sato16, HeliPaper} as well as the paramagnetic regime \cite{Roessli2002, 2009_Pappas_PRL, 2013_Janoschek_PRB, 2014_Kindervater_PRB, 2017_Pappas_PRL, 2019_Kindervater_PRX}. This work has revealed, across the entire Brillouin zone, well-defined dispersive, nonreciprocal spin waves and spin fluctuations in the ordered states and the paramagnetic regime, respectively. In comparison, for the topologically nontrivial skyrmion lattice, extensive micro-wave spectroscopy \cite{Schwarze15} and exploratory inelastic neutron scans \cite{JanoSkyrmi, weber2018non} have been reported. These experiments provide an important point of reference, but they are limited to the center of the Brillouin zone in the case of microwave spectroscopy, or are at the proof-of-principle stage for inelastic neutron scattering.

\begin{figure*}[!t]
\begin{center}
\includegraphics[width=0.5\linewidth]{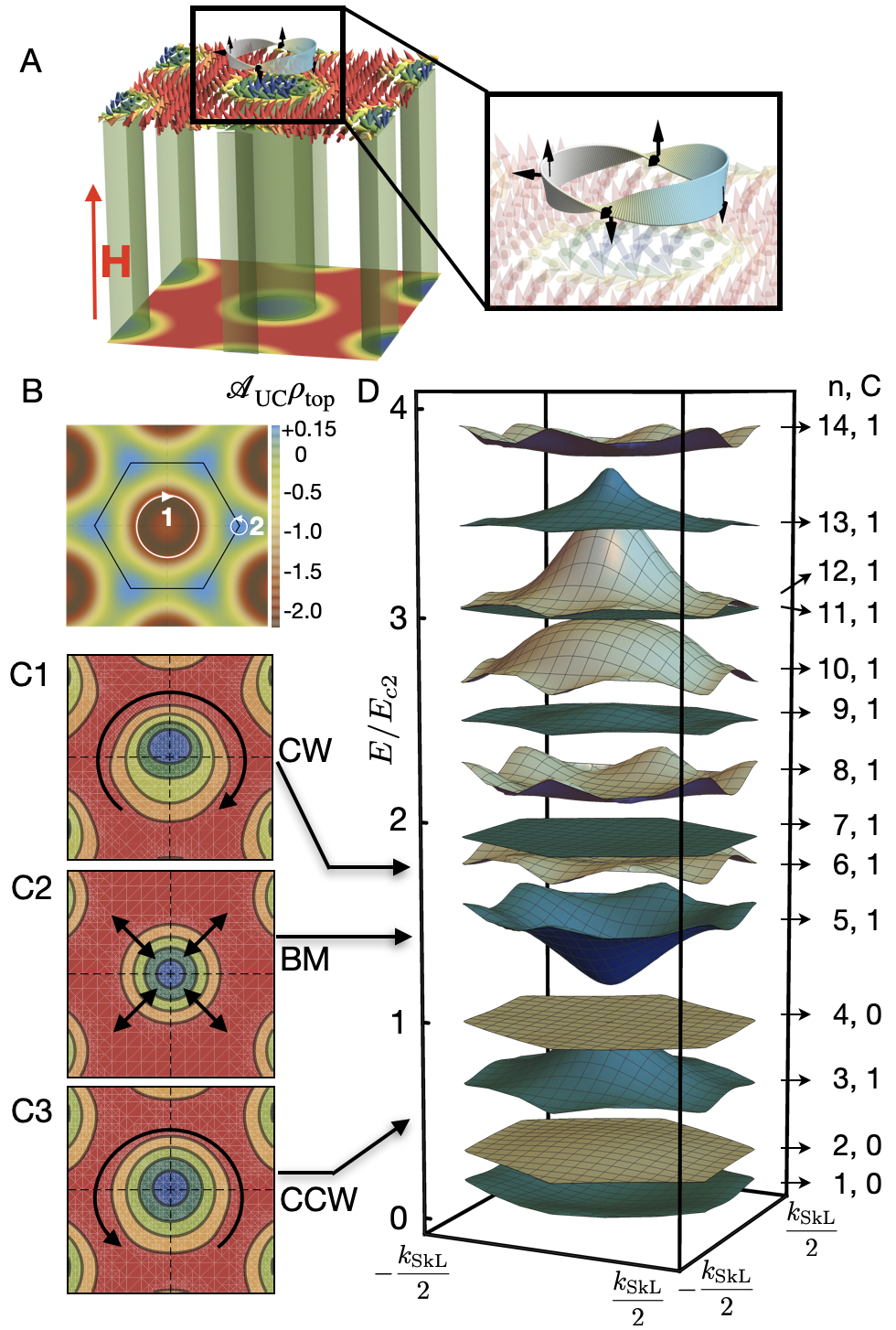}
\caption{
\textbf{Depiction of emergent Landau levels of magnons and topological magnon bands in a skyrmion lattice.} 
(A) Qualitative depiction of the skyrmion lattice in a magnetic field $H$. Inset: The classical trajectory of a magnon, illustrating the orientation of the local coordinate system (black arrows) as it tracks the local magnetization and accumulates a Berry phase. 
(B) Variation of the topological winding density, $\rho_{\rm top}$, across the skyrmion lattice, averaging to a winding number of -1 per unit cell area $\mathcal{A}_{\rm UC}$. Classical trajectories of magnons are denoted 1 and 2 (see text for details).
(C1 to C3) Key characteristics of magnons in a skyrmion lattice as observed in microwave spectroscopy. CW: clockwise mode; BM: breathing mode; CCW: counter-clockwise mode.
(D) Calculated lowest 14 bands in the skyrmion lattice of MnSi within the first Brillouin zone in units $E_{\rm c2}=g\mu_{B}\mu_0 H^{\rm int}_{\rm c2}$ \cite{Garst2017}. The dispersive character originates partly in the non-uniform topological winding density, $\rho_{\rm top}$, as illustrated in panel (B). The band index $n$ and Chern number $C$ are indicated at the right. Bands 3, 5 and 6 correspond the CCW, BM and CW mode, respectively. The Goldstone mode (GM) corresponds to $n=1$.  The Landau levels represent topological magnon bands with $C=1$. 
}\label{fig_1}
\end{center}
\end{figure*}

Our approach is best introduced by summarizing the key characteristics of the magnon spectra, as illustrated in Fig.\,\ref{fig_1}. In a semi-classical picture, magnons account for the precession of the magnetization around its equilibrium value ${\bf  M}({\bf r})$, as described locally with the help of the orthogonal unit vectors $\hat e_1 \times \hat e_2 = \hat e_3$ with $\hat e_3({\bf r}) = {\bf  M}({\bf r})/|{\bf  M}({\bf r})|$. Within this local frame of reference, the motion of magnons is influenced by the vector potential $A_i({\bf r}) = \frac{\hbar}{e} \hat e_1 \partial_i \hat e_2$ given by the spin connection known from differential geometry. The emergent local magnetic field is obtained by ${\bf B}_{\rm em} = \nabla \times {\bf A} = \frac{4\pi \hbar}{e} \hat z \rho_{\rm top}$. It is directly related to the topological skyrmion charge of the two-dimensional magnetic texture,  $\rho_{\rm top} = \frac{1}{4\pi} \hat e_3 (\partial_x \hat e_3 \times \partial_y \hat e_3)$, via the Mermin-Ho relation \cite{mermin_circulation_1976}. As the integral over a single skyrmion is quantized, $\int_{\rm UC} d^2{\bf r} \rho_{\rm top} = -1$; this amounts to a magnetic flux of $-\frac{4\pi \hbar}{e}$ per skyrmion.

Shown in Fig.\,\ref{fig_1}\,A is the hexagonal lattice of skyrmion tubes in an applied magnetic field $\bf{H}$ and the classical trajectory of a magnon, for which the local magnetization is perpendicular to $\bf{H}$. The band weaving along the trajectory depicts the orientation of the local coordinate system (black arrows), reflecting the accumulated geometric Berry phase $\oint d{\bf r} {\bf A}$. The associated topological density $\rho_{\rm top}$ is shown in Fig.\,\ref{fig_1}\,B. Although the integral $\int_{\rm UC} d^2{\bf r} \rho_{\rm top} = -1$, the density $\rho_{\rm top}$ varies substantially, resulting in a pattern of positive and negative ${\bf B}_{\rm em}$. The associated emergent Lorentz force enables localized cyclotron orbits within the magnetic unit cell, as illustrated in Fig.\,\ref{fig_1}\,B, where classical trajectories labeled "1" and "2" circulate around a minimum and a maximum of $\rho_{\rm top}$, respectively.

Magnetic resonance studies that probe the response to a uniform oscillatory magnetic field, ${\bf H}_{\rm ac}$, identified three fundamental magnon modes \cite{Mochizuki2012,Okamura:2013}, depicted in Figs.\,\ref{fig_1}\,C1 to C3. For ${\bf H}_{\rm ac}$ within the skyrmion lattice plane, clockwise (CW) and counterclockwise (CCW) gyrations of the skyrmion core are generated, whereas a breathing mode (BM) develops for ${\bf H}_{\rm ac}$ aligned parallel to the skyrmion tubes. Extensive microwave spectroscopy in MnSi, {\fcs}, and {\cso} established universal agreement with theory based on a few material-specific and sample shape-specific parameters \cite{Schwarze15}.

Going beyond this initial work, we calculated the 14 lowest-lying magnon bands of MnSi for momentum transfers ${\bf q}_\perp$ within the two-dimensional Brillouin zone of the hexagonal skyrmion lattice (SkL) as predicted in \cite{Garst2017}, where $k_{\rm SkL}$ is the magnitude of the reciprocal primitive vectors (Fig.\,\ref{fig_1}\,D). Noted on the right side are the band index $n$ and the Chern number $C$. Bands 3, 5, and 6 represent the CCW, BM, and CW modes, respectively. Bands with nonzero Chern numbers are topologically nontrivial.

The associated magnon band structure for ${\bf q}_\perp$ exhibits a nontrivial topology that is directly linked to the nontrivial real-space topology of the skyrmion lattice. For higher magnon energies, $E \gg E_{c2}$ (where $E_{c2} = g\mu_0 \mu_B H^{\rm int}_{c2}$ represents the energy associated with the internal critical field separating the conical and field-polarized phases), the spin wave equation may be approximated by a Schr{\"o}dinger equation for the Hamiltonian $\mathcal{H} = \frac{1}{2m} ({\bf p} + e {\bf A})^2$. Here, ${\bf A}$ and $m$ are the emergent vector potential and the magnon mass, respectively, obeying $\frac{\hbar^2 k_h^2}{2m} = E_{c2}$, where $k_h$ is the helix wave vector. For a topologically trivial texture, the vector potential may be removed by a gauge transformation resulting in the free magnon dispersion $E({\bf q}_\perp) = \frac{\hbar^2 {\bf q}_\perp^2}{2m}$.

In contrast, magnons in the skyrmion lattice are expected to form emergent Landau levels with a characteristic cyclotron energy $E_{\rm cycl} = \frac{\hbar e |\langle B_{\rm em} \rangle|}{m}$. Using $\mathcal{A}_{\rm UC} = \sqrt{3} a^2/2$ for the area of the magnetic unit cell with lattice constant $a = 4\pi/(\sqrt{3} k_{\rm SkL})$, this amounts to $E_{\rm cycl}/E_{c2} = \frac{\sqrt{3}}{\pi} (k_{\rm SkL}/k_h)^2  \approx \frac{\sqrt{3}}{\pi} $. Because each skyrmion contributes two flux quanta, an average density of $2\pi/\sqrt{3} \approx 3.6$ states per $E_{c2}$ is expected for each ${\bf q}_\perp$. For MnSi with $E_{c2} = 37\,\mu{\rm eV}$ at 28.5\,K, this approximately corresponds to one magnon band per $10\,\mu$eV, where the relatively flat dispersion of most bands at higher energies and their finite Chern number stem from the emergent magnetic flux ${\bf B}_{\rm em} = \nabla \times {\bf A}$ and thus $\rho_{\rm top}$ shown in Fig.~\ref{fig_1}\,B. For low energy, the scalar potential cannot be neglected, resulting in additional, topologically trivial bands.

\begin{figure*}[!t]
\begin{center}
\includegraphics[width=0.55\linewidth]{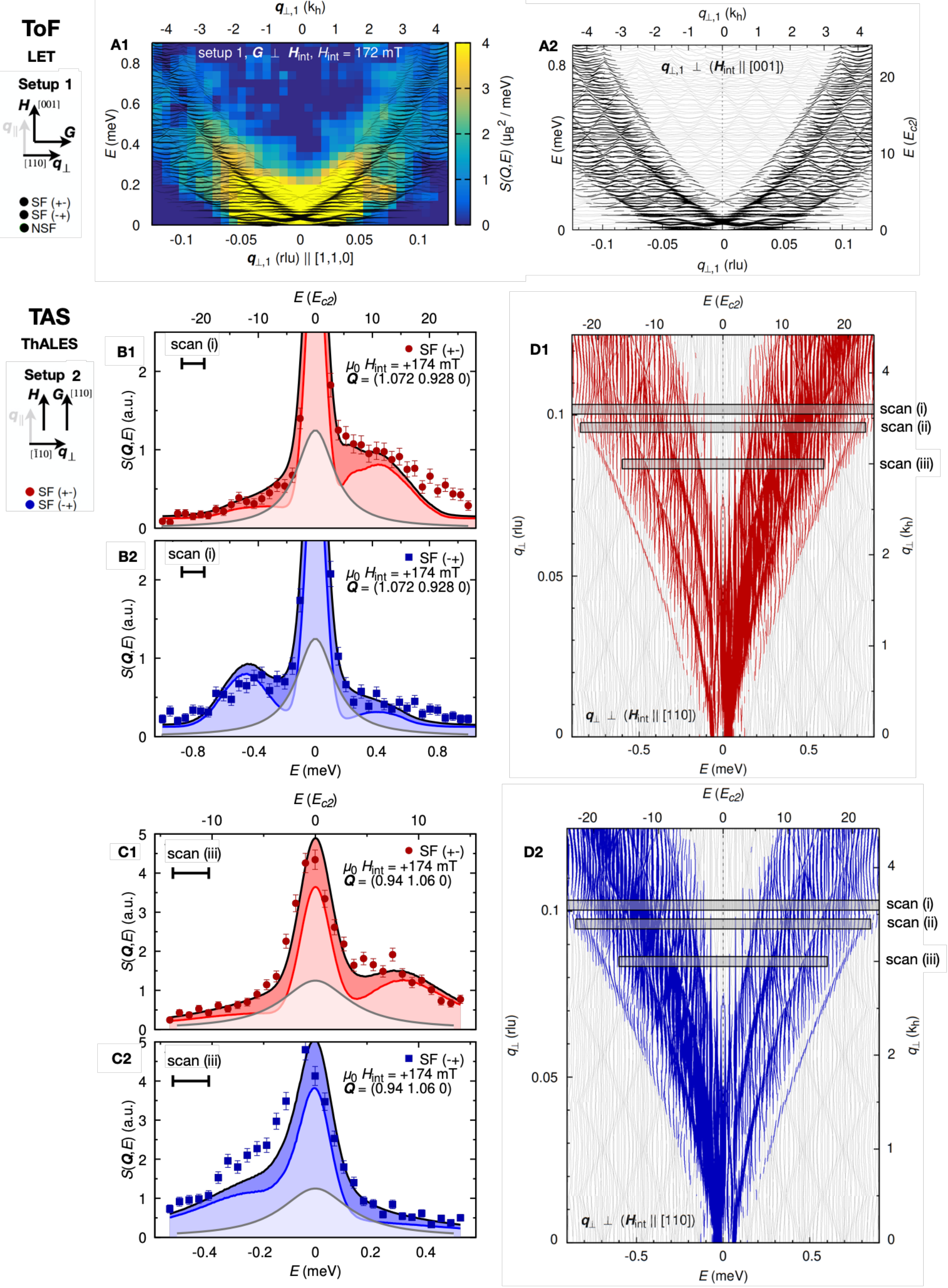}
\caption{
\textbf{Magnon spectra of MnSi for momentum transfers perpendicular to the skyrmion lattice tubes.} 
Left: Experimental data. Right: Quantitative calculations of the magnon spectra. Thin gray lines represent the magnon spectra $E({\bf q})$; black, red, and blue lines represent the magnetic response tensor, $\chi''_{ij}({\bf q},E)$. The line thickness of the latter reflects the spectral weight. Colors red/blue and green denote spin-flip and non-spin-flip processes, SF and NSF, respectively. Energy and momentum transfers are provided in two corresponding scales.
(A1, A2)  ToF intensity for momentum transfers ${\bf q}_{\perp, 1} \parallel [1, 1, 0]$. The color shading represents the experimental scattering intensity. Black lines represent the calculated magnetic response tensor shown also in (A2) (see Fig.\,S6\,B for the calculated dynamic structure factor as convoluted with
the instrumental resolution).
(B1, B2, C1, and C2) Polarized TAS intensities for selected momentum transfers and field values. Curves shown in red and blue shading represent the sum of the calculated dynamic structure convoluted with the instrumental resolution (light red/blue shading) and a quasi-elastic (QE) contribution attributed to longitudinal fluctuations (gray shading) \cite{SOM}. The same quantitative scaling factor and the same QE contribution was used for all TAS data shown in the main text and \cite{SOM} for momentum transfers perpendicular \emph{and} transverse to the skyrmion lattice ${\bf q}_{\perp}$ and ${\bf q}_{\parallel}$, respectively.
(D1, D2) Calculated magnon spectra and magnetic response tensors. The location of experimental scans shown in (B) and (C) is marked by a gray boxes [see Figs.\,S10 and S17 for further data including those of scan (ii)]. 
}
\label{fig_2}
\end{center}
\end{figure*}
\begin{figure*}[!t]
\begin{center}
\includegraphics[width=0.9\linewidth]{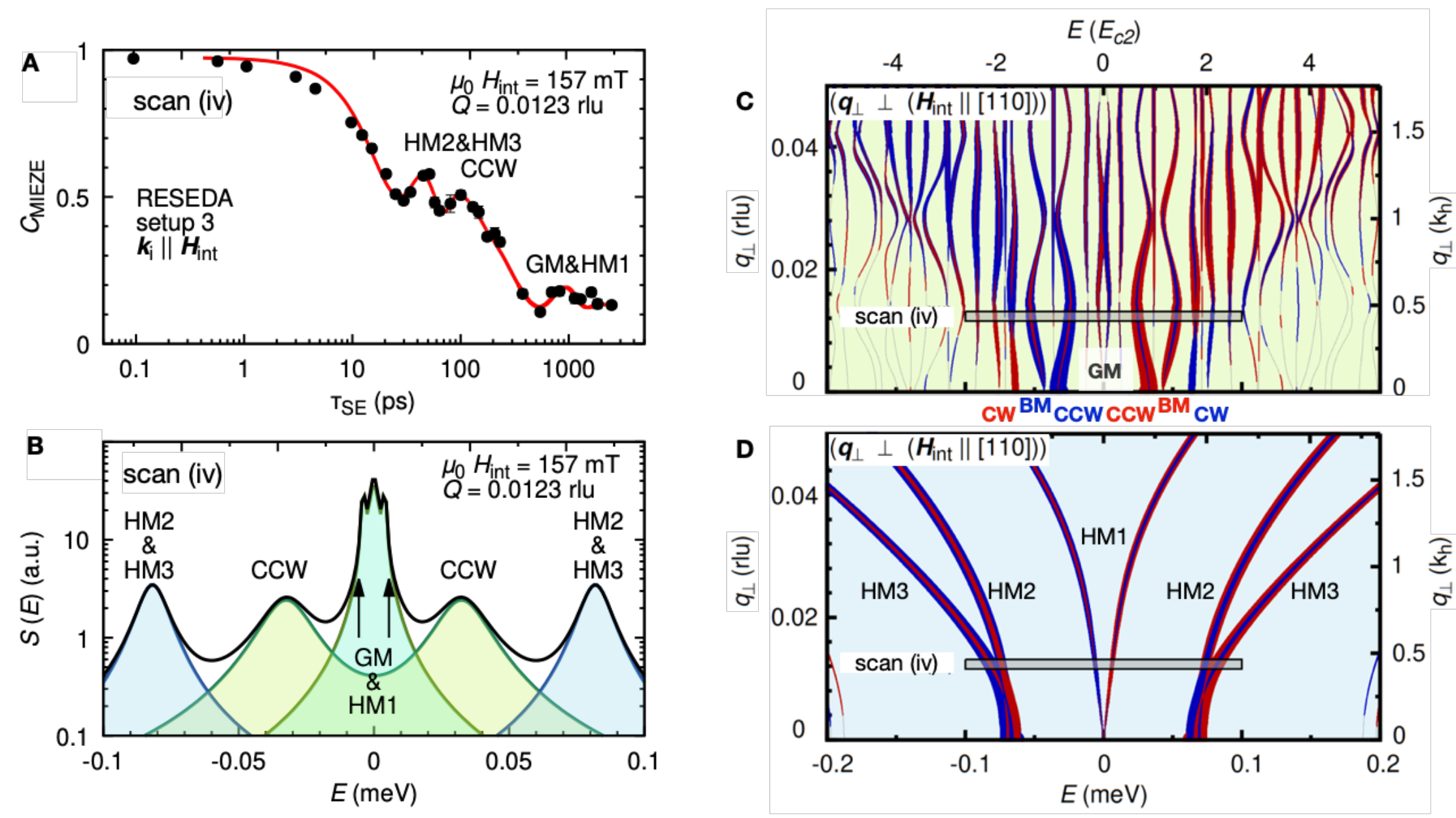}
\caption{\textbf{Neutron scattering intensity and calculated magnon spectra in MnSi in the skyrmion lattice and conical phase at very low energies}. 
MIEZE data was recorded with setup 3 \cite{SOM}.
(A) MIEZE contrast, $C_{\rm MIEZE}$, representing the intermediate scattering function, $\mathcal{S}({\bf q}, \tau)$, observed in neutron spin-echo spectroscopy at $\vert {\bf q}\vert = 0.0123\,\mathrm{rlu}$, for a large volume fraction of skyrmion phase and a small coexistent volume fraction of conical phase. The red line represents the Fourier transform of the dynamic structure factor shown in (B).
(B) Dynamic structure factor, $\mathcal{S}({\bf q},E)$, of the calculated magnon spectra (table S4 and Figs.\,S21 and S22) taking into account linewidths such that the Fourier transform (red line) matches the data in (A).
(C) Calculated magnon spectra $E({\bf q})$ (gray lines) and spectral weight of the magnetic response tensor, $\chi''_{ij}({\bf q},E)$ in the skyrmion lattice. 
(D) Calculated magnon spectra $E({\bf q})$ and spectral weight of the structure factor inferred from $\chi''_{ij}({\bf q},E)$ in the conical phase. 
In (C) and (D), color shaded lines represent $\chi''_{ij}({\bf q},E)$ in setup 2 used for polarized TAS, where red and blue denote the two spin-flip processes, SF(+-) and SF(-+), respectively. Energy and momentum transfers are provided in two corresponding scales. The gray-shaded box marked scan (iv) denotes the location where the MIEZE data were recorded. CW: clockwise mode; CCW: counter-clockwise mode; BM: breathing mode; GM: Goldstone mode; HM1 to HM3: helimagnons of the conical state.
}
\label{fig_3}
\end{center}
\end{figure*}
\begin{figure*}[!t]
\begin{center}
\includegraphics[width=0.7\linewidth]{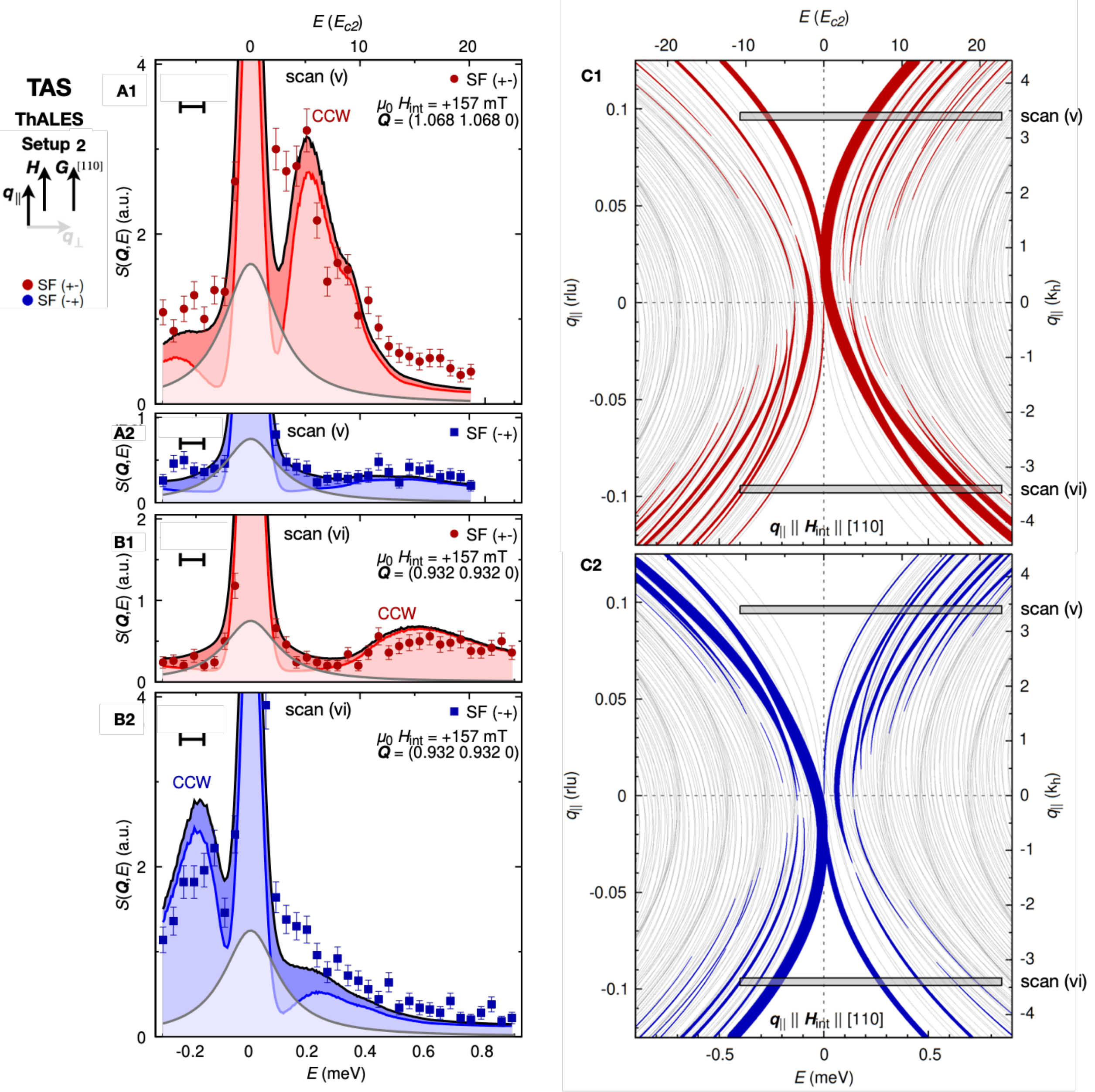}
\caption{\textbf{Polarized neutron scattering intensity and calculated magnon spectra of MnSi for momentum transfers parallel to the skyrmion lattice tubes}. 
Left: Experimental data. Right: Quantitative calculations of the magnon spectra. Thin gray lines represent the magnon spectra $E({\bf q})$; red and blue shaded lines denote the magnetic response tensor, $\chi''_{ij}({\bf q},E)$ for the two spin-flip scattering processes. The line thickness of the magnetic response tensor reflects the spectral weight. Energy and momentum transfers are provided in two corresponding scales. 
(A1, A2, B1, and B2) Polarized TAS intensities for selected momentum transfers and field values chosen to highlight well-defined, dispersive, non-reciprocal magnons. Curves shown in red and blue shading represent the sum of the calculated dynamic structure factor convoluted with the instrumental resolution (light red/blue shading) and a quasi-elastic (QE) contribution attributed to longitudinal fluctuations (gray shading) \cite{SOM}. The same quantitative scaling factor and the same QE contribution was used for all TAS data shown in the main text and \cite{SOM}, for momentum transfers perpendicular and transverse to the skyrmion lattice ${\bf q}_{\perp}$ and ${\bf q}_{\parallel}$, respectively.
(C1, C2) Calculated magnon spectra. The location of the experimental data shown in (A) and (B) is marked by a gray boxes (see Figs.\,S10 to S13 and S16 for further data).
}
\label{fig_4}
\end{center}
\end{figure*}

Polarized inelastic neutron scattering is uniquely suited to verifying the predicted excitation spectra (Fig.\,\ref{fig_1}\,D) as well as their specific character over a wide range of momentum transfers and excitation energies. In principle, this requires, however, a momentum and energy resolution much better than $k_{h}$ and the spacing of the Landau levels, respectively, as well as the identification of unambiguous characteristics accessible at low resolution. To satisfy these requirements for MnSi, which are beyond the individual limits of present-day neutron spectrometers, we combined the information inferred from three different neutron scattering methods. Shown in Figs.\,\ref{fig_2}, \ref{fig_3}, and \ref{fig_4} are the experimentally observed spectra alongside the beamline-specific theoretical predictions [see \cite{SOM} for details]. We began by surveying the spectra recorded by means of unpolarized, time-of-flight (ToF) spectroscopy (Figs.\,\ref{fig_2}\,A1 and \ref{fig_2}\,A2). Then, the distribution of excitation energies and scattering intensities across a large number of positions in parameter space was determined with cold, high-resolution, polarized triple-axis spectroscopy (TAS) (Fig.\,\ref{fig_2}\,B1 to \ref{fig_2}\,C3 and Fig.\,\ref{fig_4}). Finally, selected individual Landau levels were resolved by means of so-called modulation of intensity by zero effort (MIEZE), a type of neutron spin-echo spectroscopy (Fig.\,\ref{fig_3}).


Owing to its incommensurate, multi-$k$ nature, the magnon spectra and the magnetic
response tensor of the skyrmion lattice could not be computed in spin wave simulation packages such as SpinW. We therefore developed the necessary simulation tools. Details of the procedures on all software tools are given in  \cite{SOM}. The source code is provided as supplementary material \cite{simulation_doi}. The results of our theoretical calculations are shown on the right side of Figs.\,\ref{fig_2}, \ref{fig_3}, and \ref{fig_4}, where the bottom and top axes denote energies equivalently in units of meV and $E_{c2}$, and the left and right axes denote momentum transfers equivalently in reciprocal lattice units $\rm (r\,l\,u)$ and helical modulation lengths  $k_{\rm h}$ [see \cite{SOM} for material-specific parameters]. Shown in thin gray lines are the calculated quantitative magnon spectra $E({\bf q})$, whereas the magnetic response tensor $\chi''_{ij}({\bf q},E)$ is shown in color shading. The latter is related by means of the fluctuation-dissipation theorem (eq. S20) to the magnetic dynamic structure factor $S_{ij}({\bf q},\omega)$ measured experimentally. The red and blue shading refers to spin-flip (SF) scattering of a polarized beam corresponding purely to either SF\,$(+-)$ or SF\,$(-+)$ processes, respectively. Shown in black and green shading is SF scattering of an unpolarized beam, resulting in a mixed character$[(+-),(-+)]$ and non-spin-flip scattering (NSF) $[(++),(--)]$, respectively {\cite{SOM}}.

To unambiguously identify the salient features of the excitation spectra, we determined the magnon spectra in the skyrmion lattice plane (Figs.\,\ref{fig_2} and \ref{fig_3}) where they exhibit the emergent Landau levels. In addition, spectra for momentum transfers parallel to the skyrmion tubes were recorded (Fig.\,\ref{fig_4}), which display the characteristics of conventional nonreciprocal magnons akin those observed in the topologically trivial phases \cite{Jano10, Grigoriev15, Kugler15, Sato16, HeliPaper, JanoSkyrmi, weber2018non}. The scattering geometries were chosen carefully to further distinguish nuclear and magnetic scattering \cite{SOM}. Namely the ToF and TAS measurements were performed in the $(hk0)$ scattering plane in the vicinity of the nuclear $\left[ 110 \right]$ Bragg reflection (i.e., $\bm G_{[110]} = \left[ 110 \right]$; see the schematic depictions in Figs. 2, 3, and 4). Accordingly, the associated theoretical predictions shown in Figs.\,\ref{fig_2}, \ref{fig_3} and \ref{fig_4} were calculated for a projection onto the subspace perpendicular to the nuclear Bragg vector ${\bf G}_{[110]}$.


We begin with the magnon spectra in the skyrmion lattice plane, where the emergent Landau levels are expected. For the unpolarized ToF surveys, the magnetic field $\bm H$ was parallel to $\left[ 001 \right]$ and therefore perpendicular to both $\bm G_{[110]}$ and the scattering plane, as depicted in Fig.\,\ref{fig_2}\,A1. This is denoted as setup 1. Because the magnetic satellites were located within the scattering plane, the NSF and SF scattering contained both nuclear and magnetic contributions, additionally motivating TAS. Typical magnon spectra for ${\bf q}_{\perp}$ perpendicular to ${\bf H}$ recorded with ToF spectroscopy are shown in Fig.\,\ref{fig_2}\,A1 and Fig.\,S3. The scattering intensity is presented in yellow/blue shading. Black lines represent the calculated magnetic response tensor, $\chi''_{ij}({\bf q},E)$, where the line thickness reflects the spectral weight. The intensity compares well with the calculated spectra again shown in Fig.\,\ref{fig_2}\,A2 as explicitly demonstrated in Fig.\,S6, which is a color-shaded depiction of the calculated dynamic structure factor with a convolution of the ToF resolution.

The calculated spectra for wave vectors perpendicular to the skyrmion tubes, ${\bf q}_\perp$, shown in Fig.\,\ref{fig_2}\,A2 as gray lines, display the Landau levels stemming from the nontrivial topology of the skyrmion lattice; the calculated magnetic response tensor is shown as black lines, as in Figs.\,\ref{fig_2}\,A1. The weak dispersion of the Landau levels is caused by variations in ${\bf B}_{\rm em}$ as well as the magnon potential. Because the Landau level wave functions of magnons for ${\bf q}_\perp$ are distinctly different from the plane wave function of the neutron, the spectral weight at elevated energies gets distributed across a wide range of magnon bands. Thus, a broad distribution of spectral weight, as compared to the well-defined magnon branches along the skyrmion tubes shown in Fig.\,\ref{fig_4}, represents a key signature of the nontrivial topology.


For a quantitative comparison between theory and experiment, and to keep track of spin-flip versus non?spin-flip processes with respect to incoherent and nuclear scattering, we performed high-resolution polarized TAS. Because the predicted spacing of the Landau levels of $\sim10\,\mu{\rm eV}$ is well below the resolution of state-of-the-art TAS, we used a setup in which the magnetic field $\bm H$ was parallel to $\bm G$ (i.e., $\bm H_{[110]}$ along $\left[ 110 \right]$) such that the skyrmion lattice stabilized perpendicular to $\bm H_{[110]}$ with a pair of its satellites parallel to a $\left[ 1\bar{1}0 \right]$. This is referred to as setup 2. The polarization $\bm P$ is aligned adiabatically with $\bm H_{[110]}$ and hence $\bm G$, and thus the SF and NSF scattering were purely magnetic and nuclear, respectively.

Typical energy scans at various values of ${\bf q}_{\perp}$ are shown in Fig.\,\ref{fig_2}, B1, B2, C1, and C2. The corresponding calculated magnon spectra as well as the magnetic response tensor are shown in Fig.\,\ref{fig_2}, D1 and D2, where the gray shaded boxes denote the parameter range of the energy scans (i), (ii), and (iii) [see Figs. S10 and S17 for further data, including those of scan (ii)]. For ease of comparison with the TAS data, the calculated spectra (Fig.\,\ref{fig_2}, D1 and D2) are shown as a function of energy.

A quantitative comparison of the energy dependence of our TAS data with theory is shown in Fig.\,\ref{fig_2}\,, B1, B2, C1, and C2. Here, the light-red and light-blue shading denote magnetic SF\,$(+-)$ and SF\,$(-+)$ scattering, respectively; the light-gray shaded areas denote a quasi-elastic (QE) contribution; and the red and blue shading denote the sum of the magnon spectral weight and the QE contribution. For the magnon spectra, we convoluted the dynamic structure factor $\mathcal{S}_{ij}({\bf q},E)$ with the instrumental resolution, taking into account incoherent magnetic scattering (for clarity, the incoherent NSF scattering is not shown; see Figs. S10 to S12, S16, and S17 for depictions that display the incoherent NSF scattering). Scaling simultaneously all TAS data (i.e., those recorded for momentum transfers ${\bf q}_{\perp}$ \emph{and} ${\bf q}_{\parallel}$) by the same constant (see Fig.\,\ref{fig_2} and Fig.\,\ref{fig_4}, and Figs. S10 to S13, S16, and S17), the energy dependence of the SF scattering due to magnons (Fig.\,\ref{fig_2}, B1 and C2) is already in reasonable quantitative agreement with experiment. This agreement improves further after adding a small QE Gaussian contribution with a linewidth $\Gamma_e\approx120\,{\mu\rm eV}$. The weight of this QE scattering is essentially identical in all of the TAS data, where small differences may be attributed to the instrumental resolution. We find that the QE scattering is in good qualitative agreement with longitudinal excitations, as shown theoretically in \cite{SOM}. It is therefore not included in the model of the magnon spectra, which represent transverse spin excitations. In Figs.\,\ref{fig_2} and \ref{fig_4}, we show that the quantitative agreement between $\mathcal{S}_{ij}({\bf q},E)$ and experiment is good. Notably, the sum of the magnon spectra encoded in $\mathcal{S}_{ij}({\bf q},E)$ (red and blue solid lines) and the QE scattering (gray line) indicated in red and blue shading quantitatively agrees with the data points. Unfortunately, a similarly accurate comparison between experiment and theory taking into account the instrumental resolution is technically inaccessible for the ToF and the MIEZE data.


We resolved the lowest-lying Landau level using MIEZE neutron spin-echo spectroscopy. For the MIEZE measurements, a polarized small-angle neutron scattering configuration without polarization analysis was used, where $\bm H$ was parallel to the incident neutron beam. This is denoted as setup 3  \cite{SOM}. Probing the intermediate scattering function $\mathcal{S}({\bf q},\tau)$ (i.e., the time dependence rather than the frequency dependence of the spin correlations), MIEZE allowed simultaneous detection of several excitations with very high resolution \cite{Reseda_JPSJ}.

Shown in Fig.\,\ref{fig_3}\,A is the MIEZE contrast equivalent to $\mathcal{S}({\bf q},\tau)$, where the oscillatory decay is characteristic of several damped propagating excitations \cite{Reseda_JPSJ}. The associated dynamic structure factor, $\mathcal{S}({\bf q},E)$, with resonance energies and estimated linewidths is shown in Fig.\,\ref{fig_3}\,B. Its Fourier transform with respect to energy, $\mathcal{S}({\bf q},\tau)$, matches the experimental data well (Fig.\,\ref{fig_3}\,A, red curve). To obtain the strongest elastic and thus inelastic intensity of the skyrmion lattice, we chose a sample temperature for the MIEZE experiment for which the skyrmion phase coexisted with a small volume fraction of conical phase, as independently confirmed in TAS measurements \cite{SOM}. Calculated magnon spectra for these coexisting skyrmion lattice and conical phases are shown in Fig.\,\ref{fig_3}, C and D.

Focusing on the zone center, the calculated spectra in the skyrmion lattice (Fig.\,\ref{fig_3}\,C) reproduce the energies and spectroscopic weights of the CW, CCW, and BM modes in excellent quantitative agreement with microwave spectroscopy \cite{Schwarze15}. Further, the excitation spectrum of the conical phase for the same field strength (Fig.\,\ref{fig_3}\,D) reveals that the excitation energies of the lowest-lying helimagnons (HM1, HM2, and HM3) differ suitably from the excitation energies of the skyrmion lattice. Thus, the presence of a small volume of conical phase not only confirmed the existence of the lowest-lying Landau level, but provided an important test of the validity of the MIEZE spectroscopy.

For a direct comparison with experiment, the calculated MIEZE intensities for the skyrmion lattice and conical state, denoted scan (iv) in Fig.\,\ref{fig_3}, C and D, were evaluated numerically for setup 3 (see Table S4). Starting at the lowest energy, the Goldstone mode (GM) of the skyrmion lattice ($n = 1$) is predicted to exist at energies similar to those of HM1. This is followed by the CCW mode ($n = 3$) at  $\sim 24\,\mu{\rm eV}$ and an excitation of the skyrmion lattice at $\sim 53\,\mu{\rm eV}$ that is too weak to be seen experimentally (the BM and CW modes are also too weak to be seen). The next two helimagnons of the conical phase (HM2 and HM3) are expected at $\sim64.5\,\mu{\rm eV}$ and  $\sim 72.5\,\mu{\rm eV}$.

As shown by the red line in Fig.\,\ref{fig_3}\,A, these predictions compare well with the three magnon modes that may be discerned in our data: (i) an excitation at $E_1=4\,\mu{\rm eV}$ that may be attributed to the GM and HM1 with a linewidth $\Gamma_1=0.55\, \mu{\rm eV}$, (ii) an excitation at $E_2=30\,\mu{\rm eV}$ that may be attributed to the CCW magnon with a linewidth $\Gamma_2=10\,\mu{\rm eV}$, and (iii) an excitation at $E_3=80\,\mu{\rm eV}$ that may be attributed to HM2 and HM3 with a linewidth $\Gamma_3=6\,\mu{\rm eV}$. Thus, the predicted energies of the most intense excitations are in excellent agreement with experiment. The GM is thereby topologically trivial ($C = 0$), whereas the CCW mode, $n = 3$, is topologically nontrivial with $C = +1$.


In contrast to the skyrmion lattice plane, the spectra for momentum transfers along the skyrmion tubes, ${\bf q}_\parallel$, are not sensitive to ${\bf B}_{\rm em}$, where typical polarized TAS intensities using setup 2 are shown in Fig.\,\ref{fig_4} (see Figs. S10 to S13 and S16 for further data and the associated NSF intensities, representing purely incoherent nuclear scattering). In excellent agreement with the calculated spectra shown in Fig.\,\ref{fig_4}, C1 and C2, the TAS exhibits pronounced maxima with substantial weight. A detailed inspection of the calculated spectra \cite{SOM} allows these maxima to be attributed to the CCW mode, previously identified at the zone center using microwaves \cite{Schwarze15}. These are conventional, topologically trivial, strongly dispersive spin wave branches.

Moreover, using setup 2, polarized TAS allowed us to discern pronounced individual modes that are clearly nonreciprocal [i.e., $E(-{\bf q}_\parallel) \neq E({\bf q}_\parallel)$]. Namely for wave vectors parallel and antiparallel to ${\bf H}$, the structure factor exhibited a pronounced nonreciprocity [i.e., $\mathcal{S}_{ij}({\bf q}, E) \neq \mathcal{S}_{ij}(- {\bf q}, E)$] already evident in Fig.\,\ref{fig_4}, C1 and C2. Here Fig.\,\ref{fig_4}, A1 and B1, display the same scans under inversion of momentum transfer for the same field orientation (the same applies to the scans shown in Fig.\,\ref{fig_4}, A2 and B2). Related data under field inversion are presented in Figs. S13 and S16. This well-known nonreciprocity originates in the combination of Dzyaloshinskii-Moriya and dipolar interactions as discussed in \cite{Seki:2020aa}; it is characteristic for chiral materials \cite{tokura_nonreciprocal_2018}. In MnSi it was previously observed in the field-polarized \cite{Grigoriev15, Sato16} and conical phases \cite{weber2018non}, as well as in the paramagnetic regime \cite{Roessli2002}.


The topological magnon bands in the skyrmion lattice of MnSi we report here are highly unusual in several ways. First, the connection between the nontrivial topological winding in real space and the topological magnon bands in reciprocal space may be quantitatively captured within a universal framework based on three material-specific parameters: the energy separating the conical and field-polarized state, $E_{c2}$; the helical wave vector,  $k_{\rm h}$; and the susceptibility in the conical state $\chi^{\rm int}_{\rm con}$ \cite{SOM}. Second, the Landau levels and topological band structure are observed starting at the lowest-lying states up to high energies. Third, we expect contributions to the thermal Hall effect, which, however, will not be quantized given the bosonic character of the magnons. Fourth, an intimate analogy exists between quantum Hall phenomena and the emergent Lorentz force, Landau levels, and the topological magnon band structure caused by the nontrivial real-space topology of skyrmion lattices in chiral magnets. On the basis of the bulk/boundary correspondence, our observations in the bulk imply the existence of chiral edge states that may be technologically useful -- for instance, as a directional coupler for quantum technologies.

\clearpage

\section*{Supplementary Materials}
\noindent Materials and Methods\\
\noindent Supplementary Text\\
\noindent Fig. S1 to S22\\
\noindent Tables S1 to S4\\
\noindent References (44-80)

\clearpage
\newpage


\vline
\newpage
\noindent \textbf{Acknowledgments}\\
We wish to thank the staff at the Institute Laue Langevin, Heinz Maier-Leibnitz Zentrum, ISIS Neutron and Muon Source, Paul Scherer Institut, and SNS Oak Ridge for support. We wish to acknowledge support by E. Villard, P. Chevalier, and J. Frank, as well as the instrumental control system by J. Locatelli. We are, moreover, grateful for the IT support at the ILL cluster computing facility. We thank M. Kugler for the initial implementation of the model.
A.B. and C.P. are supported through the DFG in the framework of TRR80 (project E1, project-id 107745057), SPP 2137 (Skyrmionics) under grant number PF393/19 (project-id 403191981), ERC Advanced Grants No. 291079 (TOPFIT) and 788031 (ExQuiSid), and Germany's excellence strategy EXC-2111 390814868.
R.G. is supported by the DFG under grant No. GE 971/5-1. 
M.G. is supported by the DFG in the framework of SFB 1143 (project A07; project-id 247310070), as well as grant numbers GA 1072/5-1 (project-id 270344603) and GA 1072/6-1 (project-id 324327023). 
Work by D.M.F. and M.J. was supported by the LANL Directed Research and Development (LDRD)  program via the Directed Research (DR) project ``A New Approach to Mesoscale Functionality: Emergent Tunable Superlattices (20150082DR)''. 
Early work by M.G. and J.W. on the spin wave theory was supported by the Institute for Materials Science at Los Alamos (RR2015).
This research used resources at the  Spallation Neutron Source, a DOE Office of Science User Facility operated by the Oak Ridge National Laboratory.
The data sets collected at LET have the DOIs: 10.5286/ISIS.E.RB1620412 and 10.5286/ISIS.E.RB1720033.
The data sets collected at the ILL have the DOIs 10.5291/ILL-DATA.INTER-413, 10.5291/ILL-DATA.INTER-436, 10.5291/ILL-DATA.INTER-477, 10.5291/ILL-DATA.4-01-1597, and 10.\allowbreak5291/ILL-DATA.4-01-1621. 
Data recorded at RESEDA were collected under proposal P00745-01. 
Additional data sets (not shown) were collected at MIRA (proposals 13511 and 15633) MLZ, Garching, Germany, TASP (proposals 20181324 and 20151888) at PSI, Villigen, Switzerland, and CNCS at SNS Oak Ridge, USA.\\

\noindent \textbf{Author contributions}\\
T.W., D.M.F., and M.J. proposed and designed the ToF and TAS experiments.
T.W. proposed and designed the polarized TAS experiments.
C.P. proposed the MIEZE experiment.
A.B. grew the single crystal denoted sample 1.
T.W., D.M.F., L.B., C.F, H.G., and M.J. performed the experiments.
P.S., G.S.T., M.B., R.B., D.V., C.F, M.S., and R.G. supported the experiments at neutron spectrometers under their responsibility.
T.W., C.P., P.B., M.J., and M.G. conceived the interpretation.
J.W., and M.G. performed the theoretical calculations.
T.W., D.M.F, L.B., H.G., and C.F. performed the data analysis.
T.W. and M.G. implemented the theoretical skyrmion model for use with TAS resolution convolution.
T.W., C.P., P.B., and M.G. wrote the manuscript with input from all authors.
All authors discussed the results and reviewed the manuscript.\\

\noindent \textbf{Competing interests}\\
The authors declare that they have no competing interests. \\

\noindent \textbf{Data and materials availability} \\
Computer code for the spin wave simulations used in this paper may be found at the repository given in \cite{simulation_doi}. All data needed to evaluate the conclusions in the paper are present in the paper and/or the Supplementary Materials and available at the repository given in \cite{data_doi}. 

\clearpage

\clearpage

\clearpage

\clearpage

\newpage
\clearpage

\renewcommand\refname{Supplementary References and Notes}
\renewcommand{\theequation}{S\arabic{equation}}
\renewcommand{\thefigure}{S\arabic{figure}}
\renewcommand{\thetable}{S\arabic{table}}

\section*{\centering{Supplementary Materials}}
\newpage

\section{Materials and Methods}

\noindent
\textbf{Sample:}
Two large high-quality single-crystal MnSi were used. Both samples had a cylindrical shape. The cylinder axis of the samples was oriented along an $[001]$ direction. Further details are reported in Sec.\,\ref{sec:samples}. 

\noindent
\textbf{Neutron spectroscopy:}
A tabular summary of all neutron scattering data recorded is presented in Sec.\,\ref{sec:summary}. Time-of-flight inelastic (ToF) neutron scattering was carried out at the LET spectrometer at ISIS, Didcot, UK, and the CNCS spectrometer at the SNS at Oak Ridge National Laboratory, USA. All ToF measurements used unpolarized neutrons and were carried out with setup 2 as described in the text and in Sec.\,\ref{sec:tof}. Triple-axis spectroscopy (TAS) was carried out on three different beam-lines as described in Sec.\,\ref{sec:tas}: (i) ThALES at the Institut Laue-Langevin in Grenoble, France, (ii) MIRA at the Maier Leibniz Zentrum in Garching, Germany, and (iii) TASP at the Paul Scherrer Institut in Villigen, Switzerland. For the TAS spectroscopy setup 1 was used as described in the main text and Sec.\,\ref{sec:tas:geom}. Measurements at MIRA and TASP used unpolarized neutrons and served dominantly to establish the precise positions of the magnon spectra in reciprocal space suitable for detailed polarized measurements. ThALES was set up for longitudinal polarization analysis using the $(111)$ reflection of a polarizing Heusler monochromator and analyzer. Neutron spin-echo spectroscopy of the skyrmion dynamics was performed by means of the Modulation of IntEnsity with Zero Effort (MIEZE) setup at the instrument RESEDA at MLZ as described in Sec.\,\ref{sec:mieze}. MIEZE represents essentially an ultra-high resolution time-of-flight technique, permitting straight-forward studies under strong magnetic fields as compared with conventional neutron spin-echo spectroscopy.

\noindent
\textbf{Theoretical analysis tools:}
The magnon spectra and magnetic response tensor were calculated within linear spin wave theory as summarized in Sec.\,\ref{sec:theo}. The simulations were for a left-handed skyrmion lattice. For the determination of the excitation spectra over the wide range of excitation energies and momentum transfers addressed in our study, a theoretical framework had to be set up that allowed to identify the ground state of an incommensurate, long wavelength, multi-$k$ structure. This is not possible within present-day numerical implementations of linear spin wave theory, such as SpinW \cite{Toth_2015}. The dynamical structure factor and the associated spectral weight of typical momentum transfers accessible in neutron scattering were calculated up to very large numbers of magnetic Brillouin zones, as the modulation length is very large and the first Brillouin zone tiny. For the data analysis the dynamical structure factor was convoluted with the instrumental resolution function using the triple-axis software \textit{Takin}. For the calculations the theoretical model was implemented in C++ as a plug-in module extending \textit{Takin}. The numerical calculations required excessive computing power. They were performed on the cluster of the ILL.  More information on the procedures, as well as the full source code of all software tools, is given in Ref.\,\cite{simulation_doi}.

\clearpage
\newpage

\section{Further aspects of the experimental methods }
\label{sec:samples}

\subsection{Sample preparation and characterization}

\begin{figure}[b]
	\begin{center}
	\begin{minipage}[t]{0.03\textwidth}
		\vspace{0cm}
		\bf{A}
	\end{minipage}
	\begin{minipage}[t]{0.25\textwidth}
		\vspace{0cm}
		\includegraphics[height=0.25\textheight]{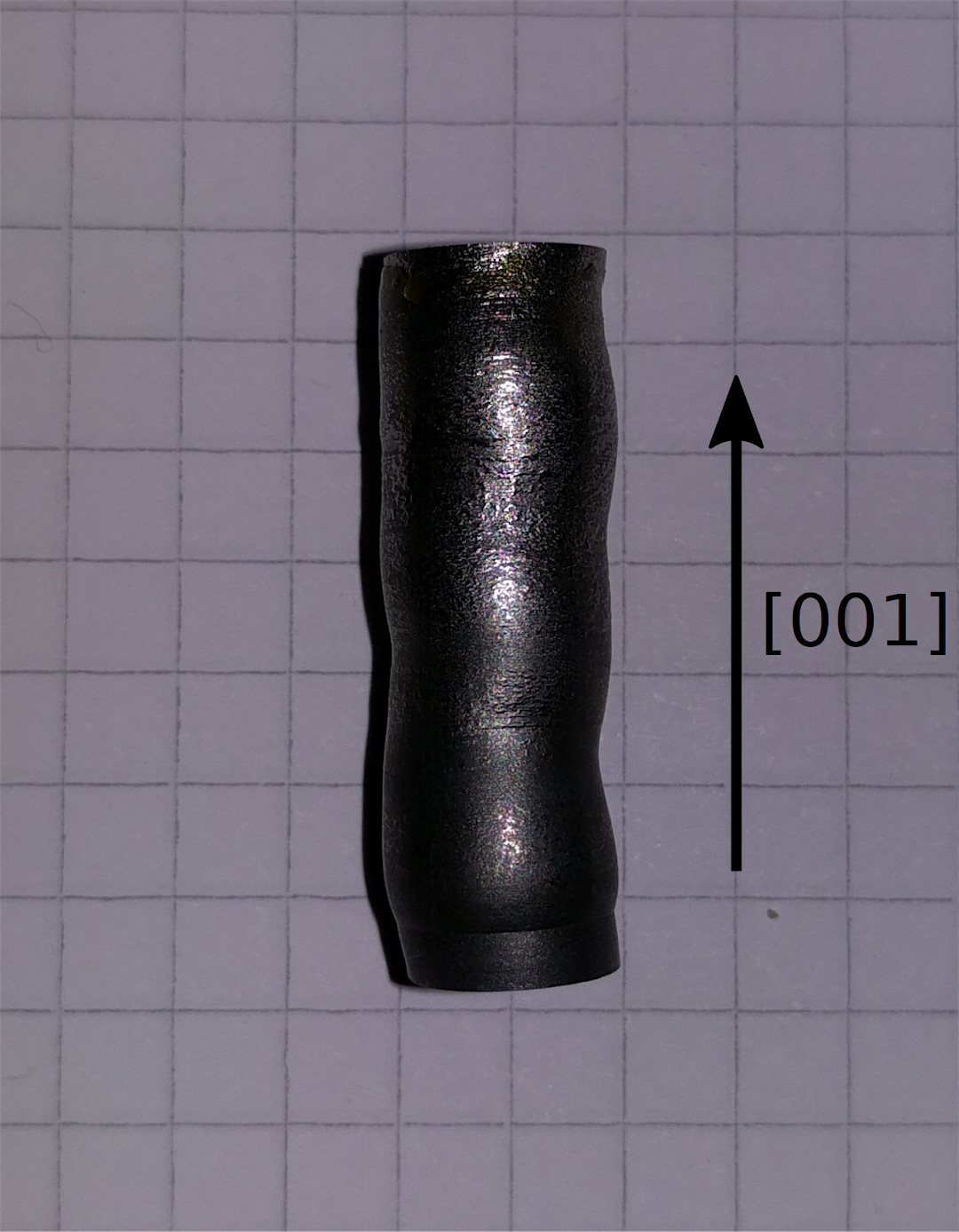}
	\end{minipage}
	\hspace{1cm}
	\begin{minipage}[t]{0.03\textwidth}
		\vspace{0cm}
		\bf{B}
	\end{minipage}
	\begin{minipage}[t]{0.25\textwidth}
		\vspace{0cm}
		\includegraphics[height=0.25\textheight]{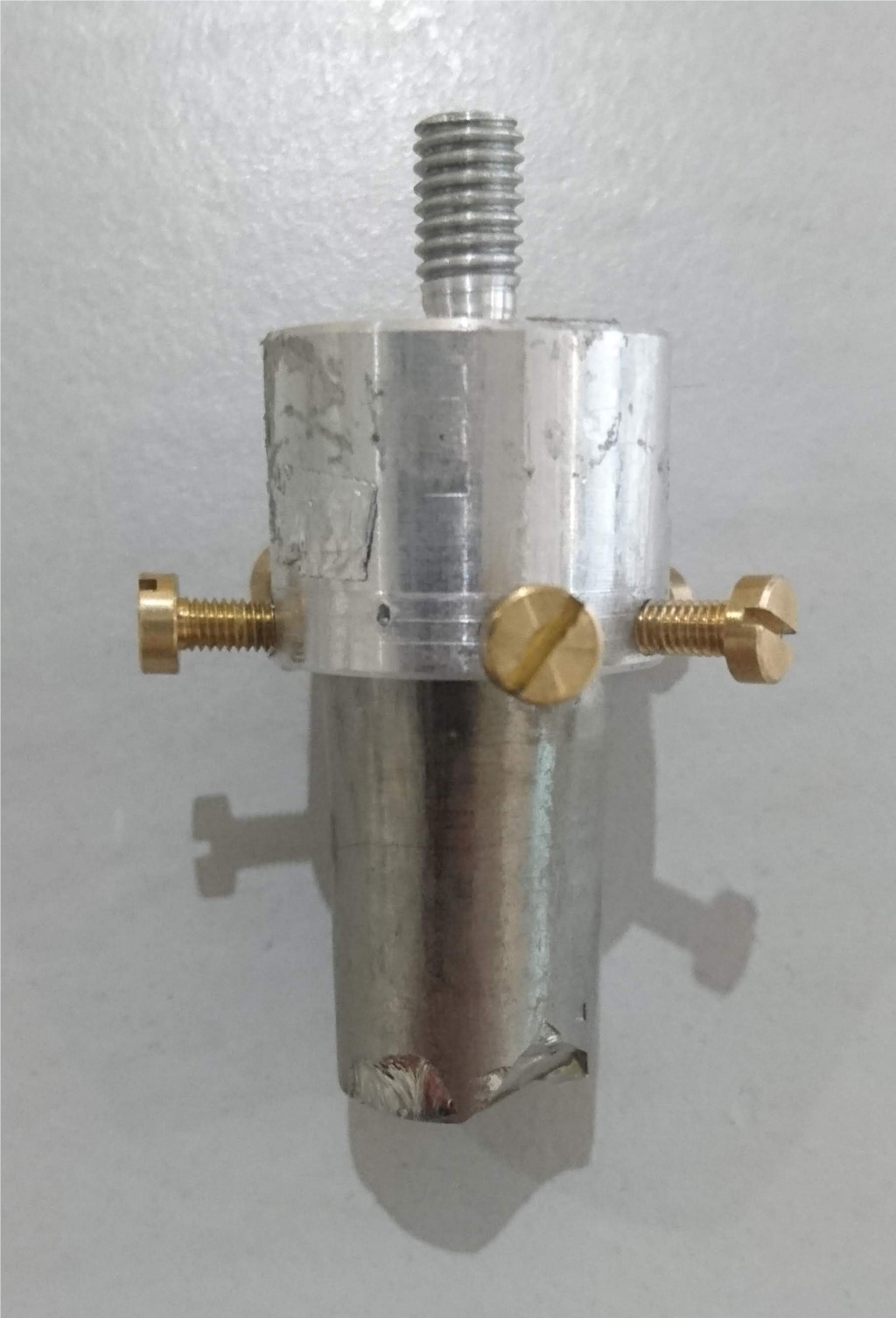}
	\end{minipage}
	\end{center}
	\caption{MnSi single crystals investigated in this study. In both crystals the cylinder axis was oriented along a $[001]$ direction. (A) Single-crystal used for triple-axis (TAS) and time-of-flight (ToF) spectroscopy denoted sample 1. (B) Single-crystal studied by means of longitudinal neutron resonance spin echo spectroscopy using Modulation of Intensity with Zero Effort (MIEZE) denoted sample 2 \cite{2019Franzb,Reseda_JPSJ}. 
\label{fig:mnsi}}
\end{figure}

Two single crystals of MnSi were used in our inelastic neutron scattering studies. For the triple-axis and the time-of-flight measurements, the same cylindrical single-crystal  was used as investigated in previous studies \cite{Bauer2010, Weber2017Field}. This crystal is denoted as sample 1 and shown in Fig.\,\ref{fig:mnsi}\,(A). The crystal has a diameter and a height of $d \approx 10\, \mathrm{mm}$ and $h \approx 30\, \mathrm{mm}$, respectively, with the cylinder axis oriented along the $[001]$ direction. It was grown by optical float-zoning under UHV-compatible conditions \cite{Bauer2010}.

For the experiments by means of longitudinal neutron resonance spin echo spectroscopy using Modulation of Intensity with Zero Effort (MIEZE) \cite{2019Franzb,Reseda_JPSJ} a cylindrical single-crystal of similar dimensions than sample 1 was used, where the diameter varied between 15.4\,mm and 16\,mm and the length between 26\,mm and 30.5\,mm. This crystal is denoted as sample 2 and shown in Fig.\,\ref{fig:mnsi}\,B. It was grown by the Czochralsky technique.

The elastic neutron scattering signal of these samples, as well as the magnetization and ac susceptibility measured on small pieces taken from sample 1 and sample 2 confirmed transition temperatures and characteristics of the magnetic phase diagram in excellent agreement with the literature \cite{Bauer2010}. Laue x-ray and neutron diffraction confirmed a high sample quality, where the structural mosaic spread of sample 1 and sample 2 were found to be $\sim$15 and $\sim$30 minutes, respectively. 

The key parameters of MnSi that define the energy and momentum range and resolution required to resolve the predicted excitation spectra in the skyrmion lattice at a temperature of 28.5\,K are summarized in Tab.\,\ref{tab:values}. It is important to emphasize the tiny absolute values of the magnetic modulation length, $k_{h}$, and associated energy scale, $E_{c2}$. To resolve the predicted magnon spectra with respect to these scales by, say, a factor of $10^{-2}$ implies, in principle, the need for a momentum resolution better $\sim 4\cdot10^{-4}\,\text{\AA}^{-1}$ and an energy resolution better  $\sim0.3\,\mu\mathrm{eV}$.


\begin{table}[hb]
\begin{center}
\begin{tabular}{|l|l|}
	\hline lattice constant & $a=4.558$ $\text{\AA}$ \tabularnewline
	\hline helix pitch in $\text{\AA}^{-1}$ & $k_{h}=0.039$ $\text{\AA}^{-1}$ \tabularnewline
	\hline helix pitch in rlu & $k_{h}^{rlu}=k_{h}\cdot\frac{a}{2\pi}=0.028$ rlu \tabularnewline
	\hline upper critical field & $\mu_0 H^{\rm int}_{\rm c2}\approx0.322\ \text{\ensuremath{\mathrm{T}}}$ \tabularnewline
	\hline energy scale & $E_{c2}=g\mu_{B}\mu_0 H^{\rm int}_{\rm c2}=37.3\,\mu\mathrm{eV}$ \tabularnewline
	\hline 
\end{tabular}
\end{center}
\caption{Key parameters of MnSi defining the momentum and energy ranges of the spectroscopic properties of the skyrmion lattice at a temperature of 28.5\,K addressed in our study, where $g=2$ and $\mu_{B}=0.0579\ \mathrm{\frac{meV}{T}}$.}
\label{tab:values}
\end{table}


For the geometries of both samples, which are similar in size, the demagnetization factors for magnetic fields applied parallel and perpendicular to the cylinder axis are \cite{Sato89}:
\begin{equation}
	N_{\parallel} \approx 0.13
	\hspace{0.3cm} \mathrm{and} \hspace{0.3cm} 
	N_{\perp} 	\approx 0.44,
\end{equation}
respectively. Further, for the cylindrical shape of the single crystals studied the demagnetizing fields are not uniform. In part of the parameter range of the skyrmion lattice, this causes a well-understood coexistence of a small volume fraction of conical phase with the skyrmion lattice phase as noticed in previous studies \cite{Muehl09}. The volume fraction of conical phase is absent when the sample shape is chosen such that the demagnetizing fields are uniform \cite{2012_Adams_PRL, adams_response_2018}.  Using a triple-axis spectrometer we confirmed that a small volume fraction of conical phase was present, when the field and temperature values were chosen such that the intensity of the skyrmion lattice were strongest. Furthermore, these tests revealed that it was also possible to choose the field and temperature such that there was no parasitic conical phase, cf. Sec.\,\ref{sec:conis}.

Special care was exercised to keep track of the magnons attributed to the fraction of conical phase, if present at all. During the time-of-flight experiments at LET and CNCS the scattering geometry was insensitive to contributions by the conical phase. 
All TAS measurements were carried out under conditions, where no conical phase was present. Namely, in the triple-axis studies at ThALES the scattering intensity at the positions of the satellites of the conical phase were measured and found to be vanishingly small. Further, the elastic scattering intensity of the skyrmion lattice peaks and their projections were measured between inelastic scans, to confirm that their intensities had remained constant. This confirmed that neither a drift in temperature nor applied field occurred during the measurements and a clear separation of the two phases could be guaranteed during the triple-axis measurements. 

In contrast, for the small-angle scattering geometry used effectively at RESEDA the elastic signal contribution of the conical phase could not be checked independently. Moreover, in order to maximize the inelastic signal of the skyrmion lattice the temperature and field were chosen such that the elastic signal of the skyrmion phase were at a maximum. At these temperatures and fields the triple axis studies had shown the presence of the small volume fraction of conical phase.  Fortunately, the energies of the magnon excitations of the skyrmion lattice and the conical phase were separated very well, permitting to distinguish them unambiguously as explained in the main text and in section \ref{sec:conis}. In fact, the spin wave excitations of the conical phase validated the observations of magnons  of the skyrmion lattice.

Unfortunately, an estimate of the volume fraction of conical phase based on the size of the inelastic signal recorded at RESEDA would require a careful integration of spectral weight well-beyond our data.  Even though the signal of the helimagnons of the conical phase appears to be comparatively large, the volume fraction of conical phase may be quite small. A similar effect was recently observed in a study of the paramagnetic to skyrmion lattice transition in MnSi, combining small angle neutron scattering with neutron spin echo spectroscopy, microwave spectroscopy and ac susceptibility measurements \cite{2019_Kindervater_PRX}.  

\subsection{Sample temperature}

For the clarification of the questions addressed in our paper, we optimized the sample temperature such that either the skyrmion lattice could be recorded without the presence of a volume fraction of conical phase or the scattering by the skyrmion lattice was maximal. Keeping this in mind, the temperatures stated for the different data sets shown in our manuscript correspond to the temperatures recorded with the actual set-up used for the measurements. No correction was applied to establish a precise correspondence of different cryogenic systems, where absolute differences may be as large as a Kelvin.

On this note differences of the temperatures stated in the figures reflect that the experimental studies reported in our manuscript comprise measurements at five different neutron scattering centres using six different beam-lines, i. e., a large number of different cryogenic systems were used. On the scale of the temperature range of the skyrmion lattice phase in MnSi for magnetic field along the $[100]$ direction, which is about 1.5\,K wide corresponding to a few percent of the absolute value of the temperature, it seemed, neither necessary nor feasible to correct for comparatively small absolute differences of temperature values arising from the calibration of thermometers and differences of the cryogenic systems used. 

\newpage
\clearpage

\section{Summary of neutron scattering data}
\label{sec:summary}

Shown in Tabs.\,\ref{tab:summary-1} and \ref{tab:summary-2} is a summary of the most important inelastic neutron scattering scans performed as part of this study. As summarized above, data were recorded at six different instruments located at five different neutron scattering centres. Denoted as "scan" in the first column is the label used in the figures to highlight the parameter range (gray box) at which data was recorded. Also listed are the neutron scattering instruments, the type of scans and polarity of the applied magnetic field, the momentum transfer, and the energy range studied. The column labelled "figure" refers to the figure numbers in the main text and the supplement showing these data. Further information on the different scans are stated in the last column labelled "comments".

\begin{table}[b]
\begin{center}
\begin{tabular}{|c|c|c|c|c|c|c|}
\hline
scan & instrument & type & $Q$ (rlu) & $E$ (meV) & figure & comments\tabularnewline
\hline
\hline
& LET &  &  &  & 2 & full mapping\tabularnewline
&  &  &  &  & S3 to S6 & \tabularnewline
\hline
\hline
(i) & ThALES & $q_{\perp}$, $H_{int}>0$ & $\left(1.072\ 0.928\ 0\right)$ & $\left[-1,\ 1\right]$ & 2 & \tabularnewline
\hline
(ii) & ThALES & $q_{\perp}$, $H_{int}<0$ & $\left(1.068\ 0.932\ 0\right)$ & $\left[-0.8,\ 0.8\right]$ & S17 & \tabularnewline
\hline
(iii) & ThALES & $q_{\perp}$, $H_{int}>0$ & $\left(0.94\ 1.06\ 0\right)$ & $\left[-0.5,\ 0.5\right]$ & 2 & \tabularnewline
\hline
(v) & ThALES & $q_{\parallel}$, $H_{int}>0$ & $\left(1.068\ 1.068\ 0\right)$ & $\left[-0.3,\ 0.8\right]$ & 4 & \tabularnewline
\hline
(vi) & ThALES & $q_{\parallel}$, $H_{int}>0$ & $\left(0.932\ 0.932\ 0\right)$ & $\left[-0.4,\ 1.1\right]$ & 4 & \tabularnewline
\hline
(vii) & ThALES & $q_{\parallel}$, $H_{int}<0$ & $\left(1.056\ 1.056\ 0\right)$ & $\left[-0.7,\ 0.8\right]$ & S16 & \tabularnewline
\hline
(viii) & ThALES & $q_{\parallel}$, $H_{int}<0$ & $\left(0.932\ 0.932\ 0\right)$ & $\left[-0.4,\ 1.1\right]$ & S16 & \tabularnewline
\hline
& ThALES & $q_{\perp}$, $H_{int}>0$ & $\left(1.06\ 0.94\ 0\right)$ & $\left[-0.1,\ 0.5\right]$ & S10 & high resol.\tabularnewline
\hline
& ThALES & $q_{\parallel}$, $H_{int}>0$ & $\left(0.939\ 0.939\ 0\right)$ & $\left[-0.4,\ 1.1\right]$ & S12 & \tabularnewline
\hline
& ThALES & $q_{\parallel}$, $H_{int}>0$ & $\left(0.94\ 0.94\ 0\right)$ & $\left[-0.4,\ 1\right]$ & S11 & \tabularnewline
\hline
& ThALES & $q_{\parallel}$, $H_{int}>0$ & $\left(1.06\ 1.06\ 0\right)$ & $\left[-0.5,\ 0.5\right]$ & S11 & \tabularnewline
\hline
& ThALES & $q_{\parallel}$, $H_{int}>0$ & $\left(1.06\ 1.06\ 0\right)$ & $\left[-0.7,\ 0.7\right]$ & S10 & high resol.\tabularnewline
\hline
\hline
(iv) & RESEDA & $q_{\perp}$, $H_{int}>0$ & $0.0123$ & $\left[-0.1,\ 0.1\right]$ & 3 & ludicr. resol.\tabularnewline
\hline
\end{tabular}
\end{center}
\caption{
Summary of inelastic neutron scattering scans at the time-of-flight spectrometer LET, the triple-axis-spectrometer ThALES, and the resonant neutron spin-echo spectrometer RESEDA. Not listed are scans used for setting up the actual measurements. See also Tab.\,\ref{tab:summary-2} }
\label{tab:summary-1}
\end{table}

\begin{table}
\begin{center}
\begin{tabular}{|c|c|c|c|c|c|c|}
\hline
scan & instrument & type & $Q$ (rlu) & $E$ (meV) & figure & comments\tabularnewline
\hline
\hline
& CNCS &  &  &  & --- & two grains \tabularnewline
\hline
\hline
& MIRA & $q_{\perp}$, $H_{int}<0$ & $\left(1.06\ 1.06\ 0\right)$ & $\left[-0.6,\ 0.6\right]$ & --- & $Q\perp H_{int}$\tabularnewline
\hline
& MIRA & $q_{\perp}$, $H_{int}<0$ & $\left(0.94\ 0.94\ 0\right)$ & $\left[-0.6,\ 0.6\right]$ & --- & $Q\perp H_{int}$\tabularnewline
\hline
& MIRA & $q_{\perp}$, $H_{int}>0$ & $\left(0.94\ 0.94\ 0\right)$ & $\left[-0.7,\ 0.2\right]$ & --- & $Q\perp H_{int}$\tabularnewline
&  &  &  &  &  & high resol.\tabularnewline
\hline
& MIRA & $q_{\perp}$, $H_{int}>0$ & $\left(0.935\ 0.935\ 0\right)$ & $\left[-1,\ 0.3\right]$ & --- & $Q\perp H_{int}$\tabularnewline
&  &  &  &  &  & high resol.\tabularnewline
\hline
& MIRA & $q_{\parallel}$, $H_{int}>0$ & $\left(1.0675\ 1.0675\ 0\right)$ & $\left[-0.6,\ 0.6\right]$ & --- & \tabularnewline
\hline
& MIRA & $q_{\parallel}$, $H_{int}<0$ & $\left(1.0675\ 1.0675\ 0\right)$ & $\left[-0.6,\ 0.6\right]$ & --- & \tabularnewline
\hline
& MIRA & $q_{\parallel}$, $H_{int}>0$ & $\left(0.9325\ 0.9325\ 0\right)$ & $\left[-0.7,\ 0.7\right]$ & --- & \tabularnewline
\hline
\hline
& TASP & $q_{\parallel}$, $H_{int}>0$ & $\left(0.92\ 1.08\ 0\right)$ & $\left[-0.4,\ 0.4\right]$ & --- & $Q\perp H_{int}$\tabularnewline
\hline
& TASP & $q_{\parallel}$, $H_{int}<0$ & $\left(0.92\ 1.08\ 0\right)$ & $\left[-0.4,\ 0.4\right]$ & --- & $Q\perp H_{int}$\tabularnewline
\hline
& TASP & $q_{\parallel}$, $H_{int}>0$ & $\left(0.93\ 1.07\ 0\right)$ & $\left[-0.4,\ 0.4\right]$ & --- & $Q\perp H_{int}$\tabularnewline
\hline
& TASP & $q_{\parallel}$, $H_{int}<0$ & $\left(0.93\ 1.07\ 0\right)$ & $\left[-0.4,\ 0.4\right]$ & --- & $Q\perp H_{int}$\tabularnewline
\hline
& TASP & $q_{\perp}$, $H_{int}>0$ & $\left(1.068\ -0.932\ 0\right)$ & $\left[-0.2,\ 0.6\right]$ & --- & \tabularnewline
\hline
& TASP & $q_{\perp}$, $H_{int}>0$ & $\left(0.932\ 1.068\ 0\right)$ & $\left[-0.4,\ 0.8\right]$ & --- & \tabularnewline
\hline
& TASP & $q_{\perp}$, $H_{int}>0$ & $\left(0.932\ -1.068\ 0\right)$ & $\left[-0.5,\ 0.5\right]$ & --- & \tabularnewline
\hline
& TASP & $q_{\parallel}$, $H_{int}>0$ & $\left(0.944\ 0.944\ 0\right)$ & $\left[-0.4,\ 0.8\right]$ & --- & \tabularnewline
\hline
& TASP & $q_{\parallel}$, $H_{int}<0$ & $\left(0.944\ 0.944\ 0\right)$ & $\left[-0.4,\ 0.8\right]$ & --- & \tabularnewline
\hline
& TASP & $H_{int}<0$ & $\left(0.89\ 1\ 0\right)$ & $\left[-0.4,\ 0.8\right]$ & --- & \tabularnewline
\hline
& TASP & $H_{int}>0$ & $\left(0.89\ 1\ 0\right)$ & $\left[-0.4,\ 0.8\right]$ & --- & \tabularnewline
\hline
& TASP & $H_{int}>0$ & $\left(0.9038\ 1\ 0\right)$ & $\left[-0.4,\ 0.8\right]$ & --- & \tabularnewline
\hline
\end{tabular}
\end{center}
\caption{
Summary of inelastic neutron scattering scans at the time-of-flight spectrometer CNCS, and the triple-axis-spectrometers TASP and MIRA. Not listed are scans used for setting up the actual measurements. See also Tab.\,\ref{tab:summary-1} }
\label{tab:summary-2}
\end{table}

\newpage
\clearpage

\section{Time-of-flight neutron spectroscopy}
\label{sec:tof}

Time-of-flight spectroscopy was carried out with two main objectives in mind. First, to obtain surveys of the excitation spectra in the skyrmion lattice phase as described in Sec.\,\ref{sec:ToF:scatt}. Second, to normalize the inelastic scattering intensity quantitatively as described in Sec.\,\ref{sec:ToF:calib}. 

\subsection{Scattering geometry and supplementary data
\label{sec:ToF:scatt}}

Time-of-flight inelastic neutron scattering was carried out at the LET spectrometer at ISIS, Didcot, UK \cite{LET}, and the CNCS spectrometer at the SNS at Oak Ridge National Laboratory, USA \cite{CNCS}. All measurements used unpolarized neutrons and were carried out with a setup as described below. Unfortunately the beam-time at CNCS did not produce useful scientific data, as the sample only after the experiment was found to contain two large crystallographic grains (neither sample 1 nor sample 2 were available at CNCS). However, the measurements at CNCS clearly demonstrated, that the required high momentum resolution may be achieved in time-of-flight spectroscopy, motivating the measurements at LET.

\begin{figure*}[thb]
	\begin{centering}
	\includegraphics[width=0.75\textwidth]{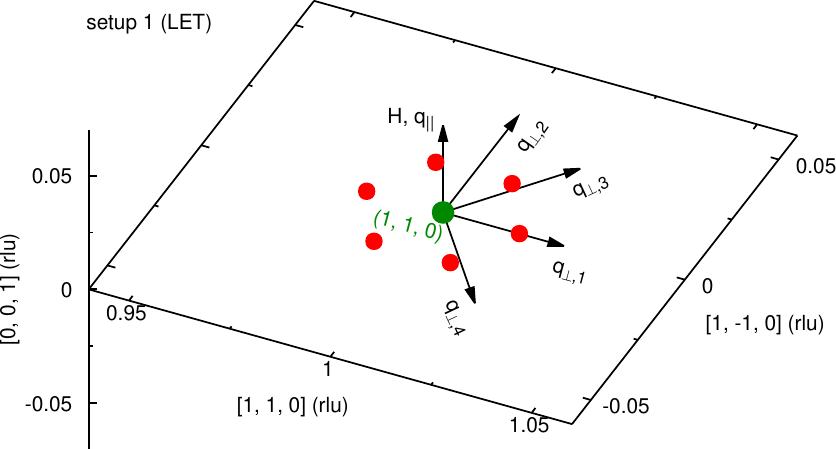}
	\caption{Qualitative depiction of setup 1 conceived for the time-of-flight measurements (see Fig.\,\ref{fig:scanpos3d} for setup 2 used in triple-axis scattering and Fig.\,\ref{fig:reseda_schematic} for setup 3 used in MIEZE spectroscopy). The green circle indicates the $(110)$ nuclear Bragg peak. It is surrounded by the six magnetic satellites of the skyrmion lattice (red circles) as stabilized by a magnetic field ${\bf H}$ perpendicular to the  $(hk0)$-plane along the vertical $[001]$ direction. Measurements were conducted in the $(hk0)$ scattering plane. Data shown in Fig.\,2\,(A1) of the main text displays the behaviour for inelastic scattering for $\bm{q}_{\perp, 1}$ along $[110]$.}
	\label{fig:scanpos3d-let}
	\end{centering}
\end{figure*}

The experimental setup used for the time-of-flight studies is shown in Fig.\,\ref{fig:scanpos3d-let}. We refer to this scattering geometry as setup 1. Data were recorded in the skyrmion lattice phase of MnSi around the $(110)$ nuclear Bragg peak in an applied magnetic field directed along the vertical $[001]$ axis. In the experiments at LET a magnetic field $\mu_0 H = 180\,{\rm mT}$ was applied at a sample temperature $T = 28.1\,{\rm K}$. For these measurements the spectrometer was set up with an incident neutron energy of $E_i =  3.2\,{\rm meV}$. This corresponds to an energy resolution of $80\,\mu{\rm eV}$ at the elastic line.

The locations of the traces in momentum space along $[110]$ and $[1\bar{1}0]$ are denoted $\bm{q}_{\perp, 1}$ and $\bm{q}_{\perp, 2}$, respectively. While $\bm{q}_{\perp, 1}$ cuts across a skyrmion lattice peak, $\bm{q}_{\perp, 2}$ is located between two skyrmion lattice peaks. Additional scans along $[100]$ and $[010]$ are denoted $\bm{q}_{\perp, 3}$ and $\bm{q}_{\perp, 4}$, respectively. The experimental data recorded at LET for $\bm{q}_{\perp, 1}$ are shown in Fig.\,2 in the main text.  

\begin{figure}[htb]
	\begin{center}
	\includegraphics[width=0.95\textwidth]{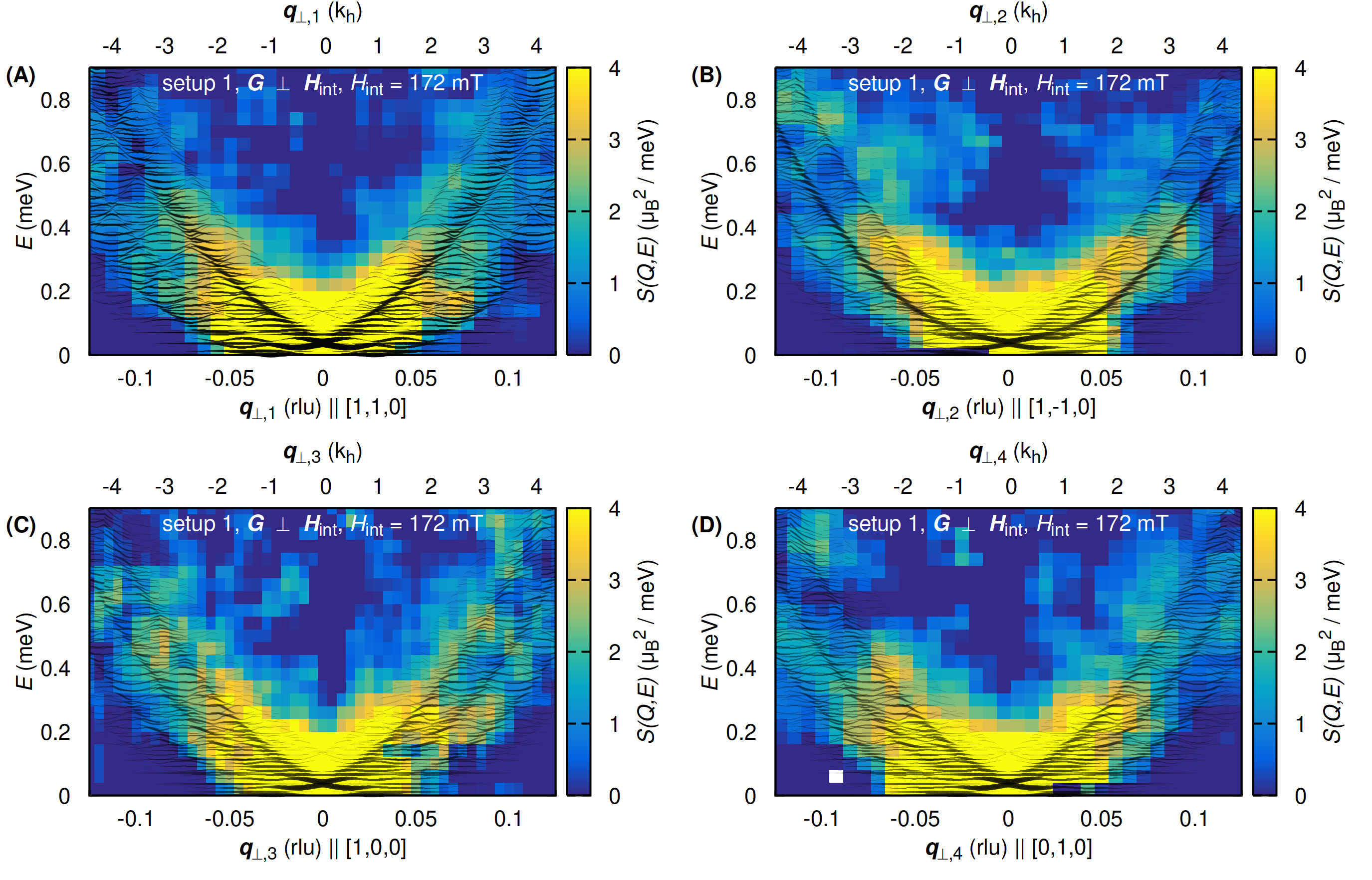}
	\end{center}
	\caption{Contour maps of inelastic data as measured at LET for momentum transfers $\bm{q}_{\perp,i}$ with $i = 1, 2, 3, 4$ along $[110]$, $[1\bar{1}0]$, $[100]$ and $[010]$, respectively. Data for $i=1$ as defined in Fig.\,\ref{fig:scanpos3d-let} are shown in Fig.\,2\,(A1) in the main text. The maps display reciprocal positions $\bm{q}_{\perp}$, which are within the skyrmion plane, i.e., in the plane perpendicular to the applied magnetic field $\bm{H}$. In all panels black lines represent the magnetic response tensor predicted theoretically, where the line-thickness denotes the calculated spectral weight (cf. Sec.\,\ref{sec:theo}). 
	\label{fig:let_exp}}
\end{figure}

Further slices of the data recorded at LET that are complementary to those presented in Fig.\,2 of the main text are shown in Fig.\,\ref{fig:let_exp}. They depict normalized contour maps of the $(\bm{q}_{\perp}, E)$-plane as obtained by slicing the full four-dimensional $S\left( \bm{Q}, E \right)$ data sets differently. Data shown in Fig.\,\ref{fig:let_exp} display the behaviour for directions $\bm{q}_{\perp, 2}$ along $[1\bar{1}0]$, as well as $\bm{q}_{\perp, 3}$ and $\bm{q}_{\perp, 4}$ along $[100]$ and $[010]$, respectively. The typical binning of the transverse momentum $\bm{q}_{\perp}$ and energy was $\triangle q = 0.005\,{\rm rlu}$ and $\triangle E = 0.035\,{\rm meV}$, respectively. The data was integrated for momenta $\bm{q}_\parallel$ in the range from -0.015 to 0.015 rlu. 

\begin{figure}[htb]
	\begin{center}
	\includegraphics[width=0.95\textwidth]{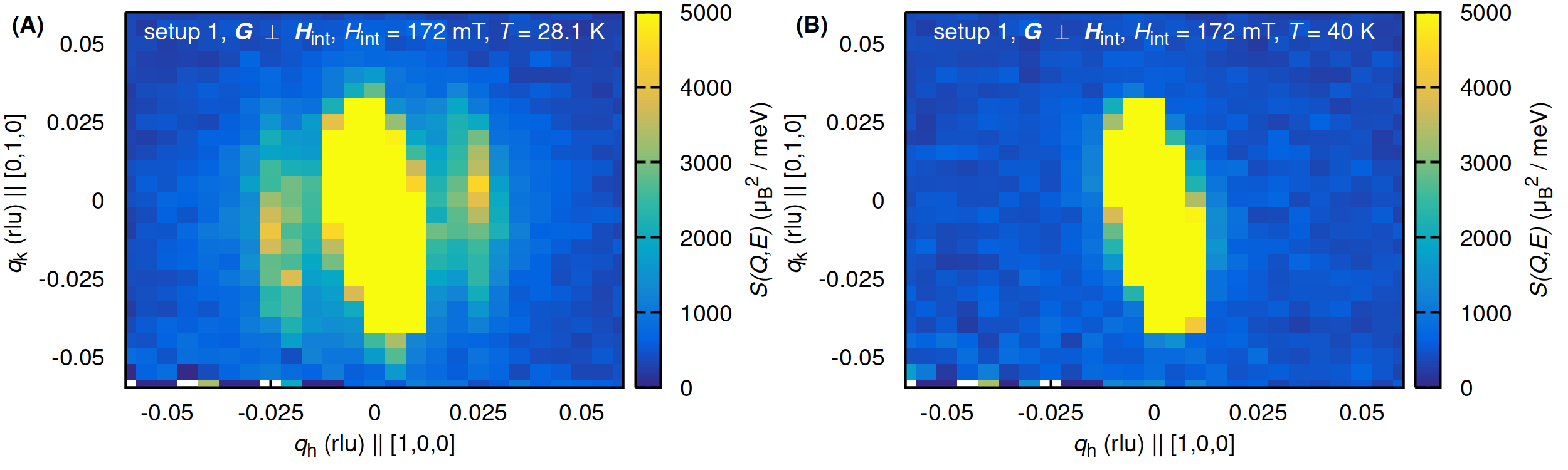}
	\end{center}
	\caption{Constant-energy slices for $E = 0$ and two different temperatures as a function of momentum in the skyrmion plane, i.e. perpendicular to the applied magnetic field $\bm{H}$. Data were determined by integrating over an energy interval $E = [-0.1 \ldots +0.1]\,{\rm meV}$ and wave vector interval $q_l = [-0.075 \ldots +0.075]\,{\rm rlu}$. (A) Constant-energy slice for $E = 0$ in the skyrmion lattice phase at 28.1\,K. The magnetic satellites are clearly visible. (B) Constant-energy slice for $E = 0$ in the paramagnetic state at 40\,K. Only the nuclear (110) Bragg peak is visible reflecting the momentum resolution.}
\label{fig:let_exp_Q}
\end{figure}

\begin{figure}[htb]
	\begin{center}
	\includegraphics[width=0.55\textwidth]{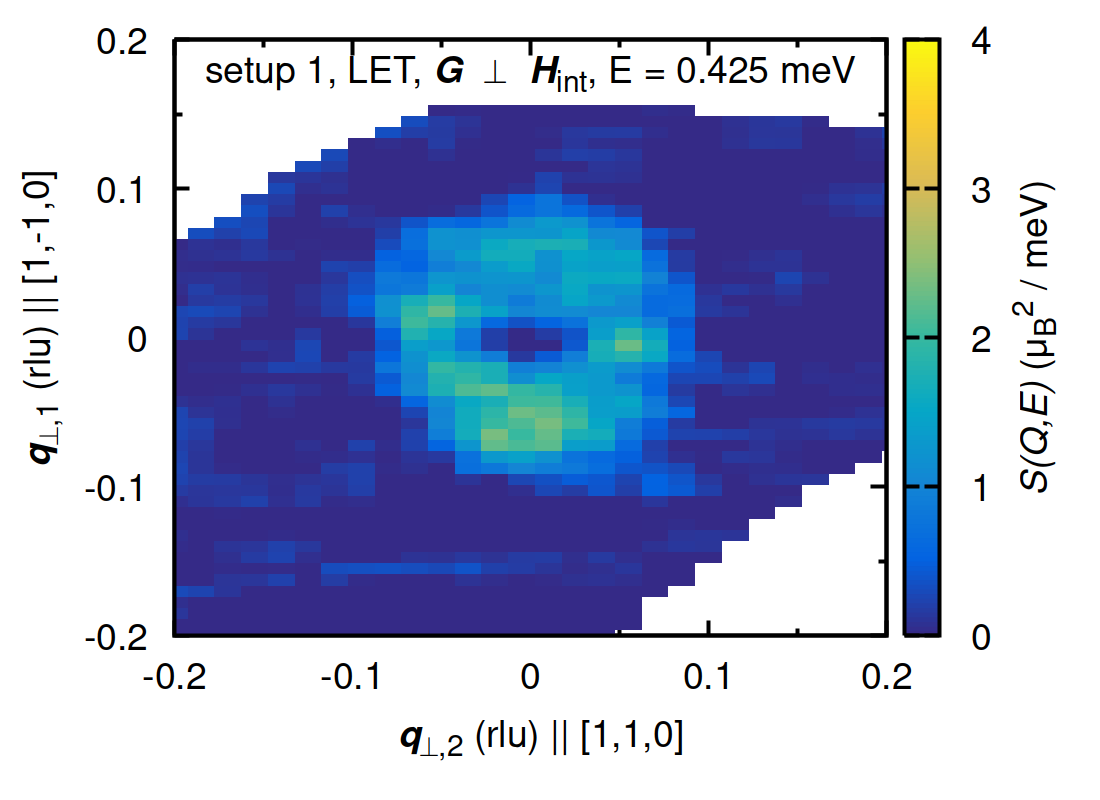}
	\end{center}
	\caption{Constant-energy slice for $E = 0.425$\,meV as a function of momenta in the skyrmion plane, i.e. perpendicular to the applied magnetic field $\bm{H}$. The data was obtained by means of an integration about the energy interval $E = [0.4 \ldots 0.45]$ meV.}
	\label{fig:let_exp_E}
\end{figure}

We note, that the time-of-fight measurements also permitted to infer the behavior at constant energy. Examples are shown in Fig.\,\ref{fig:let_exp_Q} and Fig.\,\ref{fig:let_exp_E}, which display constant-energy slices as a function of $\bm{q}_{\perp, 1}$ and $\bm{q}_{\perp, 2}$. Fig.\,\ref{fig:let_exp_Q} displays a constant energy slice at $E=0$ in the skyrmion phase at 28.1\,K [panel \ref{fig:let_exp_Q}\,(A)] and in the paramagnetic state at 40\,K [panel \ref{fig:let_exp_Q}\,(B)]. These data were determined by integrating over an energy interval $E = [-0.1 \ldots +0.1]\,{\rm meV}$ and a wave vector interval $q_l  = [-0.075 \ldots +0.075]\,{\rm rlu}$. The binning in $q_h$ and $q_k$ is $0.005\,{\rm rlu}$. The scattering intensity associated with the nuclear (110) Bragg peak reflects the high momentum resolution. At 28.1\,K the presence of the magnetic Bragg peaks of the skyrmion lattice can be resolved. This underscores recent advances in state-of-the-art time-of-flight neutron spectrometers. Fig.\,\ref{fig:let_exp_E} displays a constant energy slice at $E = 0.425$\,meV, obtained by integrating the data in the interval $E = [0.4 \ldots 0.45]\,{\rm meV}$.

\subsection{Normalization of the scattering intensity
\label{sec:ToF:calib}}

To normalize the magnetic scattering for a quantitative comparison with theory, the data were corrected for non-magnetic contributions such as coherent and incoherent nuclear scattering, as well as spurious effects and multiple scattering due to the sample holder and the cryomagnet following the well-established method described in Ref.\,\cite{Janoscheke15}. For this procedure, four data sets were collected. First, we determined the scattering intensity $I_{\rm SkL, 28.1\,K}$ of the sample at $T = 28.1\,{\rm K}$ within the skyrmion phase. Second, a reference data set, $I_{\rm para, 40\,K}$, was recorded at $T = 40\,{\rm K}$ in the paramagnetic phase well above the ordering temperature to isolate the nuclear scattering. Following this, two more background data sets were collected at the same temperatures as the two data sets with sample, however, using an empty sample holder, i.e., $I_{\rm back, 28.1K}$ and $I_{\rm back, 40K}$, respectively. Using these four data sets the magnetic contribution was obtained as
\begin{equation}
	I_{\rm mag} = \left(I_{\rm SkL, 28.1\,K} - I_{\rm back, 28.1\,K} \right) - \left(I_{\rm para, 40\,K} - I_{\rm back, 40\,K} \right).
\end{equation}
Prior to the subtraction, all data sets were corrected for the Bose factor. 

The correction procedure we used takes into account that certain parts of the cryostat may be at a different temperature than the sample. Thus, corrections for temperature dependent effects of the cryostat and the sample have to be carried out independently.

The dynamical structure factor was normalized by means of the known scattering intensity of a vanadium sample of 5.84 g measured under the same instrumental conditions. The normalized data  was smoothed using the smooth function in \textit{Horace}, which is based on a hat function and three bins on either side of the point of reference. The data reduction and absolute normalization were performed using the software package \textit{Mantid}  \cite{Mantid}. For the data treatment the \textit{Horace} software \cite{Horace} was used. 

In principle, the normalized magnetic time-of-flight scattering determined this way may be used to normalize also the triple-axis data, keeping in mind that the quantitative uncertainty of the normalization of the time-of-flight data is $\sim 30\,\%$ \cite{Janoscheke15}. In turn, the normalization of the time-of-flight data paves the way towards a parameter-free comparison with theory. However, the scattering intensity depends sensitively on the size of the ordered moment. As the skyrmion phase is located at the border of the paramagnetic state, the ordered moment  may differ by a factor of two between the different measurements. Thus, in order to gauge the scattering intensity appropriately, in principle, the sample temperature must be tracked very carefully beyond the accuracy of the standard experimental set-up provided by the sample environment available in our studies. Keeping these constraints in mind, we confirmed that our data are consistent with a parameter-free comparison between theory and experiment.


\subsection{Alternative comparison of the time-of-flight data with theory}

\begin{figure}[htb]
	\begin{center}
	\includegraphics[width=0.95\textwidth]{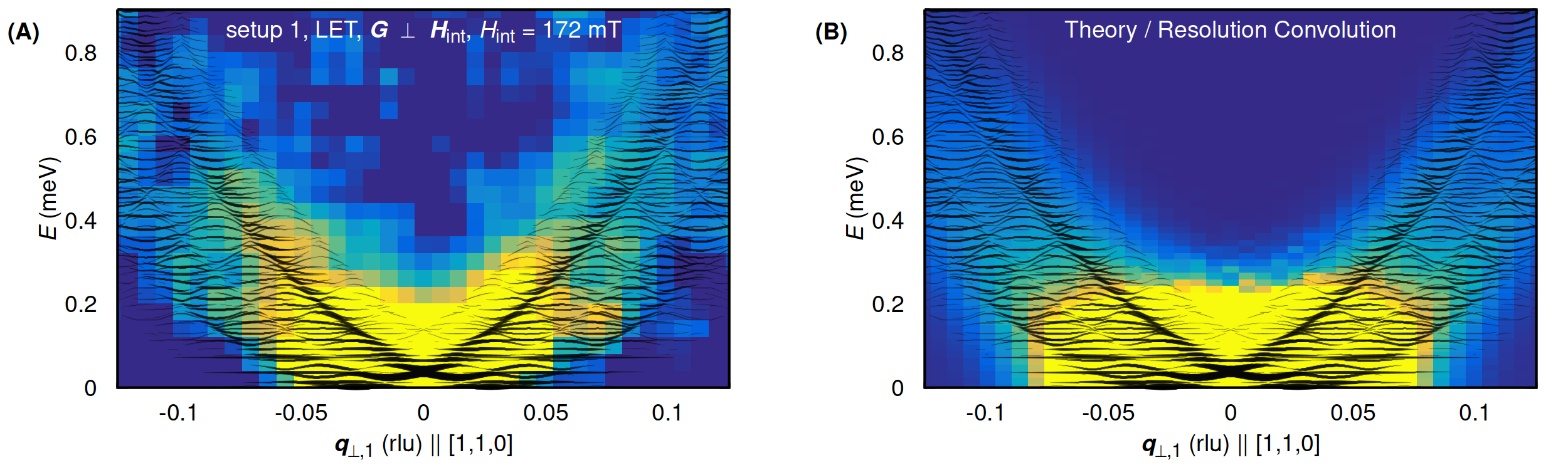}
	\end{center}
	\caption{Comparison of the experimentally determined time-of-flight data as depicted by means of a color scale and, shown in panel (A), with the calculated dynamical structure factor as convoluted with the instrumental resolution and depicted on the same color scale, as shown in panel (B). Panel (A) is identical to Fig.\,4\,(A) in the main text. In both panels black lines represent the magnetic response tensor predicted theoretically, where the line-thickness denotes the calculated spectral weight.}
	\label{fig:let_convo_110}
\end{figure}

For an alternative graphical comparison of the calculated spectral weight (thin black lines in Fig.\,\ref{fig:let_exp}) with the experimental time-of-flight data, we convoluted the theoretical results with the instrumental resolution function of LET to produce a color map. This approach compares with the treatment of the triple-axis data described below. An example is shown in Fig.\,\ref{fig:let_convo_110}, where we used the software package \textit{Takin}, \cite{takin_2, takin_1, Takin2017, Takin2016} and its implementation of the time-of-flight resolution calculation method of Violini, \textit{et al.} \cite{Violini2013}. 

Reproduced in Fig.\,\ref{fig:let_convo_110}\,(A) is Fig.\,2\,(A1) of the main text, showing the experimental data in the form of an intensity map. Also shown are the predicted spectral weights of the transverse excitations (thin black lines), ignoring the possible presence of longitudinal contributions attributed to the quasi-elastic scattering seen in triple axis spectroscopy as explained below. For comparison shown in Fig.\,\ref{fig:let_convo_110}\,(B) is the calculated intensity map as inferred from the theoretically predicted spectral weights, i.e., the thin black lines, as convoluted by the resolution function of LET. 

All experimental observations are well reproduced in the theoretically predicted intensity map. No malign resolution-based effects, e.g., apparent shifts in energy or linewidth anomalies, are observed in the convoluted spectra.  However, it is important to emphasize that the color scales in panels (A) and (B) cannot be compared in detail, as the Violini method used for the calculation of the instrumental resolution represents a fairly crude approximation which does not account for instrument-specific details such as the double chopper system of LET. In addition, Violini does not feature the scaling factor $R_0$ and one needs to refer to the bare resolution volume instead.

\newpage
\clearpage


\section{Polarized neutron triple-axis spectroscopy}
\label{sec:tas}

Polarized neutron triple-axis spectroscopy (TAS) was performed in order to obtain direct spectroscopic evidence of the broad density of magnon modes perpendicular to the skyrmion lattice plane and its polarization dependence as compared with the well-defined individual excitations parallel to the skyrmion lattice. In addition these measurements served to provide evidence for the non-reciprocal character of the excitation spectra. Polarized triple-axis measurements allow to separate magnetic and nuclear scattering unambiguously. Finally, these measurements offer a straightforward way to rule out the presence of spurious scattering, this way generating data sets amenable for a critical comparison with theory.

\subsection{Scattering geometry and instrumental set-up}
\label{sec:tas:geom}

Triple-axis spectroscopy was carried out on three different beam-lines: (i) ThALES \cite{thales} at the Institut Laue-Langevin in Grenoble, France, (ii) MIRA \cite{MIRAnew} at the Maier Leibniz Zentrum in Garching, Germany, and (iii) TASP \cite{Tasp96} at the Paul Scherrer Institut in Villigen, Switzerland. Measurements at MIRA and TASP used unpolarized neutrons and served predominantly to establish the precise positions of the magnon spectra in reciprocal space suitable for detailed polarized measurements. This concerned, in particular, the identification of magnons featuring sufficient scattering intensity in comparison with the background, as well as discrimination of spurious scattering. Extensive inelastic polarized scattering measurements were performed at ThALES. 

\begin{figure*}[ht]
	\begin{centering}
	\includegraphics[width=0.85\textwidth]{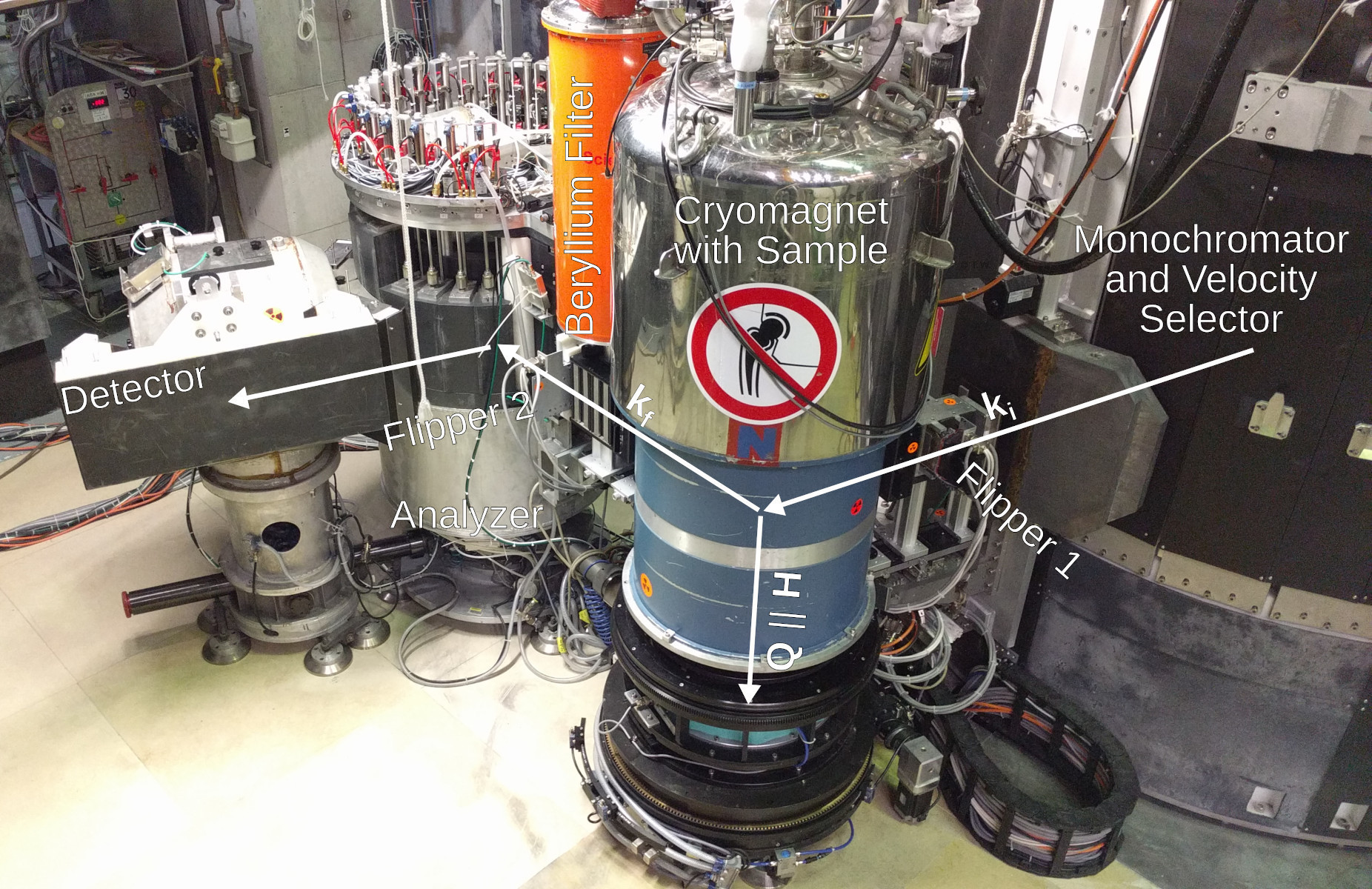}
	\caption{
	The triple-axis spectrometer ThALES at the ILL. The polarization of the incident and scattered neutrons was analyzed using Heusler single crystals. The polarization of the neutrons was maintained by guide fields. The superconducting magnet provided a horizontal magnetic field at the sample position.}
	\label{fig:thales}
	\end{centering}
\end{figure*}

The main body of the triple axis measurements were performed using a similar setup, focussing on the vicinity of the nuclear $(110)$ Bragg peak with the magnetic field applied either along $[001]$, $[110]$, or $[1\bar{1}0]$.
ThALES was set up for longitudinal polarization analysis using the $(111)$ reflection of a polarizing Heusler monochromator and analyzer \cite{Moon69}. A photograph of the instrument is shown in Fig.\,\ref{fig:thales}. Spin flippers were placed before and after the sample. The incident neutron beam was focused on the sample using variable horizontal and fixed vertical focussing. Higher order neutrons were removed by means of a neutron velocity selector and, when needed, by a Beryllium filter. Scattered neutrons with a fixed energy of either $E_{f,1} = 3.5\ \mathrm{meV}$ or $E_{f,2} = 4.06\ \mathrm{meV}$ were selected with a vertically focussing analyzer. 

Unless stated otherwise, all of the data at ThALES were recorded with a configuration: "open - open - sample - 30' - open". Exceptions are shown in Fig.\,\ref{fig:elast}\,(C) and (D) and Fig.\,\ref{fig:thales_exp3_highres}, which were recorded with a configuration: "open - 30' - sample - 30' - open". We note that at ThALES it is, in general, not possible to collimate the beam additionally before the sample as the advanced focussing system will cause a collapse of the intensity otherwise. 

\begin{figure*}[ht]
	\begin{centering}
	\includegraphics[width=0.75\textwidth]{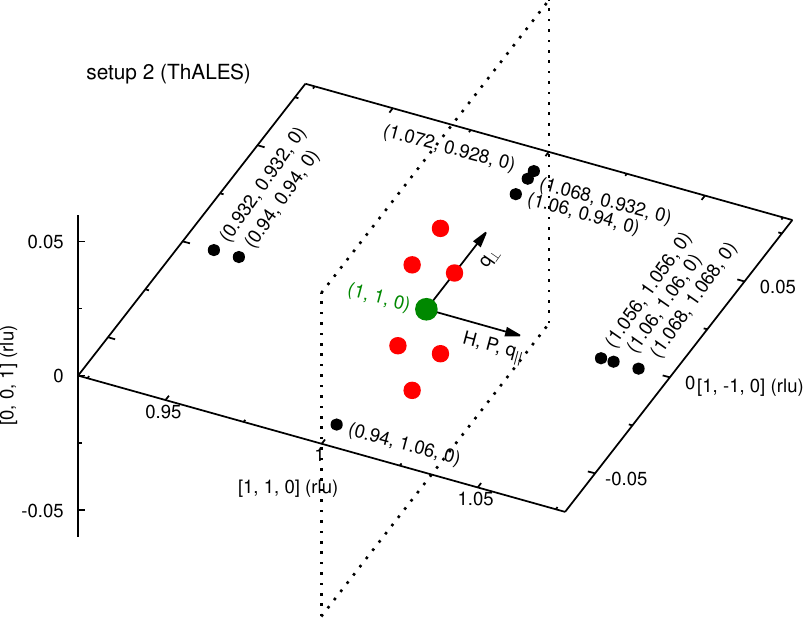}
	\caption{Qualitative depiction of setup 2 conceived for the polarized triple-axis measurements. The green circle indicates the $(110)$ nuclear Bragg peak. Measurements were conducted in the $(hk0)$ scattering plane. The horizontal magnetic field was applied along the $[110]$ direction, stabilizing the skyrmion lattice in a plane perpendicular to $[110]$ as indicated by the dotted rhomboid. The six magnetic satellites associated with the skyrmion lattice in the plane perpendicular to $[110]$ are shown in red. Magnetic anisotropies align the skyrmion lattice such that one of the magnetic Bragg peaks is oriented along $\left[ 1\bar{1}0 \right]$. The positions in reciprocal space, where energy scans were carried out, are marked in black. }
	\label{fig:scanpos3d}
	\end{centering}
\end{figure*}

\begin{figure}[h]
	\begin{centering}
	\includegraphics[width=0.49\textwidth]{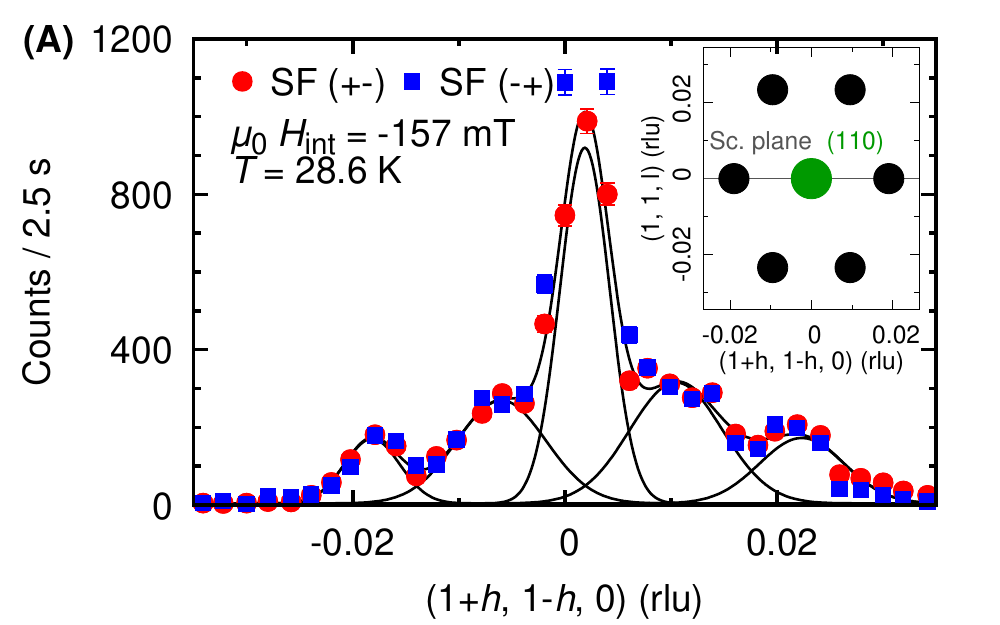}
	\includegraphics[width=0.49\textwidth]{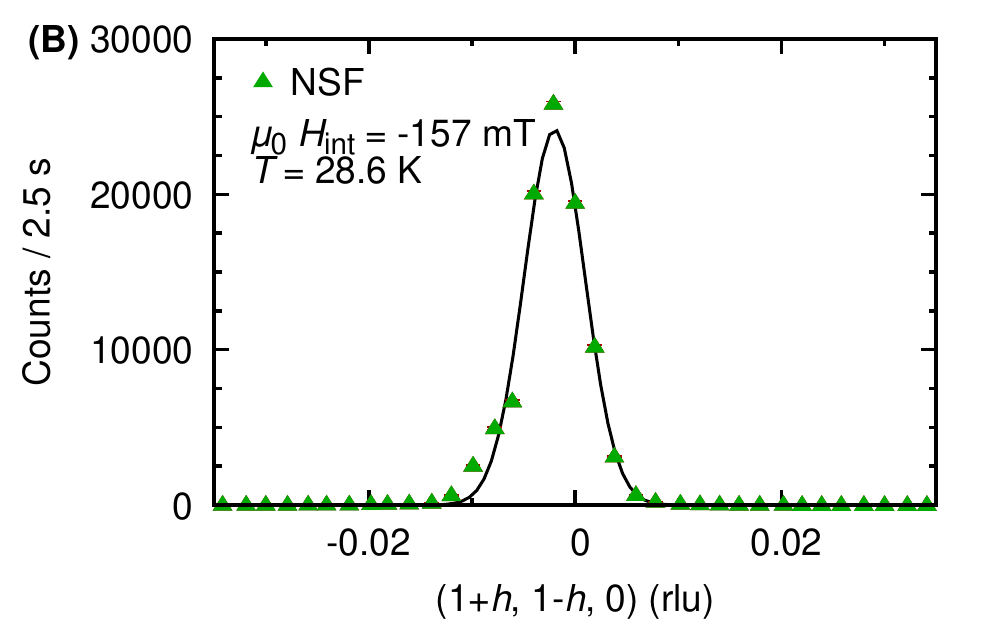}
	\includegraphics[width=0.49\textwidth]{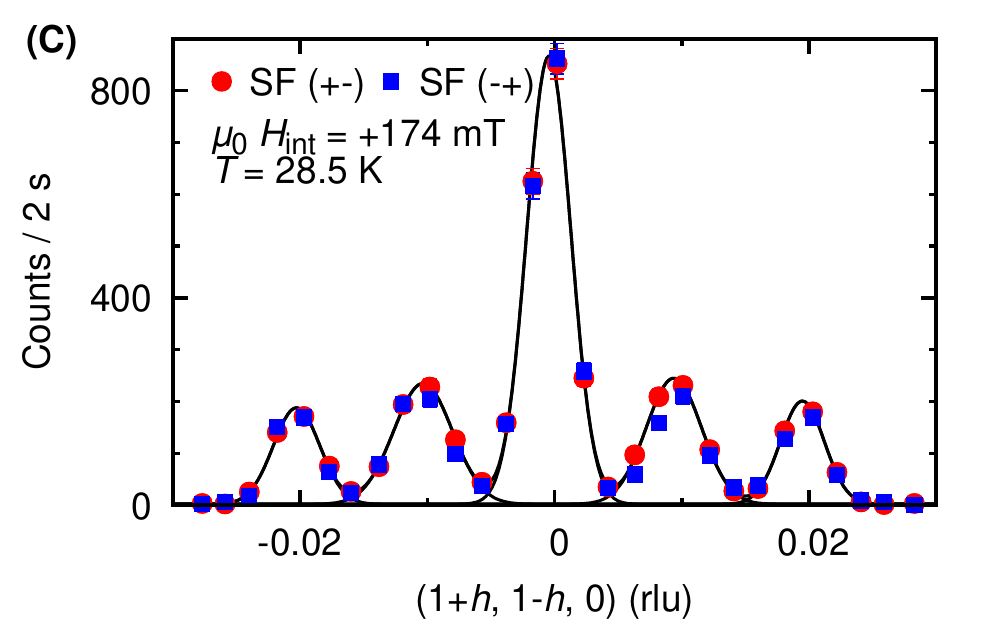}
	\includegraphics[width=0.49\textwidth]{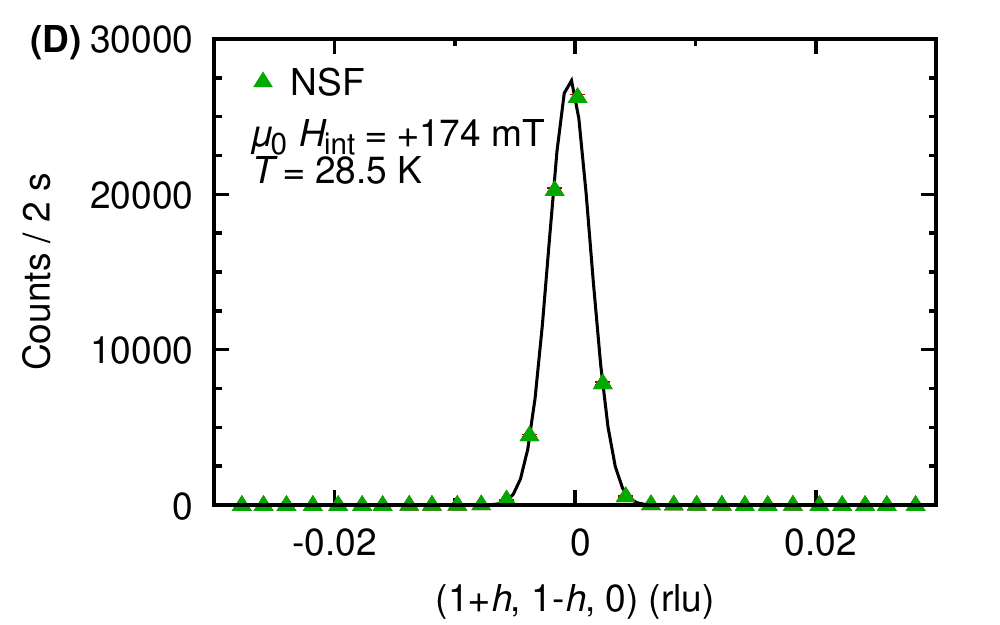}
	\caption{Elastic scans along the $\left[ 1\bar{1}0 \right]$ direction. The inset of panel (A) depicts the reciprocal lattice of the skyrmion crystal as oriented perpendicular to the scattering plane. The skyrmion lattice plane is defined by the $\left[ 1\bar{1}0 \right]$ and $\left[ 001 \right]$ directions with one of its reciprocal lattice vectors pointing along $\left[ 1\bar{1} 0 \right]$. Panels (A) and (C): Scattering by the two magnetic skyrmion satellites and projections of the four out-of-plane satellites may be distinguished in the spin-flip channels "SF (+-)" and "SF (-+)", where the latter reflects the coarse vertical resolution. The non-spin-flip channel, "NSF", exhibits purely nuclear $\left( 110 \right)$ scattering as shown in panels (B) and (D). Data shown in panels (C) and (D) were recorded with a 30 minute collimators before and after the sample providing very high resolution.}
	\label{fig:elast}
	\end{centering}
\end{figure}

We refer to the scattering geometry described in the following and shown in Fig.\,\ref{fig:scanpos3d} as setup 2. It was dominantly used for the triple-axis studies at ThALES, MIRA and TASP.  The key idea motivating setup 2 concerned, that the magnetic field, $\bm{H}$, the neutron polarization $\bm{P}$ and the total momentum transfer $\bm{Q}=\bm{G}+\bm{q}$ were parallel to each other. This way the origin of the spin-flip (SF) and non-spin-flip (NSF) scattering were decoupled, such that the SF and the NSF scattering were purely magnetic and nuclear, respectively. 

In setup 2 measurements were performed in the $(hk0)$ scattering plane in the vicinity of the $\bm{G} = \left( 110 \right)$ nuclear Bragg reflection. A magnetic field $\bm H_{[110]}$ was applied within the scattering plane along the $\left[ 110 \right]$ direction \cite{magnet}. The applied field defined the direction of the polarization $\bm P$ ($P \simeq91\%$) of the neutrons at the sample position, and, stabilized the skyrmion lattice within the plane perpendicular to $\bm H_{[110]}$, thus spanning a skyrmion plane defined by the $\left[ 1 \bar{1} 0 \right]$ and $\left[ 001 \right]$ high-symmetry directions of the nuclear lattice as shown in Fig.\,\ref{fig:scanpos3d} and in the inset of Fig.\,\ref{fig:elast}. The static structure factor of the hexagonal skyrmion lattice was dominated by six Bragg peaks at wavevectors with a magnitude $k_{\rm SkL} = 0.027$ rlu $= 0.037$ \AA$^{-1}$. As a consequence, the skyrmion lattice is described well in terms of a superposition of three helices \cite{adams_response_2018}.

In setup 2 a reciprocal lattice vector of the skyrmion lattice is expected to point in the $\left[ 1\bar{1}0 \right]$ direction \cite{adams_response_2018}.  This orientation of the skyrmion lattice was confirmed by elastic scans along the $\left[ 1 \bar{1} 0 \right]$ direction where typical data are shown in Fig.\,\ref{fig:elast}. The two magnetic satellites within the scattering plane gave rise to scattering in the spin-flip channels as shown in Fig.\,\ref{fig:elast}\,(A) and \ref{fig:elast}\,(C). Due to the coarse vertical momentum resolution scattering by the four out-of-plane satellite peaks was also picked up [cf. inset of Fig.\,\ref{fig:elast}\,(A)]. 

The additional scattering intensity at zero momentum transfer, seen in Fig.\,\ref{fig:elast}\,(A) and (C), represented a 'spill-over' of the nuclear NSF scattering that originates in the finite flipping ratio of the polarizers. It is important to emphasize that this spill-over was vanishingly small in the inelastic scans reported in our manuscript. It is helpful to note that the superposition of single-handed helices forming the skyrmion lattice leads to spin-flip scattering ``SF (+-)'' and ``SF (-+)'' at identical positions in reciprocal space. The strong non spin-flip-peak (NSF) is purely nuclear and due to the $(110)$ nuclear Bragg reflection, where the corresponding data are shown in Fig.\,\ref{fig:elast}\,(B) and (C).


\subsection{Complementary triple-axis data}
\label{sec:tas:supp}

\begin{figure}[t]
	\begin{center}
	\includegraphics[width=0.49\textwidth]{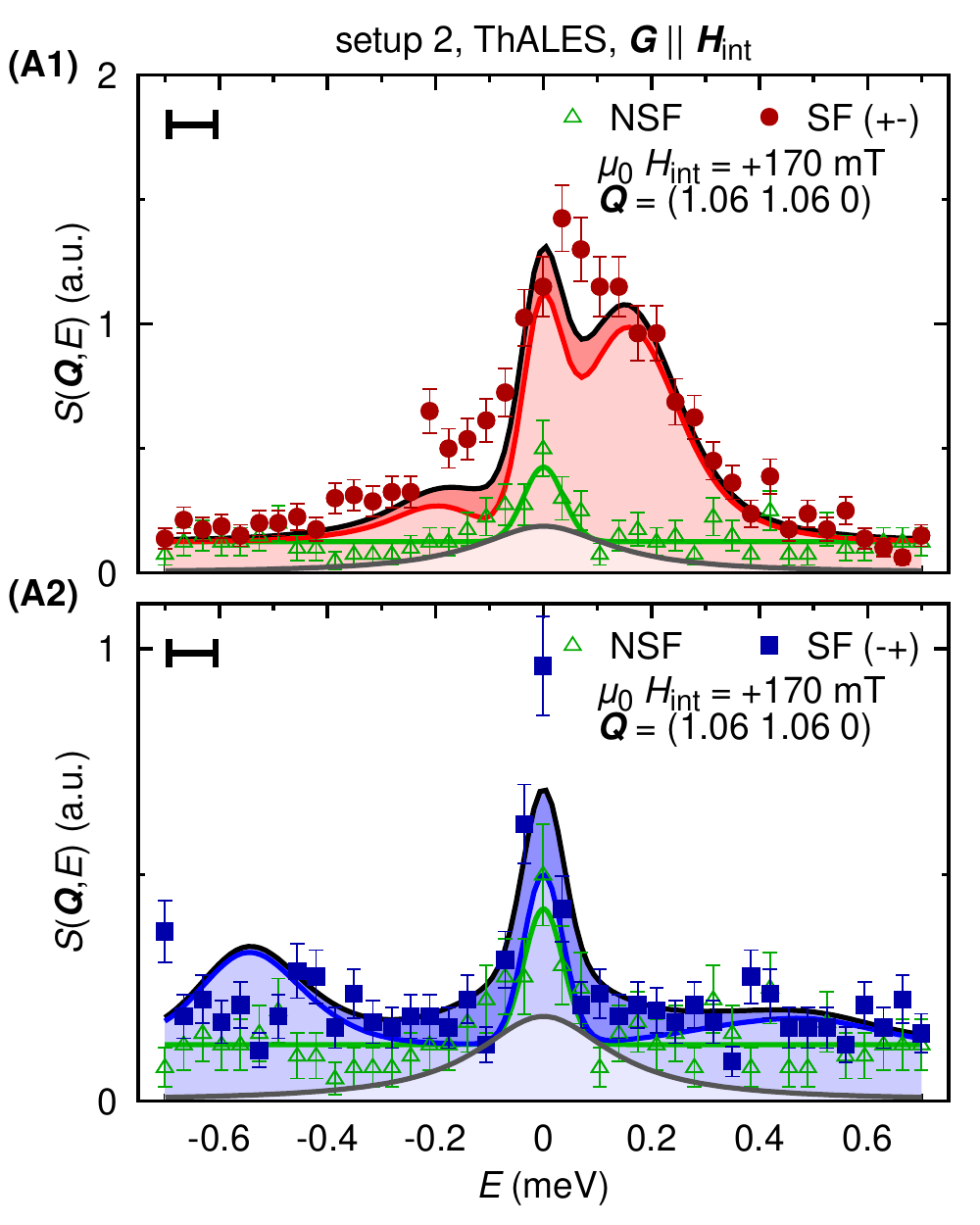}
	\includegraphics[width=0.49\textwidth]{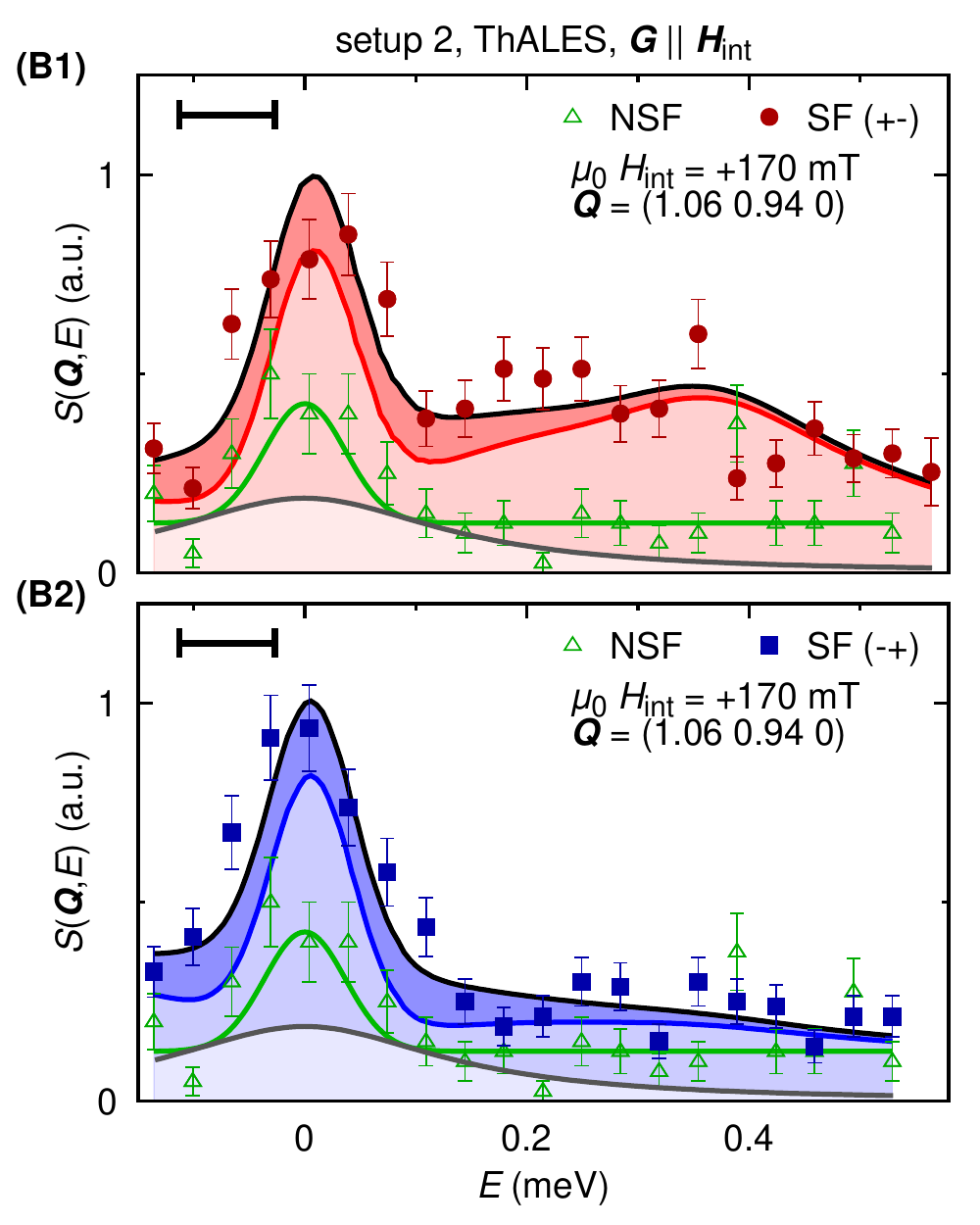}
	\end{center}
	\caption{High-resolution data of magnon spectra in the skyrmion lattice phase of MnSi. The resolution was improved by means of collimators (30 min) before and after the sample. Panels (A1), (A2) and (B1), (B2) show data for momentum transfers $\bm{q}_{\parallel}$ parallel to the reciprocal lattice vector $\bm{G} = \left(110\right)$ and $\bm{q}_{\perp}$ perpendicular to $\bm{G}$, respectively. Solid lines show the results of the model as convoluted with the resolution function of ThALES. The thick solid bars shown at the top left corner of each panel indicate the energy resolution for the energy range shown here. The gray shaded areas represent a quasi-elastic contribution that may be attributed to longitudinal fluctuations of the skyrmion lattice (see Secs.\,\ref{sec:quasielastic} and \ref{subsec:longfluc} for details). The black line represents the sum of the theoretical predictions comprising spin waves and longitudinal fluctuations.}
	\label{fig:thales_exp3_highres}
\end{figure}

\begin{figure}[t]
	\begin{center}
	\includegraphics[width=0.49\textwidth]{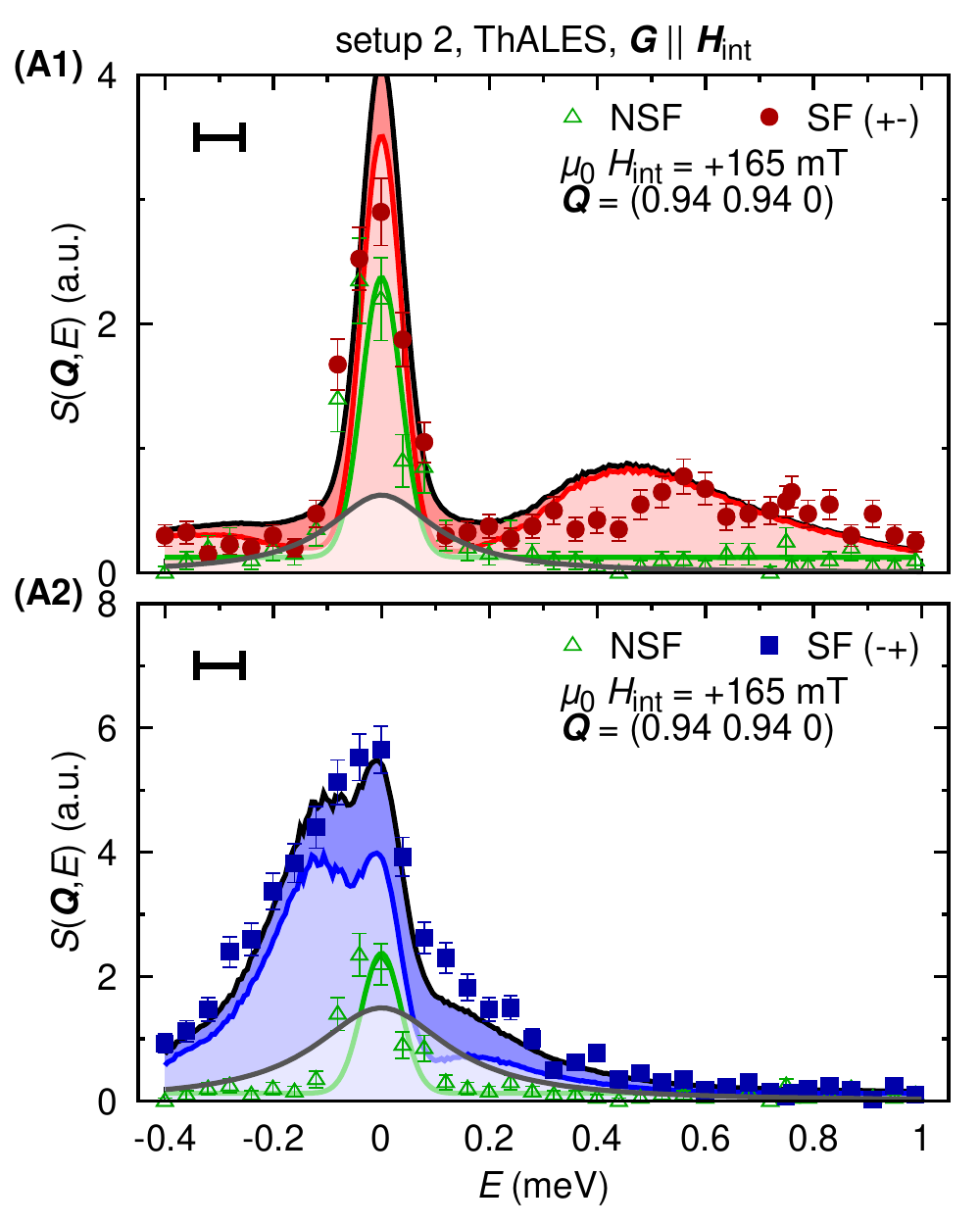}
	\includegraphics[width=0.49\textwidth]{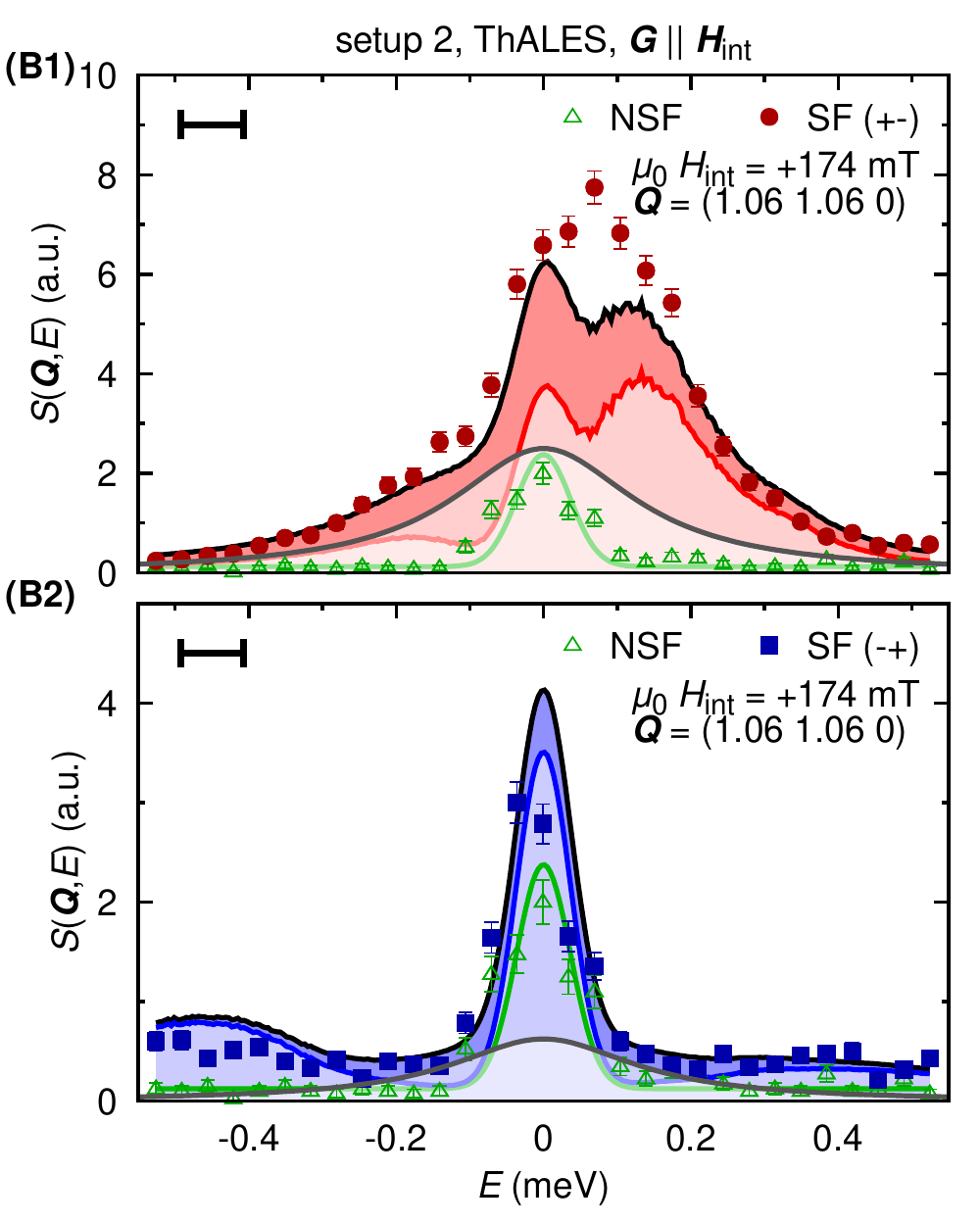}
	\end{center}
	\caption{Magnon spectra in the skyrmion lattice phase of MnSi for momentum transfers $\bf{q}_{\parallel}$ parallel to the reciprocal lattice vector $\bm{G} = \left(110\right)$ and a magnetic field $\bm{H} \parallel \left[110\right]$. The data exhibit strong evidence of the non-reciprocity of the excitations. A single collimator with a maximum horizontal beam divergence of 30 min was installed between the sample and the analyzer. The solid lines show simulations of the model convoluted with the resolution function of ThALES. The thick solid bars shown at the top left corner of each panel indicate the energy resolution. The gray shaded areas represent a quasi-elastic contribution that may be attributed to longitudinal fluctuations of the skyrmion lattice (see Secs.\,\ref{sec:quasielastic} and \ref{subsec:longfluc} for details). The black line represents the sum of the theoretical predictions comprising spin waves and longitudinal fluctuations.}
	\label{fig:thales_exp3}
\end{figure}

\begin{figure}[t]
	\begin{center}
	\includegraphics[width=0.49\textwidth]{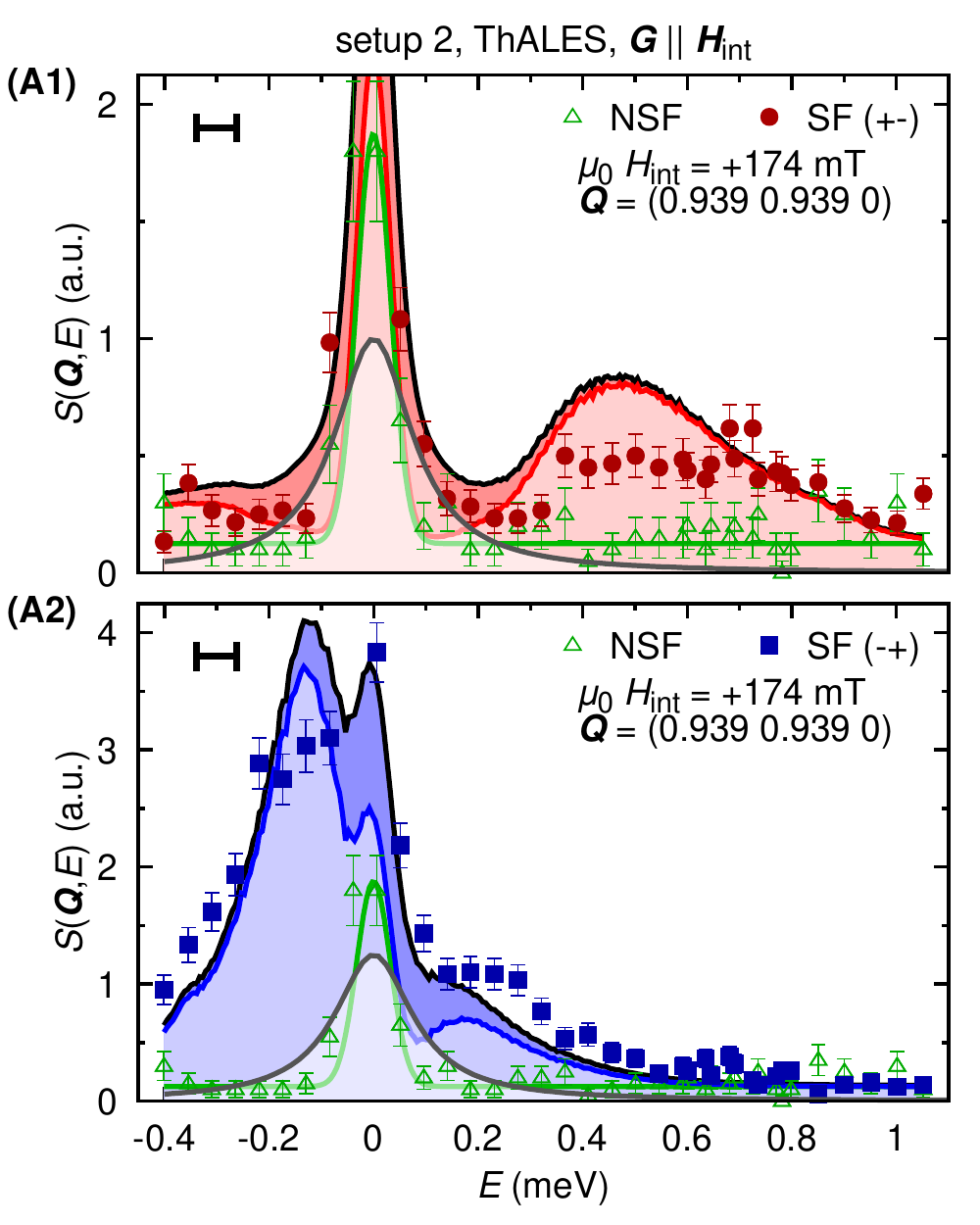}
	\end{center}
	\caption{Magnon spectra in the skyrmion lattice phase of MnSi.
	A single collimator with a maximum horizontal beam divergence of 30 min was installed between the sample and the analyzer. The solid lines show simulations of the model convoluted with the resolution function of ThALES. The thick solid bars shown at the top left corner of each panel indicate the energy resolution. The gray shaded areas represent a quasi-elastic contribution that may be attributed to longitudinal fluctuations of the skyrmion lattice (see Secs.\,\ref{sec:quasielastic} and \ref{subsec:longfluc} for details). The black line represents the sum of the theoretical predictions comprising spin waves and longitudinal fluctuations.}
	\label{fig:thales_exp3_main}
\end{figure}

\begin{figure}[h]
	\begin{center}
	\includegraphics[width=0.75\textwidth]{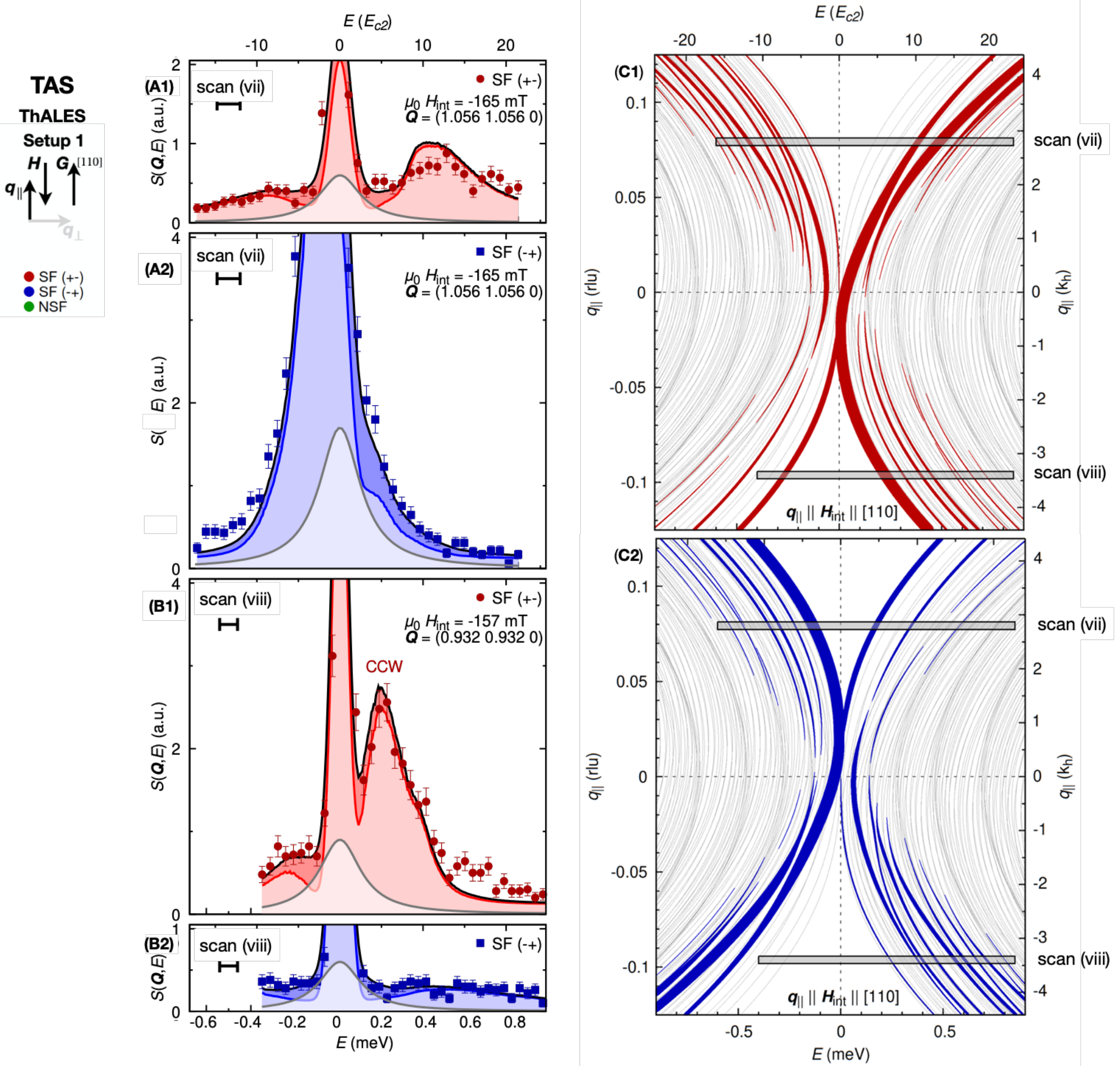}
	\end{center}
	\caption{Polarized neutron scattering intensity and calculated magnon spectra of MnSi for momentum transfers parallel to the skyrmion lattice tubes under negative magnetic fields. Data shown here contrast those recorded under positive field presented in Fig.\,4 in the main text. See text for further details.
(A1, A2) and (B1, B2) Experimental data of scan (vii) and scan (viii), respectively.
(C1, C2) Calculated magnon spectra and spectral weight for $\rm SF(+-)$ and $\rm SF(-+)$ scattering, where the parameter range of scans (vii) and (viii) are marked by gray boxes.
	}
	\label{fig:non-recip-neg}
\end{figure}

Shown in Figs.\,2 and 4 of the main text are the key results observed in polarized triple-axis spectroscopy. Additional data recorded at  ThALES are shown in Figs.\,\ref{fig:thales_exp3_highres} to \ref{fig:non-recip-neg}. As for the figures shown in the main text, the solid lines represent the theoretical model (outlined in section \ref{sec:theo}) as convoluted with the instrumental resolution function using the same parameters as in the main text, see section \ref{sec:tas:analysis} for more details. Note that the same normalization constant was used for all simulations, notably the comparison with the experimental data shown in Figs.\,2 and 4 in the main text as well as Figs.\,\ref{fig:thales_exp3_highres} to \ref{fig:non-recip-neg}. Here, $\bm{q} = \bm{Q}-\bm{G}$ denotes the reduced momentum transfer. 

For the high-resolution measurements shown in Fig.\,\ref{fig:thales_exp3_highres} we found a magnon line width of $\Gamma = 77\, \upmu\mathrm{eV}$ (half-width at half-maximum). Similar lifetimes may be expected for all magnon modes. Moreover, this value compares well with previous studies of magnons in the conical phase at a comparable magnitude of the magnetic field \cite{Weber2017Field}.

\subsection{Analysis of triple-axis data}
\label{sec:tas:analysis}

For a comparison of our experimental data with the theoretical model (see Sec.\,\ref{sec:theo} for details) we convoluted the results of the theoretical model, $S\left( \bm{Q}, E \right)$, with the instrumental resolution function $R\left( \bm{Q} - \bm{Q}_0, E - E_0 \right)$. The resolution of a triple-axis spectrometer may be described well by Gaussian transmission functions of the instrument's neutron-optical components. The full resolution function of the instrument takes the form of a four-dimensional Gaussian distribution in momentum and energy space (see Ref.\,\cite{Popovici1975} for details):
\begin{equation}
R\left( \bm{Q} - \bm{Q}_0, E - E_0 \right) \ = \
R_0 \cdot R_N \cdot \exp \left[ -\frac{1}{2} \cdot \left(\begin{array}{c}
\bm{Q} - \bm{Q}_0\\
E - E_0
\end{array}\right)^T \cdot M \cdot \left(\begin{array}{c}
\bm{Q} - \bm{Q}_0\\
E - E_0
\end{array}\right) \right],
\label{eq:tasreso}
\end{equation}
where the $4\times4$ matrix $M$ is known as the resolution matrix. $M$ represents the inverse covariance between the components of the momentum transfer vector $\bm{Q} - \bm{Q}_0$ and the energy transfer $E - E_0$, where $\bm{Q}_0$ and $E_0$ denote the nominal values. The covariance describes the deviation of the neutron beam with momentum vector $\bm{Q}$ and energy $E$ from the nominal values $( \bm{Q}_0, E_0)$. $R_0$ and $R_N$ represent scaling factors of the resolution and the Gaussian normalization, respectively. $R_N$ is related to the resolution volume and ensures that $\int d^3\bm{Q} \int dE \  R\left( \bm{Q} - \bm{Q}_0, E - E_0 \right) = R_0$.

\begin{figure}[ht]
	\begin{center}
	\includegraphics[width=0.75\textwidth]{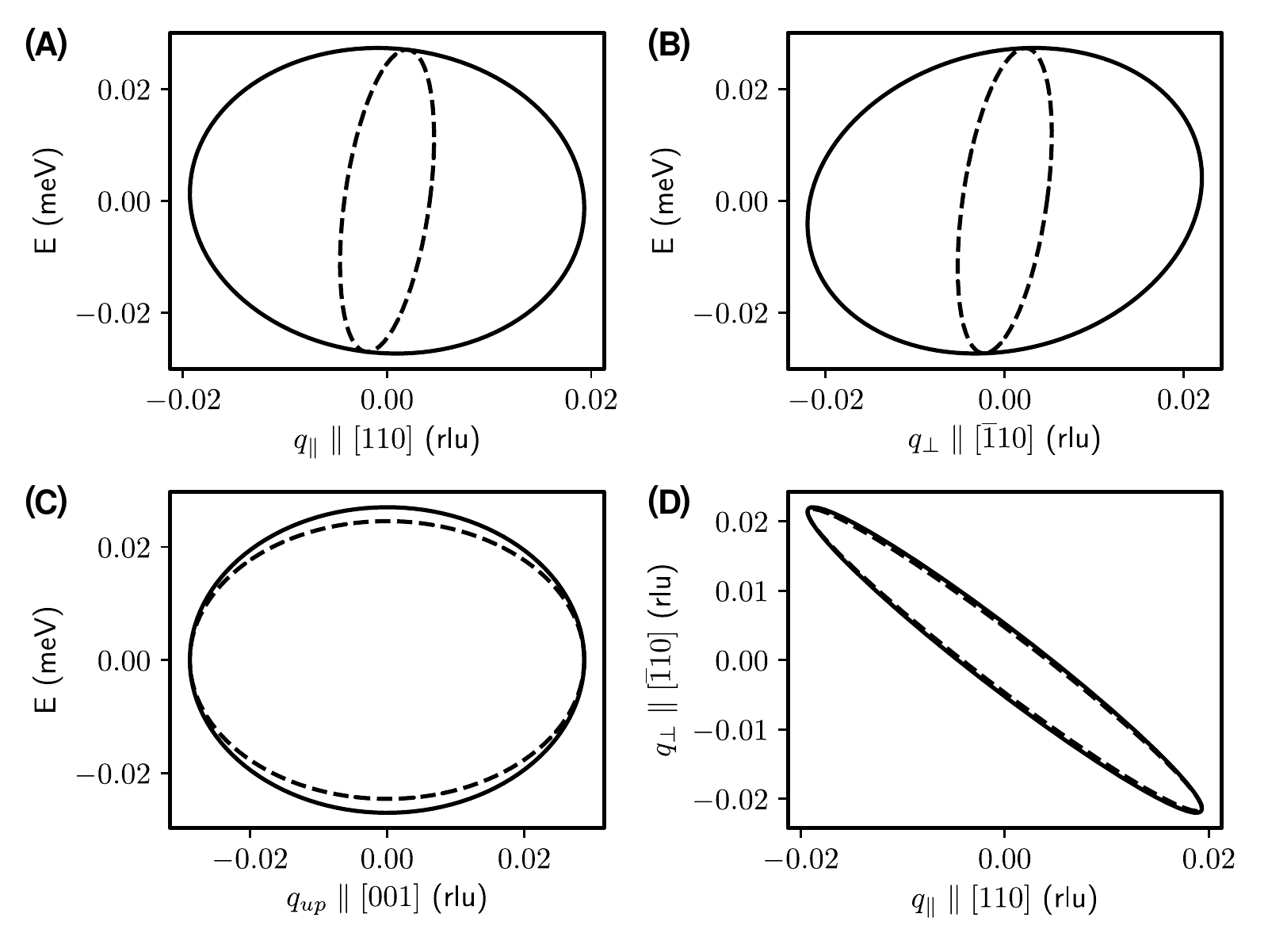}
	\end{center}
	\caption{Calculated resolution function of setup 2 as implemented at ThALES for the $(110)$ Bragg peak of MnSi and $k_f$ = $1.3$ \AA$^{-1}$. The solid lines depict the HWHM contour lines of projections of the four-dimensional resolution function $R$ [cf. Eq.\,\eqref{eq:tasreso}] into the plane defined in each panel. The dashed lines denote a cut through the four-dimensional resolution function $R$ by this plane.}
	\label{fig:thales_reso13}
\end{figure}

Geometrically, $M$ can be interpreted as an ellipsoid denoting the $1\,\sigma$ contour surface of the four-dimensional Gaussian distribution $R$. Typical results of a resolution calculation for setup 2 at ThALES using the well-known Popovici method \cite{Popovici1975} are shown in Fig.\,\ref{fig:thales_reso13}. Here the projection of the four-dimensional ellipsoid onto a specific plane is depicted by a solid line, whereas a cut of the ellipsoid with the plane is depicted by a dashed line.

For a given resolution function, the neutron intensity as effectively measured at the nominal position $\left(\bm{Q}_0, E_0\right)$ of the triple-axis spectromenter is given by the convolution integral:
\begin{equation}
I\left( \bm{Q}_0, E_0 \right) = \intop d^3\bm{Q} \intop dE \   R\left( \bm{Q} - \bm{Q}_0, E - E_0 \right) \cdot S\left(\bm{Q}, E \right).
\label{eq:tasconvo}
\end{equation}
For the data treatment applied in our study, we used a Monte-Carlo approach to evaluate the integral displayed in Eq.\,\eqref{eq:tasconvo} numerically. The Monte-Carlo integration effectively probes the dynamical structure factor $S\left(\bm{Q}, E \right)$ with Gaussian-distributed random events $\left(\bm{Q}, E \right)$ and sums up the results using the resolution function $R$ as weighting factor for the individual events.

Technically speaking, uniformly distributed one-dimensional random numbers were generated for each component of $\bm{Q}$ and $E$ using the Mersenne Twister 19937 function \cite{matsumoto1998}. These components are converted into one-dimensional Gaussian distributions using a Box-M\"uller transformation \cite{golder1976}, with the Gaussian sigmas corresponding to the ellipsoidal principal axes lengths of the resolution matrix $M$. The four-dimensional Gaussian distribution of $\bm{Q}$ vectors and energy transfers obtained this way were transformed back to the eigensystem of $M$ and may then be interpreted as probabilities for neutron events with wave vector $\bm{Q}$ and energy $E$.

For practical reasons, we pre-calculated the dynamical structure factors on a four-dimensional grid comprising the $\left(\bm{Q}, E \right)$ coordinates centered around the $\left(110\right)$ Bragg peak. In turn, for the actual resolution-convolution we only needed to calculate the instrumental resolution for each coordinate and query the grid for the corresponding dynamical structure factor. The pre-calculation of the grid was performed at the Linux cluster computing facility at the ILL.

Taken together, the resolution denoted by the black bars shown in the figures displaying the experimental data represent the incoherent energy width of the triple-axis spectrometer. Experimentally, the resolution function may be measured using a perfectly isotropic scatterer such as vanadium and the Bragg reflections of MnSi. Theoretically, it is given by the one-dimensional Gaussian width obtained after three subsequent projections of the four-dimensional resolution ellipsoid along all three $Q$ directions as explained above. Graphically, the resolution may be estimated taking the maximum extent in $Q$ of the projected resolution contour lines shown in Fig.\,\ref{fig:thales_reso13}.

Our resolution calculation and data analysis software, \textit{Takin}, \cite{takin_2, takin_1, Takin2017, Takin2016} is free and open-source. Its C++ code as well as that of its associated library \cite{tlibs} are available online at \href{https://code.ill.fr/scientific-software/takin}{https://code.ill.fr/scientific-software/takin}. The source code repository for the skyrmion calculation plugin for \textit{Takin}, which was used to calculate the spectra and dynamical structure factors presented in our study may be found at \href{https://code.ill.fr/scientific-software/takin/plugins/mnsi}{https://code.ill.fr/scientific-software/takin/plugins/mnsi}. A compilation of all source codes as used for the work reported in this paper is available online \cite{simulation_doi}.


\subsection{Quasi-elastic scattering 
\label{sec:quasielastic}}

For the comparison with theoretical prediction, all of the experimental data were scaled by the same constant. In principle, this scaling parameter may be determined quantitatively by means of the normalization with respect to a Vanadium standard as measured by time-of-flight spectroscopy, cf. Sec.\,\ref{sec:ToF:calib}. Thus the comparison with theory, in principle, can be made parameter-free, where the combined uncertainties may be as large as a factor two to three.

The theoretical analysis of the triple-axis data takes into account magnetic and nuclear scattering. The magnetic scattering, which by design of our experiments, is purely spin-flip (SF) in character, comprises transverse spin excitations, notably the spectrum of spin waves where the theory is described in Sec.\,\ref{subsec:spinwaves}, longitudinal spin fluctuations where the theory is described in Sec.\,\ref{subsec:longfluc}, and incoherent spin-flip scattering. Just as  the SF scattering is purely magnetic, the non-spin-flip (NSF) processes are purely nuclear.

The agreement between theoretical prediction and experiment improves considerably, when empirically adding a quasi-elastic contribution as indicated in Figs.\,2 and 4 of the main text, as well as Figs.\,\ref{fig:thales_exp3_highres} to \ref{fig:non-recip-neg}. For this quasi-elastic scattering we find essentially the same linewidth for all triple-axis data sets, while the quantitative strength of the scattering varies somewhat. We attribute these differences of scattering strength to resolution effects.

The additional quasi-elastic scattering was only observed in the spin-flip (SF) scattering channels and, consequently, it is of magnetic origin. It is qualitatively consistent with longitudinal spin fluctuations of the skyrmion lattice as discussed in Sec.\,\ref{subsec:longfluc}, where a Lorentzian contribution that originates in the six magnetic Bragg peaks may be expected. The polarization setup at ThALES is sensitive to these fluctuations. In turn, we modelled the longitudinal fluctuations by the Lorentzian function
\begin{equation}
	S_{qe}\left(E\right) \ \propto\  \frac{\Gamma_{qe}} {E^2 + \Gamma_{qe}^2},
\end{equation}
where we found that a linewidth $\Gamma_{qe} = 0.12 \pm 0.02\, \mathrm{meV}$ reproduces the data well. We note that $\Gamma_{qe}$ is too subtle to make quantifiable assertions from the time-of-flight data (Sec.\,\ref{sec:tof}) and too large to be discernible in the spin-echo measurements (Sec.\,\ref{sec:mieze}).

\subsection{Spurious scattering }
\label{sec:spuris} 

The orientation of the four-dimensional resolution ellipsoid is strongly influenced by the scattering configuration of the triple-axis spectrometer. As shown in panels (B) and (D) of Fig.\,\ref{fig:thales_reso13}, it is mostly the tilting of the ellipsoid in the $(q_\perp,E)$- and the $(q_\parallel,q_\perp)$-plane that is affected by the configuration of the spectrometer. For instance, for small values of the transverse reduced momentum $q_{\perp} \lesssim 0.03\, \mathrm{rlu}$ the overlap of the tails of the resolution function with the Bragg peaks may be so strong that spurious scattering emerges in the form of peaks at finite $E$- transfers, also known as Bragg tails, suggestive of inelastic scattering. Note that $\bm{q}$ is defined with respect to to the nuclear lattice vector $\bm{G}$ as $\bm{q} = \bm{Q}-\bm{G}$ and $\bm{q}_{\perp}$ relates to the $\left[ 1 \bar{1} 0 \right]$ directions perpendicular to $\bm{G}$.

\begin{figure*}[t]
	\begin{centering}
	\includegraphics[width=0.45\textwidth]{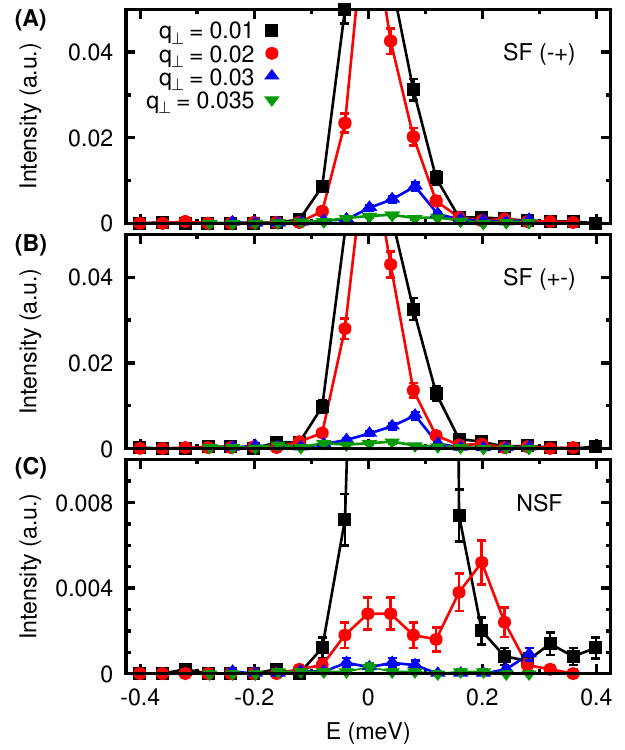}
	\caption{Typical spurious scattering associated with Bragg tails, where $\bm{q}_{\perp}$ are in $\rm (rlu)$. (A), (B) Bragg tails emanating from the magnetic satellites as observed in the spin-flip (SF) channels. The intensity decreases quickly as  $\bm{q}_{\perp}$ exceeds $0.02\,{\rm (rlu)}$ (blue triangles). (C) Bragg tail due to the nuclear $(110)$ Bragg peak as observed in the non-spin-flip (NSF) channel. With increasing $\bm{q}_{\perp}$ the peak moves to increasing $E$.}
	\label{fig:tails}
	\end{centering}
\end{figure*}

In our experiments we made great efforts to identify such spurious scattering and to avoid it. For instance, Fig.\,\ref{fig:tails} shows the spurious magnetic and nuclear Bragg tails, which are observed in the spin-flip and in the non-spin-flip channels, respectively. The tails in the spin-flip channels ``SF (+-)'' and ``SF (-+)'' are due to the overlap of the resolution function with the Bragg peaks of the skyrmion lattice, while the scattering appearing in the  ``NSF'' channel shows the Bragg tail due to the nuclear (110) peak. The spin-flip signal is strongest for reduced momentum transfers $\bm{q}_\perp$ at and below $\bm{q}_{\perp} = (0.02\, \overline{0.02}\, 0)$ because of the close proximity to the magnetic satellites. With increasing $\bm{q}_{\perp}$, the Bragg tails move to higher energy.

\subsection{Further figures and data}
\label{sec:main:nsf} 

\begin{figure}[h]
	\begin{center}
	\includegraphics[width=0.8\textwidth]{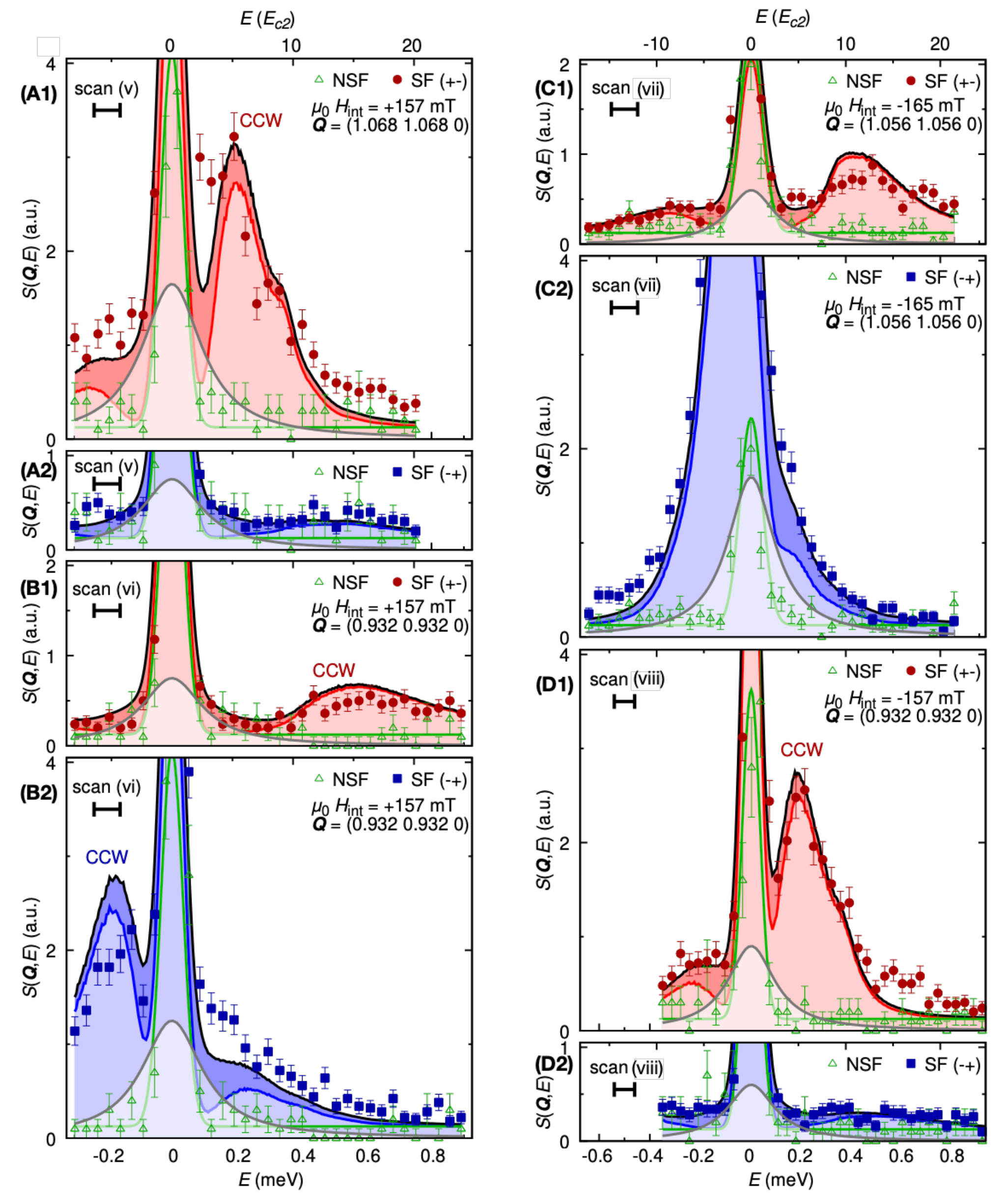}
	\end{center}
	\caption{Reproduction of the experimental data shown in Fig.\,4 and Fig.\,\ref{fig:non-recip-neg} of the main text and the supplement, respectively. In addition the NSF scattering is shown. Due to the experimental set-up used to record these data the NSF channel was sensitive to incoherent nuclear scattering only. The NSF scattering exhibits a Gaussian lineshape.}
	\label{fig:main_fig3_nsf}
\end{figure}

\begin{figure}[h]
	\begin{center}
	\includegraphics[width=0.32\textwidth]{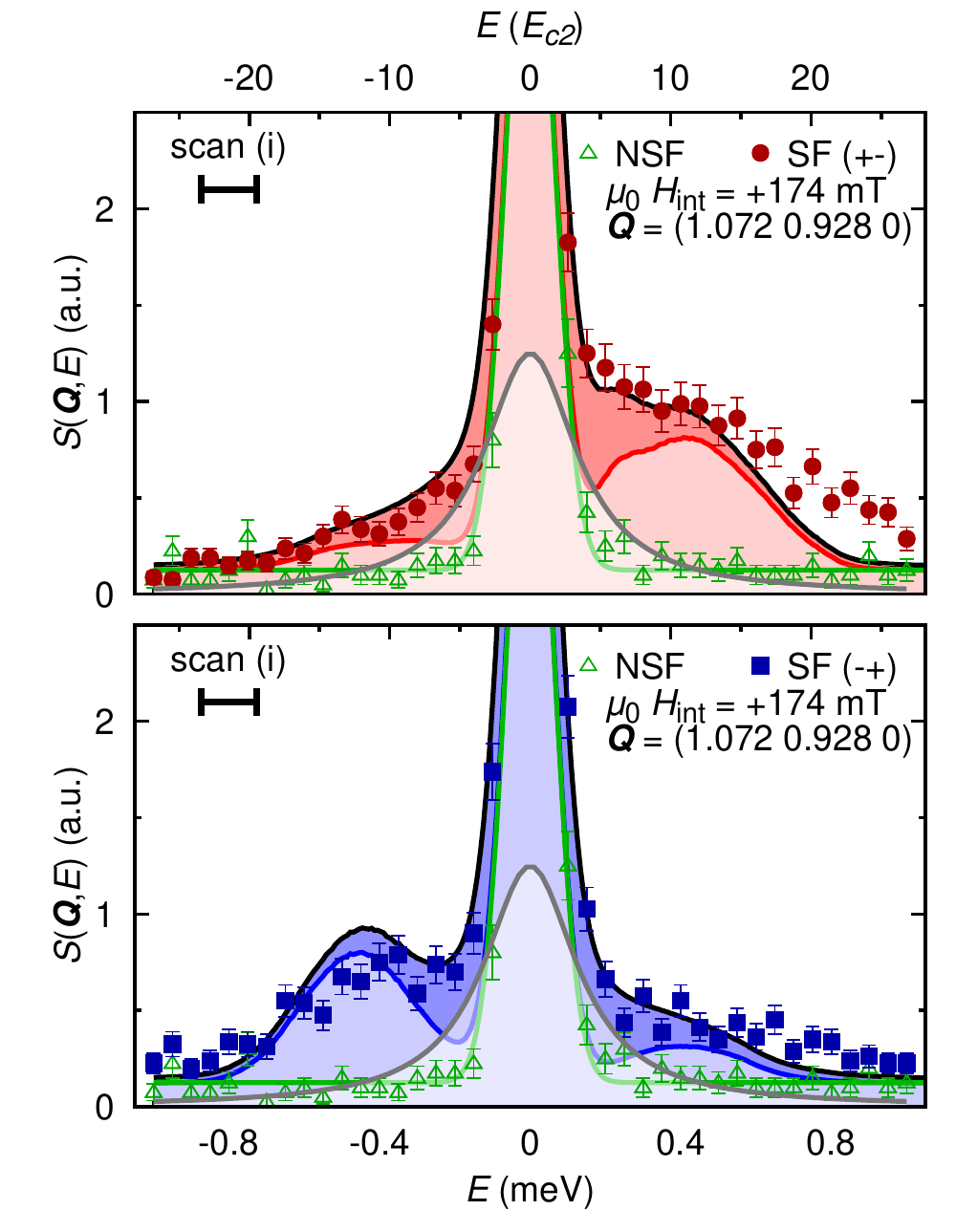}
	\includegraphics[width=0.32\textwidth]{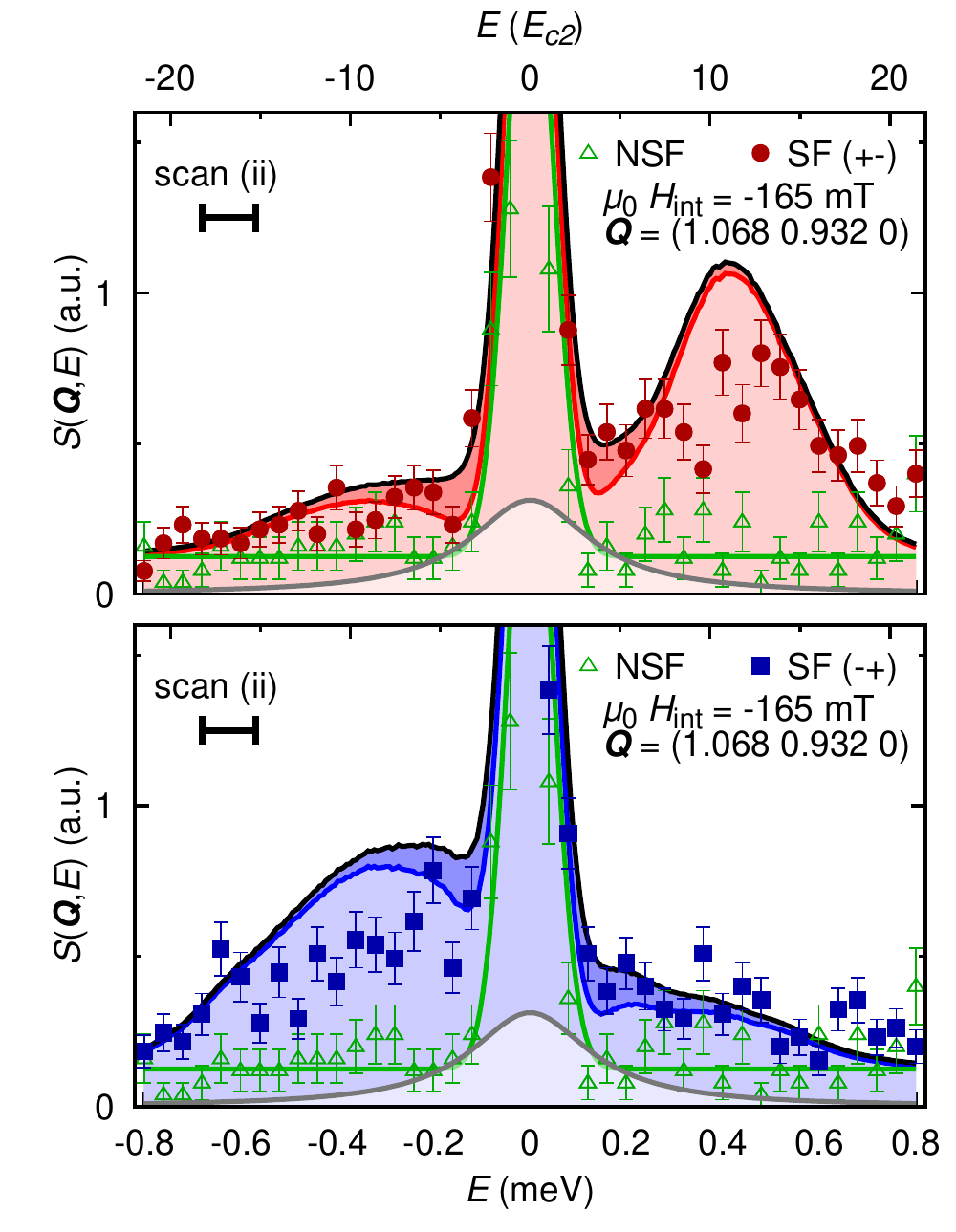}
	\includegraphics[width=0.32\textwidth]{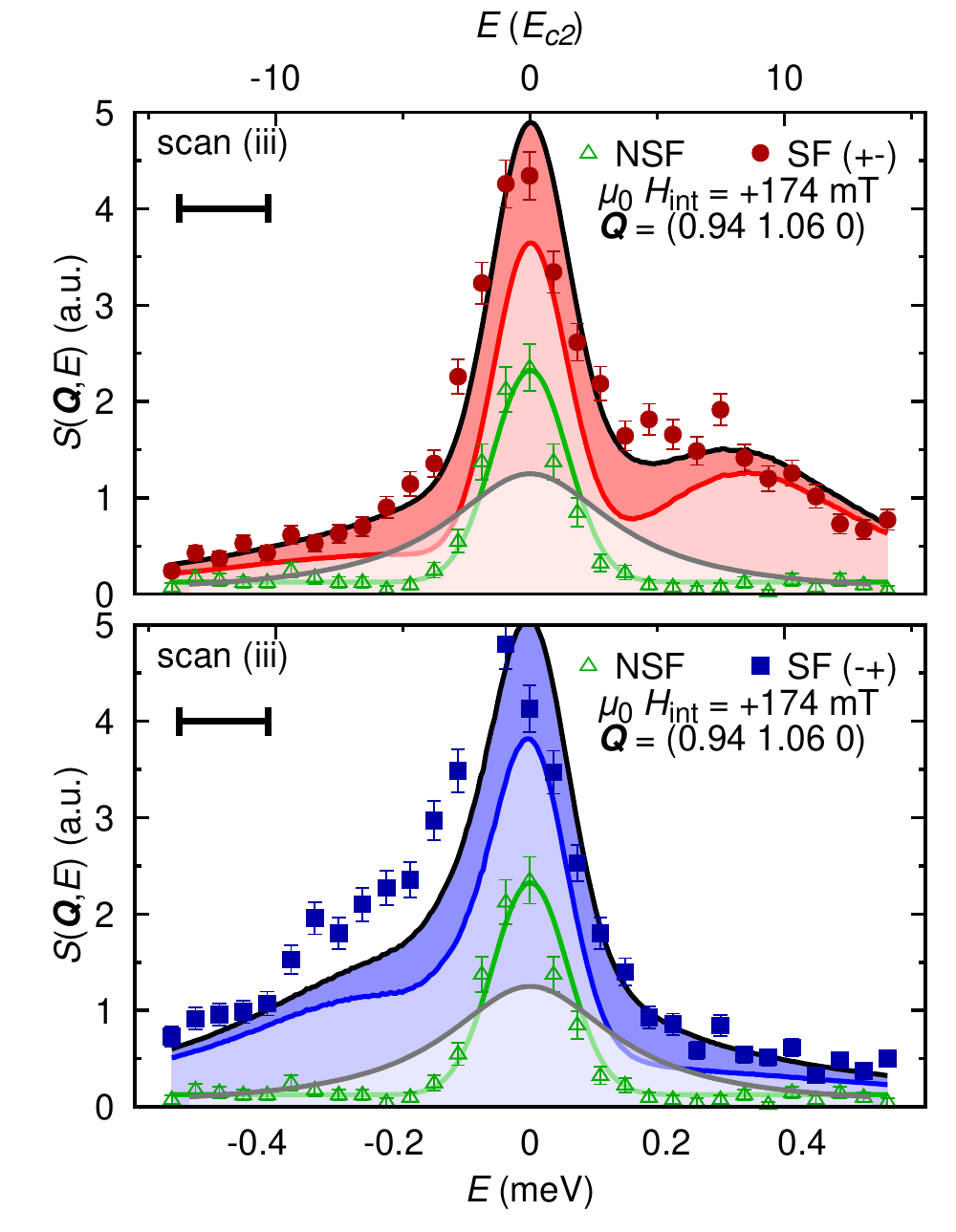}
	\end{center}
	\caption{Reproduction of Figs.\,2\,(B1, B2), and 2\,(C1, C2) shown in the main text, depicting scans (i) and scans (iii) in the skyrmion lattice plane. Also shown are the data recorded in scan (ii). In comparison to the figures presented in the main text NSF scattering is also shown. For the experimental set-up used to record these data the NSF channel was sensitive to incoherent nuclear scattering only. The NSF scattering exhibits a Gaussian lineshape.}
	\label{fig:main_fig4_nsf}
\end{figure}

Reproduced in Figs.\,\ref{fig:main_fig3_nsf} and \ref{fig:main_fig4_nsf} are the data shown in Figs.\,2\,(B1, B2), 2\,(C1, C2), 4\,(A1, A2), and 4\,(B1, B2) in the main text. Also shown is the data recorded in scan (ii) as denoted in Fig.\,2\,(D1, D2) of the main text. In comparison to the presentation in the main text also depicted is the scattering contribution observed in the NSF channel which is not shown in the main text for the sake of clarity. For setup\,2 used to record these data, the NSF channel was sensitive to incoherent nuclear scattering only. In all data sets recorded, the NSF channel exhibited a Gaussian line-shape as expected.


\newpage
\clearpage

\section{Neutron resonance spin-echo spectroscopy 
\label{sec:mieze}}

\subsection{Operational aspects and scattering geometry}
 
Neutron spin-echo spectroscopy of the skyrmion dynamics was performed by means of the Modulation of IntEnsity with Zero Effort (MIEZE) setup at the instrument RESEDA at MLZ \cite{2015Franz, 2019Franzb}. MIEZE represents essentially an ultra-high resolution time-of-flight technique, permitting straight-forward studies under strong magnetic fields as compared with conventional neutron spin-echo spectroscopy.

\begin{figure}[b]
	\begin{center}
	\includegraphics[width=0.65\textwidth]{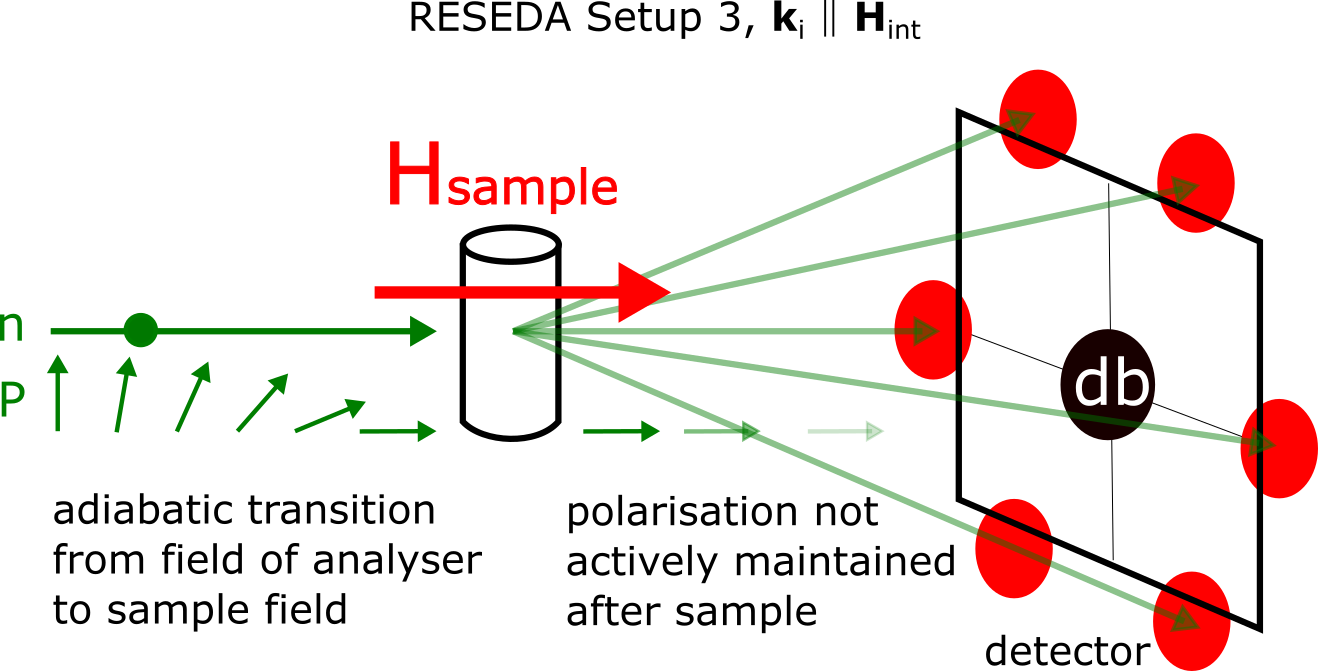}
	\end{center}
	\caption{Schematic depiction of setup 3 conceived for the neutron resonance spin-echo measurements at RESEDA. Not shown are details associated with the resonance spin-flippers. A detailed account may be found in Refs.\,\cite{2019Franzc, 2019Franzb, Reseda_JPSJ}. At the entry of the instrument the polarization is perpendicular to the direction of flight. It effectively undergoes an adiabatic transition up to the sample position, such that it is parallel to the applied field at the sample during the scattering process. After the sample it is not necessary to maintain the beam polarization. The scattering intensity at the detector corresponds effectively to that of a small-angle neutron scattering setup using polarized neutrons without polarization analysis.}
	\label{fig:reseda_schematic}
\end{figure}

In the MIEZE setup, all spin manipulations are carried out upstream of the sample position. An analyser in front of the sample converts the time dependent rotating polarisation into a sinusoidal modulation of intensity which is recorded with very high time resolution using a two-dimensional detector \cite{Cascade2011}.
 
Changing the modulation frequency, the dynamical properties of the sample may be studied over an exceptionally wide range in Fourier times. Furthermore, MIEZE as implemented at RESEDA is optimized for the investigation of small-angle scattering using a pinhole collimation system upstream of the sample.  As illustrated schematically in Fig.\,\ref{fig:reseda_schematic} the scattering configuration corresponds effectively to a small-angle instrument using polarized neutrons without polarization analysis after the sample.

\begin{figure}
	\begin{center}
	\includegraphics[width=0.6\textwidth]{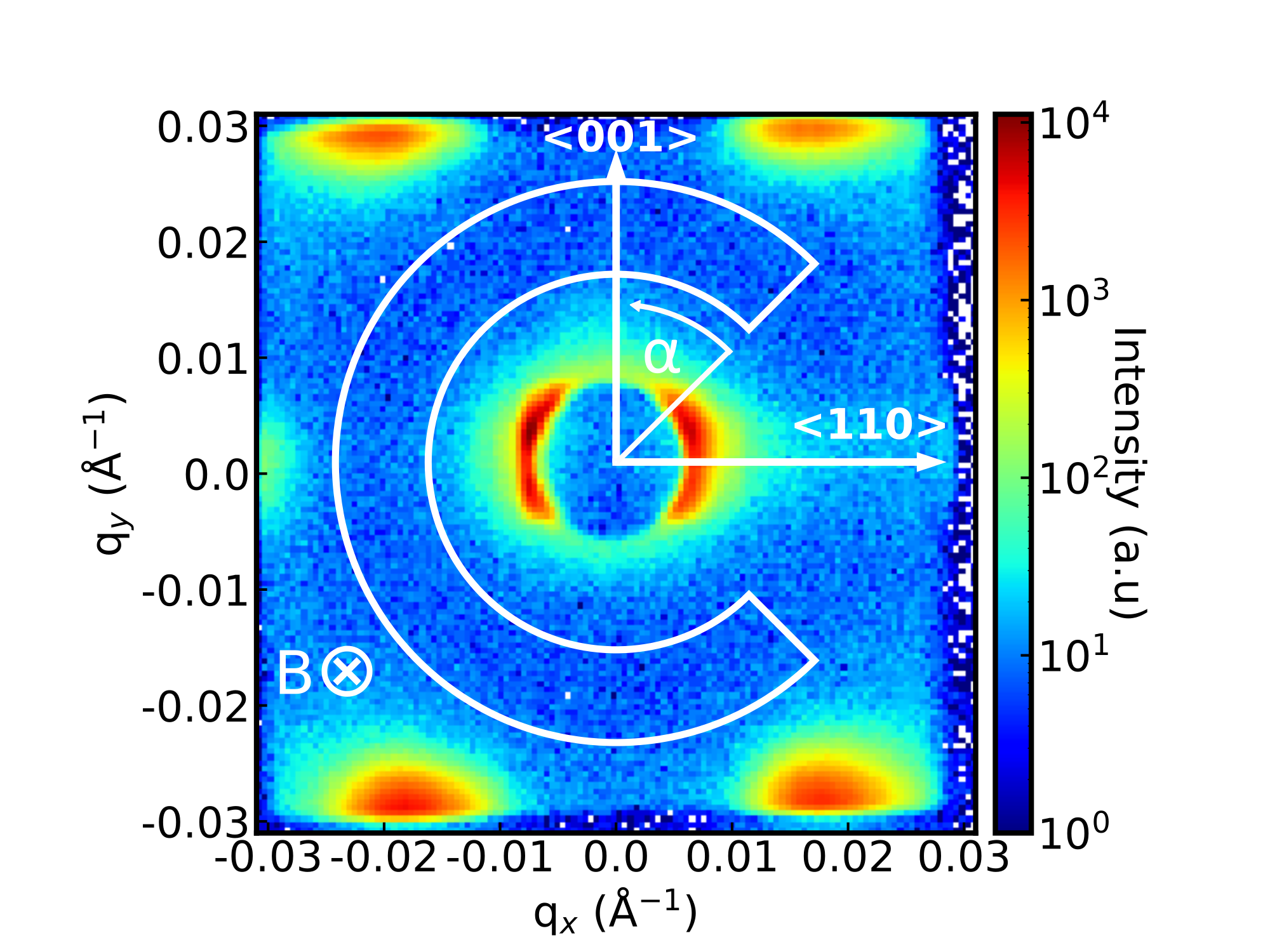}
	\end{center}
	\caption{Elastic scattering around $(000)$ measured at RESEDA in the MIEZE configuration. Five of the six skyrmion peaks may be observed on the detector. Indicated by a thin white line is the sector of integration used for the determination of the signal contrast of the MIEZE data. Also marked are the direction of the magnetic field and the crystallographic orientation of the sample. The azimuthal angle $\alpha$ increases counter-clockwise in the mathematically positive sense as shown.}
	\label{fig:reseda_elast}
\end{figure}

The MIEZE setup at RESEDA was implemented in the longitudinal neutron resonance spin echo (LNRSE) configuration \cite{2004Haussler}. In this setup the static magnetic fields of the resonant flippers are oriented parallel to the neutron beam with the resonating fields perpendicular to the static field orientation. In the longitudinal MIEZE setup the small angle background is strongly reduced as compared to transverse MIEZE as the amount of aluminium in the neutron beam is reduced from $\sim$cm to below mm. Moreover, the longitudinal setup permits the implementation of a field subtraction coil to reduce the effective field integral, this way boosting the dynamic range at small Fourier times, i.e., effectively larger energy transfers \cite{2019Franzc}. As the limit of vanishingly small Fourier times leads beyond the spin-echo approximation, we wish to emphasize that the dynamical range covered in our study of MnSi is extremely well-behaved and perfectly consistent with the spin-echo approximation.

For the MIEZE measurements sample 2 was oriented with the cylinder axis vertical and the magnetic field parallel to the neutron beam, respectively. In turn, a $\langle 001\rangle$ axis was vertical. The orientation of the cylinder was chosen such that a $\langle 110\rangle$ was horizontally perpendicular to the beam as depicted in Fig.\,\ref{fig:reseda_schematic}.

Data were recorded for a temperature of 28.5\,K and an applied field of 180\,mT [Fig.\,\ref{fig:reseda_elast}\,(A)], where the largest elastic scattering intensity of the skyrmion peaks was found. The field was applied along the wave vector $\bm{k}_i$ of the incident neutrons. The energy of the incident neutrons was $E_i\,=\,2.272$\,meV. For the data treatment, all counts in a sector on the position-sensitive detector were integrated as shown in Fig.\,\ref{fig:reseda_elast}\,(B). The sector was chosen such that a small angular segment featuring parasitic scattering due to the cryostat was not covered. For what follows the detector segment is described by an azimuthal angle $\alpha$, starting at one end and increasing counter-clockwise in the mathematically positive sense. The total contrast was normalized to the signal of a graphite sample that was measured to determine the resolution function.

\subsection{Contributions of the Skyrmion lattice and the conical phase 
\label{sec:conis}}

As explained in Sec.\,\ref{sec:samples}, for the temperature and magnetic field featuring the highest scattering intensity of the skyrmion lattice peaks, a small volume fraction of conical phase was observed in triple-axis spectroscopy in coexistence with the skyrmion lattice. This may be attributed in parts to demagnetization effects and an inhomogeneous internal magnetic field born out of the cylindrical sample shape. In turn, in the triple-axis measurements great care was exercised to chose the temperature and field such that no conical volume fraction was present in order to guarantee a clean signal.  At ThALES, this was ensured by verifying two conditions before and after each inelastic scan. First, the six elastic skyrmion satellites and their projections were measured to confirm that their intensities remained constant, thus ensuring the absence of drifts in temperature or field. Second, scattering at the positions of the satellites of the conical phase was confirmed to be absent along the direction of the applied field. This way, data collection in a phase-pure skyrmion lattice was guaranteed.

For the spin-echo measurements at RESEDA, we decided to maximise the scattering by magnons in the skyrmion lattice as the expected signal was small. In turn, for these conditions a small volume fraction of conical phase was expected as previously observed in our triple-axis measurements as well as small angle scattering \cite{2019_Kindervater_PRX}. Based on the scattering geometry it was not possible to confirm the absence of the conical peaks in the same set-up, as the position of satellites associated with the conical phase were along the field direction and thus parallel to the neutron beam.

\begin{figure}[htb]
	\begin{center}
	\includegraphics[width=1.0\textwidth]{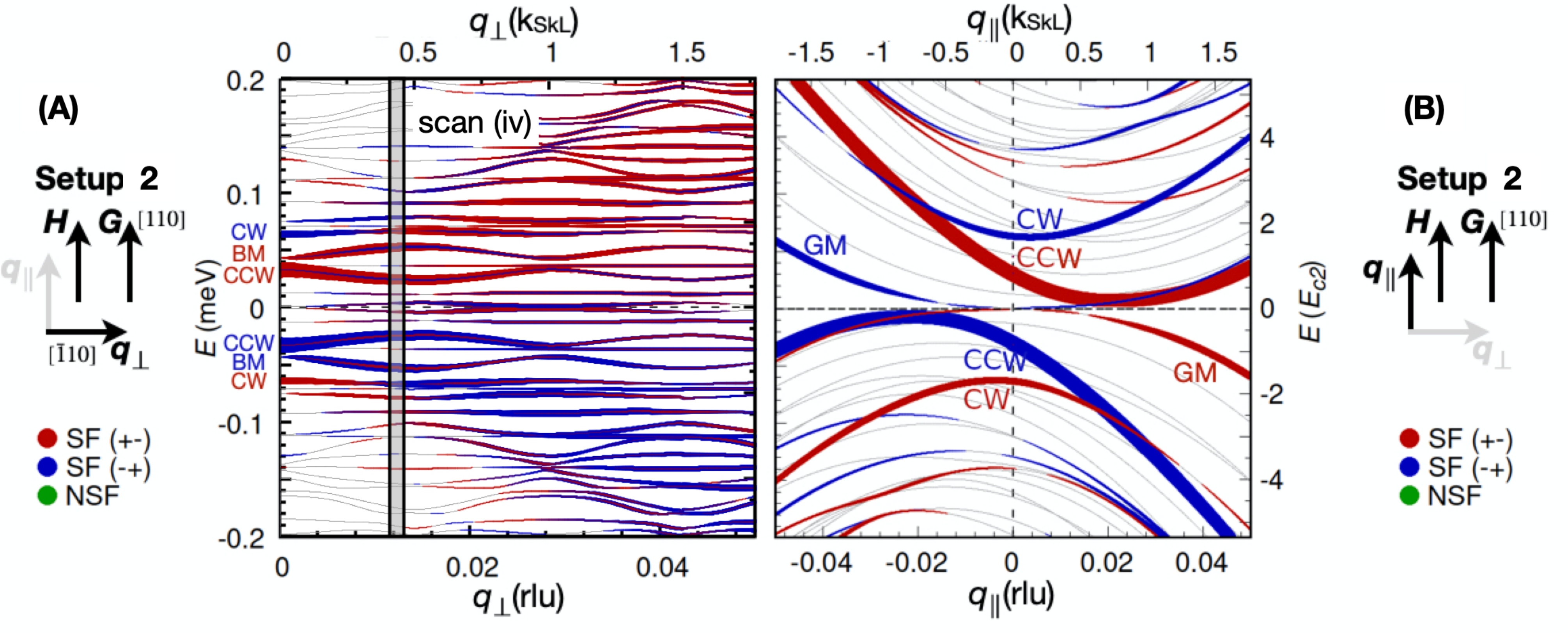}
	\end{center}
	\caption{Quantitative prediction of the magnon spectra, $E({\bf q})$, and spectral weight of the magnetic response tensor, $\chi''_{ij}({\bf q},E)$ in MnSi at low energies. The simulation is for a left-handed skyrmion lattice and for setup 2 used in TAS. Energy scales of all panels are equivalently depicted in units meV and $E_{c2}$ as shown to the left- and right-hand-side, respectively.  Momentum transfers are given in two corresponding scales at the top and bottom of the panels. Shown by thin gray lines are the magnon spectra. Denoted in color shading is $\chi''_{ij}({\bf q},E)$, where the size of the symbols reflects the spectral weight. The color shading denotes spin-flip and non-spin-flip scattering, SF and NSF, respectively.  Gray-shaded box marked "scan (iv)" denotes the location where MIEZE data was recorded using setup 3. CW: clock-wise mode; CCW: counter-clock-wise mode; BM: breathing mode; GM: Goldstone mode.
(A)  Calculated magnon spectra for momentum transfers ${\bf q}_\perp$ perpendicular to the skyrmion tubes. 
(B) Calculated magnon spectra for momentum transfers ${\bf q}_\parallel$ parallel to the skyrmion tubes.
	}
	\label{fig:calc_low_E_spectra}
\end{figure}
\begin{figure}[htb]
	\begin{center}
	\includegraphics[width=1.0\textwidth]{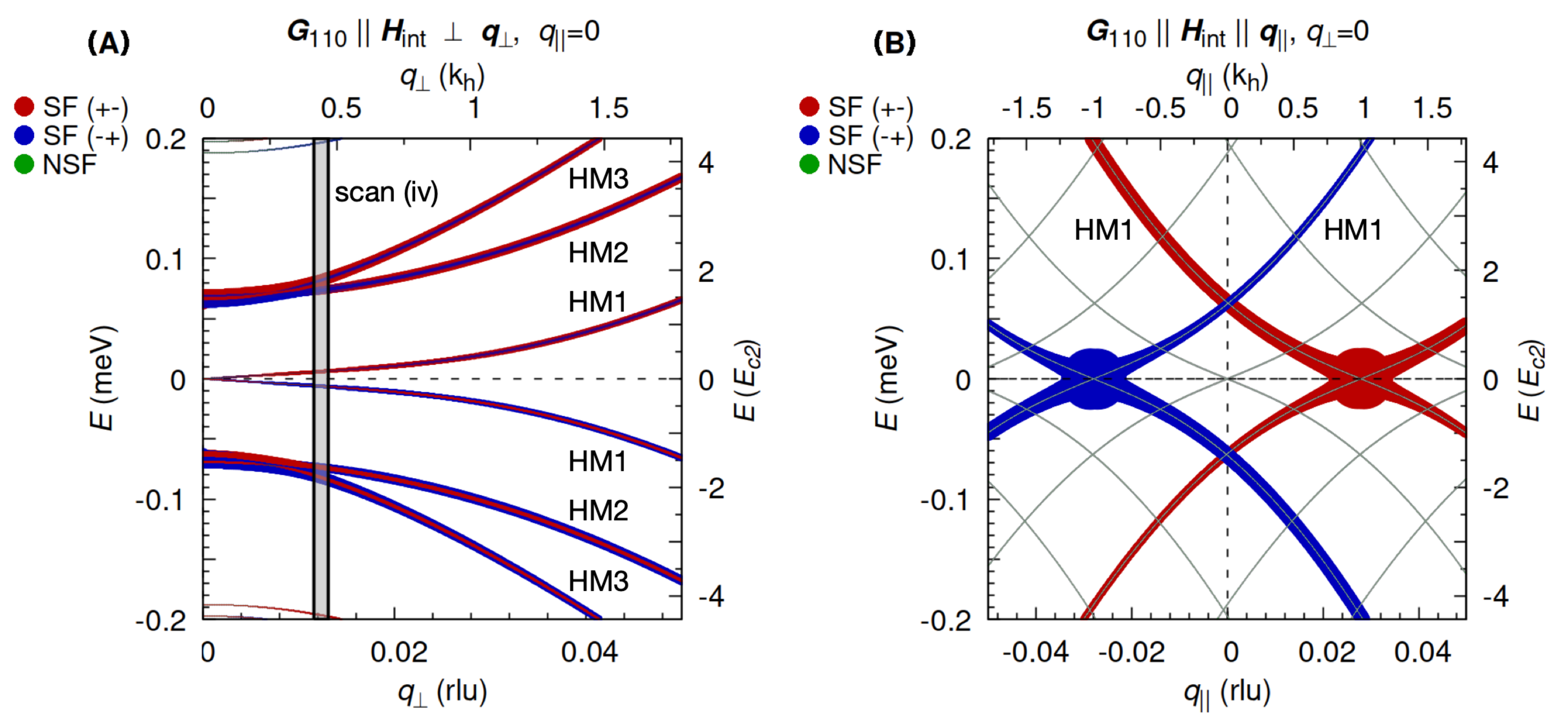}
	\end{center}
	\caption{Calculated magnon dispersion in the conical phase of MnSi \cite{HeliPaper, Weber2017Field}. Panels A and B show the dispersion branches perpendicular ($q_\perp$) and parallel ($q_\parallel$) to the helical wavevector $k_h$, respectively. The temperature $T = 27.5$ K and the magnetic field $\mu_0H = 170$ mT are selected to be in close proximity to the skyrmion phase. Panel A corresponds to the setup used at RESEDA for the measurement of the dispersion in the skyrmion phase. The gray shaded area labelled scan (iv) marks the parameter range where the MIEZE data was recorded, cf. Fig.\,3 in the main text. According to Tab.\,\ref{tab:reseda_skx_heli}, the energies of the magnons in the skyrmion phase are significantly smaller than the energy of the magnons in the conical phase. They can thus be easily distinguished. See also Fig.\,\ref{fig:calc_low_E_spectra}.
	}
	\label{fig:conis}
\end{figure}

\begin{figure}[htb]
	\begin{center}
	\includegraphics[width=0.6\textwidth]{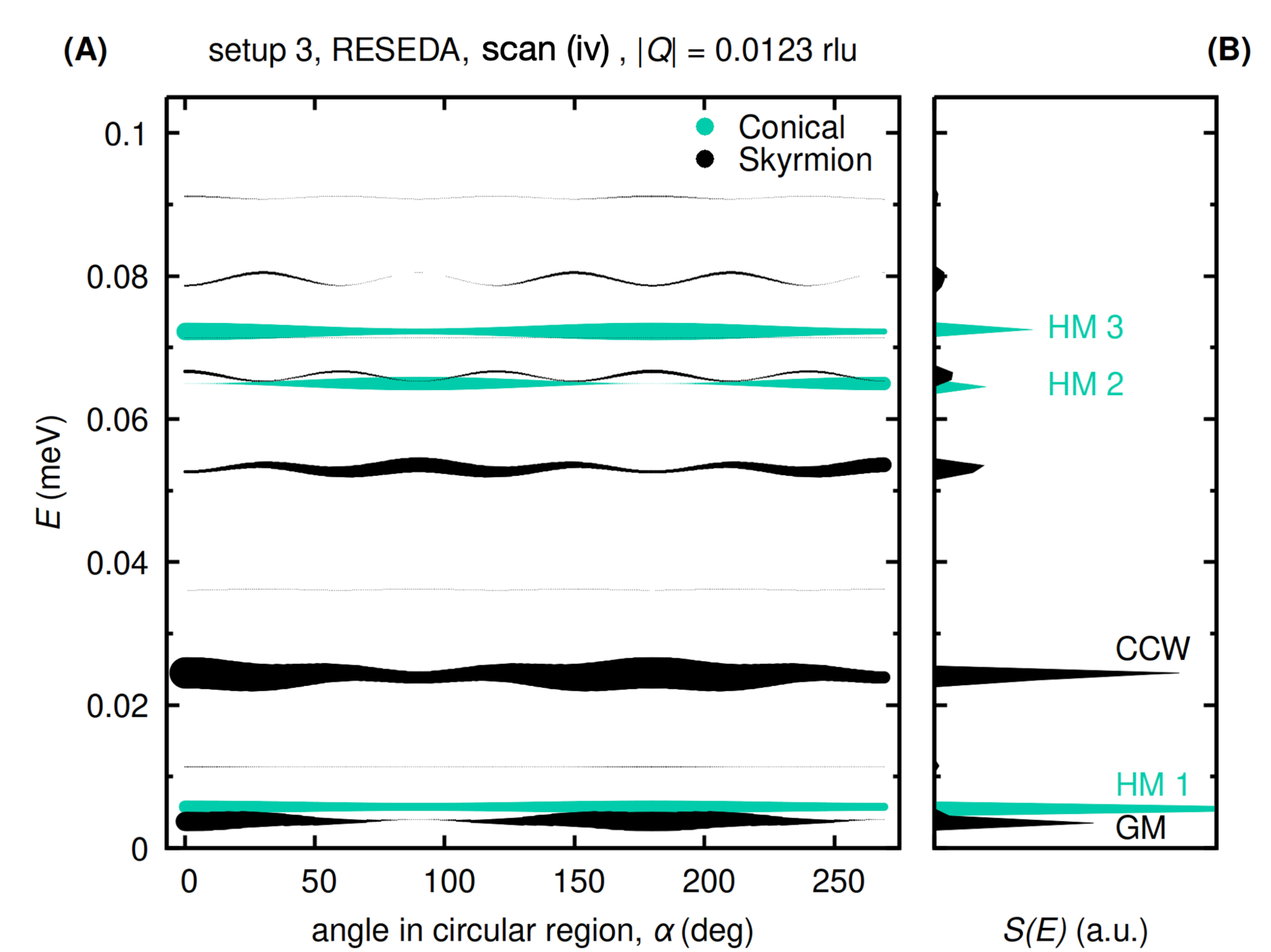}
	\end{center}
	\caption{Calculated excitation energies of magnons in the skyrmion lattice and the conical phase \cite{HeliPaper, Weber2017Field} in the detector segment shown in Fig.\,\ref{fig:reseda_elast}, corresponding to a momentum transfer $\vert Q\vert = 0.0123\,{\rm rlu}$. The plot takes into account the Bose factor. The instrumental resolution is taken into account without the radial resolution-limit and the precise $R_0$ value of the resolution ellipsoid. (A) Excitation energies as a function of azimuthal angle in the detector segment. (B) Integrated intensities in the detector segment. The labels denote GM: Goldstone mode in the sykrmion lattice ($n=0$, $C=0$), CCW: counter-clockwise mode in the skyrmion lattice ($n=3$, $C=+1$), HM1 through HM3: helimagnons in the conical phase. See also Fig.\,1\,(D) in the main text.}
	\label{fig:reseda:energy}
\end{figure}

At low $q$, the magnons from both phases appear at clearly separated energies. To illustrate this point, shown in Figs.\,\ref{fig:calc_low_E_spectra} and \ref{fig:conis} are the calculated dispersions of the skyrmion lattice and the conical state for a temperature and field where the skyrmion phase is the ground state. These figures correspond to Figs.\,3\,(C) and (D) in the main text. The location at which the MIEZE data was recorded with setup 3 is marked in gray shading. Compared to the dispersion of the skyrmion lattice a significant signal contribution of the conical state appears at very low energies and relatively high energies only, which can be identified easily. Thus the HM modes of the volume fraction of conical phase underscore the validity of the signal contributions by the skyrmion phase. 

Shown further in Fig.\,\ref{fig:reseda:energy}\,(A) are the calculated semi-quantitative scattering intensities of the magnons in the skyrmion lattice and the conical phase as a function of the azimuthal angle $\alpha$ of the detector segment shown in Fig.\,\ref{fig:reseda_elast} and defined in the previous section. Integrating these intensities over the detector segment without correction for the radial resolution and without taking into account the precise $R_0$ value of the resolution ellipsoid, a rough guide of the expected scattering intensities was obtained as shown in Fig.\,\ref{fig:reseda:energy}\,(B). These intensities provided an important point of reference when fitting the experimental data.

Summarized in Tab.\,\ref{tab:reseda_skx_heli} are the calculated values of the imaginary parts of the magnetic response tensor, $\chi''_{j}$, and the dynamical structure factor, $S_{j}$, of the skyrmion lattice ($j=SkL$) and the conical phase ($j=con$) as shown in Fig.\,\ref{fig:conis} and Fig\,3 of the main text, respectively. As the values are shown for certain energy bins (first column), certain excitation modes may appear in several bins. Further, shown in the fourth column is the Bose factor. For direct comparison with experiment, the results of spin wave theory were integrated for the semi-circular detector segment of the Cascade detector shown in Fig.\,\ref{fig:reseda_elast} as explained above, adding the dynamical structure factor for discrete points along a scan path and binning them for different energies. In addition the Bose factor was taken into account. 

Shown in the column labeled $S_{SkL}$ are the predicted intensities of magnons of the skyrmion lattice, while the column labeled $S_{con}$ displays the intensities predicted for the conical state. The energies of the most intense excitations are in excellent agreement with experiment.  At the lowest energy the Goldstone mode of the skyrmion lattice, marked GM ($n=1$), and a helimagnon of the conical phase, marked HM1, are expected at similar energies. These are followed by the counter-clockwise mode, marked CCW ($n=3$) of the skyrmion lattice around $\sim 24\,\mu{\rm eV}$. An excitation of the skyrmion lattice around $\sim 53\,\mu{\rm eV}$ appears to be too weak to be seen experimentally. Finally, two more helimagnons (HM2 and HM3) of the conical phase are expected at $\sim64.5\,\mu{\rm eV}$ and  $\sim 72.5\,\mu{\rm eV}$. The GM corresponds thereby to the lowest-lying band $n=1$, while the CCW mode corresponds to the lowest Landau level, $n=3$, where the latter is topologically non-trivial with a Chern number $C=+1$ (cf. Fig.\,1 in the main text).

As described in the main text, three magnons may be discerned in the experimental data: (i) An excitation at $E_1=4\,\mu{\rm eV}$ with a linewidth $\Gamma_1=0.55\, \mu{\rm eV}$ that we attribute to the GM and HM1. (ii) An excitation at $E_2=30\,\mu{\rm eV}$ with a linewidth $\Gamma_2=10\,\mu{\rm eV}$ that we attribute to the CCW magnon. (iii) An excitation at $E_3=80\,\mu{\rm eV}$ with a linewidth $\Gamma_3=6\,\mu{\rm eV}$ that we attribute to HM2 and HM3.


\begin{table}
	\begin{center}
	\begin{tabular}{|c|c|c|c|c|c|c|}
	\hline
	$E$ (meV)  & $\chi''_{SkL}$ (a.u.) & $\chi''_{con} (a.u.)$& Bose & $S_{SkL}$ (a.u.) & $S_{con}$ (a.u.)     & label, sky. band no. \tabularnewline \hline \hline
	0.0035     & {\bf 0.0038}  & 0	&  702 	& {\bf 2.65} 	& 0         	&  GM, $n=1$   \tabularnewline \hline
	0.0045     & 0.00061	& 0 		&  546 	& 0.33        	& 0         	& GM, $n=1$      \tabularnewline \hline
	0.0055     & 0         	& {\bf 0.017}  &  447 	& 0			& {\bf 7.77} &  HM1   \tabularnewline \hline
	0.0115     & 0.00039	& 0       	&  214 	& 0.083	& 0         	&       \tabularnewline \hline
	0.0235     & 0.019     & 0       	&  105 	& 1.97		& 0         	& CCW, $n=3$       \tabularnewline \hline
	0.0245     & {\bf 0.040}	& 0   &  101 	& {\bf 4.05}	& 0	&  CCW, $n=3$   \tabularnewline \hline
	0.0365     & 0.00044	& 0      	& 67.8 	& 0.03	& 0         	&       \tabularnewline \hline
	0.0525     & 0.013	& 0      	& 47.3 	& 0.56	& 0         	&       \tabularnewline \hline
	0.0535     & 0.020	& 0       	& 46.4 	& 0.95	& 0         	&       \tabularnewline \hline
	0.0645     & 0		& 0.023   	& 38.6 	& 0		& 0.91      & HM2 \tabularnewline \hline
	0.0655     & 0.0087	& 0       	& 38 		& 0.33	& 0         	&       \tabularnewline \hline
	0.0665     & 0.0075	& 0       	& 37.4 	& 0.28	& 0         	&       \tabularnewline \hline
	0.0715     & 0.00017	& 0       	& 34.9 	& 0.0059	& 0         	&       \tabularnewline \hline
	0.0725     & 0		& {\bf 0.049}  & 34.4 	& 0		& {\bf 1.7} &  HM3   \tabularnewline \hline
	0.0785     & 0.0033	& 0      	& 31.8 	& 0.11	& 0         	&       \tabularnewline \hline
	0.0795     & 0.0062	& 0       	& 31.4 	& 0.19	& 0         	&       \tabularnewline \hline
	0.0805     & 0.0057	& 0       	& 31 		& 0.18	& 0         	&       \tabularnewline \hline
	0.0905     & 0.0020	& 0       	& 27.6 	& 0.056	& 0         	&       \tabularnewline \hline
	0.0915     & 0.0022	& 0       	& 27.3 	& 0.061	& 0         	&       \tabularnewline \hline
	\end{tabular}
	\end{center}
	\caption{
	Calculated values of the integrated imaginary part of the magnetic response tensor, $\chi''$, and dynamical structure factor $S_j$, for the skyrmion lattice ($j=SkL$) and conical helix ($j=con$) \cite{HeliPaper, Weber2017Field, Kugler15} under similar fields and temperatures. For comparison with the data recorded at RESEDA, the Bose factor (Bose) and a weighting associated with the semi-circular sector of the detector shown in Fig.\,\ref{fig:reseda_elast}\, are taken into account. The last column designates the nature of the excitations for which an identification is possible. For the skyrmion lattice the column includes the band number, $n$, as shown in Fig.\,1 of the main text. GM ($n=1$) and CCW ($n=3$) denote the Goldstone mode and the counter-clockwise (CCW) mode of the skyrmion phase, respectively, while HM1 and HM2 denote helimagnons of the conical phase. The same labels are also used in Fig.\,3\,(A) and 3\,(B) in the main text as well as Figs.\,\ref{fig:calc_low_E_spectra} to \ref{fig:reseda:energy}. These excitations represent also the most intense modes, in excellent agreement with experiment. The relevance of the Bose factor boosting the intensity [cf. Eq. \eqref{eq:structfact}] is most pronounced for the Goldstone mode "GM". }
	\label{tab:reseda_skx_heli}
\end{table}

\newpage
\clearpage

\section{Theory of neutron spectroscopy in the skyrmion lattice }
\label{sec:theo}

\renewcommand\vec{\mathbf}

In this section we present the details of the theoretical evaluation of the inelastic neutron scattering cross-section due to spin wave excitations of the magnetic skyrmion lattice in chiral magnets. The presentation follows previous work in which we focussed on microwave resonances \cite{Schwarze15} and microwave spin-wave spectroscopy \cite{Seki:2020aa} of magnon excitations. For the determination of the excitation spectra over the wide range of excitation energies and momentum transfers addressed in our study, two theoretical challenges had to be addressed. First, a theoretical framework had to be set up that allowed to identify the ground state of an incommensurate, long wavelength, multi-$k$ structure. This is not possible within present-day numerical implementations of linear spin wave theory, such as SpinW \cite{Toth_2015}. Second, the dynamical structure factor and the associated spectral weight of typical momentum transfers accessible in neutron scattering had to be calculated up to very large numbers of Brillouin zones, as the modulation length is very large and the first Brillouin zone tiny.

The account of the theoretical work presented in the following proceeds as follows. In section \ref{subsec:TheoryBasics}  the energy functional  for chiral magnets and the wave equation that determines the spin wave excitations is presented. In section \ref{subsec:SkX1} we discuss the linear spin wave theory for the skyrmion lattice in chiral magnets, where section \ref{subsub:HighEnergy} addresses the high-energy limit in which the spin wave equation reduces to a Schr\"odinger equation in the presence of an emergent orbital magnetic field. Finally, section \ref{subsec:InScattering} focusses on the inelastic scattering cross section using polarized neutrons including the intermediate scattering function relevant for the neutron spin-echo technique.

\subsection{Linear spin wave theory for cubic chiral magnets}
\label{subsec:TheoryBasics}

\subsubsection{Free energy and equation of motion}

To lowest order in spin-orbit coupling, cubic chiral magnets are described by the
free energy functional $\mathcal{F} = \mathcal{F}_0 + \mathcal{F}_{\rm dip}$ for the dimensionless magnetization vector $\vec m$. The first term is given by
\begin{align} \label{F0}
\mathcal{F}_0 &= \int d{\bf r} \Big[ \frac{J}{2} (\nabla_j \vec m_i)^2 + D \vec m (\nabla \times \vec m) - \mu_0 M_s H \vec m_z + \lambda (\vec m^2 - 1)^2\Big]
\end{align}
where $J$ is the exchange interaction, $D$ is the DM interaction, $M_s$ is the saturation magnetization, $\mu_0$ is the magnetic field constant, and $H$ is the magnetic field applied along the $z$-axis. For right-handed magnetic systems $D>0$; for left-handed magnetic systems $D<0$. For the crystals in our experiments $D<0$. Moreover, we limit ourselves to a description of  spin waves, i.e., transverse fluctuations of the magnetization well within the ordered phase. In practice, we use a fixed value $\lambda = 160 000$ so that the length $|\vec m({\bf r},t)|$ varies spatially, for example, by less than half a per mille within the skyrmion crystal phase. Below, we will shortly comment on the signatures expected for longitudinal fluctuations.
The dipolar energy functional reads
\begin{align} \label{Fdip}
\mathcal{F}_{\rm dip} &= \int \frac{d{\bf k}}{(2\pi)^3} \frac{1}{2} \vec m_i(-{\bf k},t) \chi^{-1}_{{\rm dip},ij}({\bf k}) \vec m_j({\bf k},t)
\end{align}
with the Fourier transform $\vec m({\bf k},t) = \int d{\bf r} e^{- i{\bf k}{\bf r}} \vec m({\bf r},t)$. For wavevectors much larger than the inverse linear size of the sample $|{\bf k}|\,\gg\,1/L$, the susceptibility is given by $\chi^{-1}_{{\rm dip},ij}({\bf k}) = \mu_0 M_s^2 \frac{{\bf k}_i {\bf k}_j}{{\bf k}^2}$. For zero wavevector, $\chi^{-1}_{{\rm dip},ij}(0) = \mu_0 M_s^2 N_{ij}$,  it is determined by the demagnetization matrix $N_{ij}$ with unit trace tr$\{N_{ij}\} = 1$.

The equation of motion that governs the magnetization dynamics is given by
\begin{align} \label{EoM}
\partial_t \vec m = - \gamma \vec m \times \vec B_{\rm eff}
\end{align}
with the effective field $\vec B_{\rm eff} = -\frac{1}{M_s} \frac{\delta \mathcal{F}}{\delta \vec m}$ and the gyromagnetic ratio $\gamma = g \mu_B/\hbar>0$. The effective magnetic field vanishes, i.e., $\vec B_{\rm eff}|_{\vec m_0} = 0$, when it is evaluated with the equilibrium magnetization $\vec m_0({\bf r})$. To first  order in the deviation $\delta \vec m({\bf r},t) = \vec m({\bf r},t) - \vec m_0({\bf r})$ the effective field $\vec B_{{\rm eff},i}({\bf r},t) =  -\frac{1}{M_s}  \int d{\bf r'} \chi_{ij}^{-1}({\bf r},{\bf r'}) \delta \vec m_j({\bf r'},t)$ is determined by the susceptibility $\chi_{ij}^{-1}({\bf r},{\bf r'}) = \delta^2 \mathcal{F}/(\delta \vec m_i({\bf r})\delta \vec m_j({\bf r'}))|_{\vec m_0}$ that is to be evaluated with the equilibrium magnetization $\vec m_0({\bf r})$. Expanding the equation of motion \eqref{EoM} in first order in $\delta \vec m$, we obtain the wave equation
\begin{align} \label{SpinwaveEq}
\partial_t \delta \vec m({\bf r},t) = \frac{\gamma}{M_s} \vec m_0({\bf r}) \times \int d{\bf r'} \chi^{-1}({\bf r},{\bf r'}) \delta \vec m({\bf r'},t).
 \end{align}
This equation provides the spin wave spectrum and its eigenvectors on the level of linear spin wave theory.

Due to the soft-spin constraint $\lambda$ in Eq.~\eqref{F0} the vector field $\vec m(\vec r,t)$ is to a very good approximation a unit vector so that the deviation $\delta \vec m$ will be locally orthogonal to $\vec m_0$. We  therefore choose a local basis set of unit vectors $\hat e_i(\vec r)$ with $\hat e_1 \times \hat e_2 = \hat e_3 \equiv \vec m_0/|\vec m_0| \approx \vec m_0$ and introduce the chiral vectors $\hat e_{\pm} = \frac{1}{\sqrt{2}} (\hat e_1 \pm  i \hat e_2)$ that obey $\hat e_3 \hat e^\dagger_3 + \hat e_+ \hat e^\dagger_+ + \hat e_- \hat e^\dagger_- = \mathds{1}$. We expand the deviations $\delta \vec m(\vec r, t) = \sqrt{\frac{\gamma \hbar}{M_s}}(\hat e_+(\vec r) \psi_+(\vec r, t) + \hat e_-(\vec r) \psi_-(\vec r, t))$ where the prefactor has been introduced for convenience so that the square of the magnon wavefunctions, $|\psi_+|^2$ and $|\psi_-|^2$, has the dimension of a probability density. Using the relation $i \vec m_0 \times \hat e_\pm \approx (\pm1) \hat e_\pm$, and projecting the spin wave equation onto the local frame of reference we obtain
\begin{align} \label{SpinwaveEq2}
i \alpha \hbar \partial_t \psi_\alpha({\bf r},t) =   \int d{\bf r'} \tilde \chi^{-1}_{\alpha \beta}({\bf r},{\bf r'})   \psi_\beta({\bf r'},t),
\end{align}
where $\alpha,\beta = \pm 1$, and we abbreviated
\begin{align}
\tilde \chi^{-1}_{\alpha \beta}({\bf r},{\bf r'}) =  \frac{\gamma \hbar }{M_s}    \Big[ \hat e^\dagger_\alpha(\vec r) \chi^{-1}({\bf r},{\bf r'}) \hat e_\beta(\vec r')\Big] .
\end{align}

\subsubsection{Parameters for MnSi}
\label{subsubsec:Parameters}

The theoretical framework presented here depends on a few physically transparent, phenomenological parameters that are well-established for MnSi through independent measurements (see also table\,\ref{tab:values}). Experimentally accessible combinations of parameters are the helical wavevector $k_h = D/J$, the internal critical field $\mu_0 H^{\rm int}_{c2} = D^2/(J M_s)$ separating the conical from the field-polarized phase, and the susceptibility in the conical phase $\chi_{\rm con}^{\rm int} = \frac{\mu_0 M_s^2}{J k_h^2}$. The critical field also determines the characteristic energy scale $E_{c2} = g \mu_B \mu_0 H^{\rm int}_{c2}$. For MnSi at 28.5 K the parameters are given by $k_h \approx 0.039$/\AA\, $\mu_0 H^{\rm int}_{c2} \approx 0.35$\,T and $\chi_{\rm con}^{\rm int} \approx 0.34$. For the magnitude of the primitive reciprocal lattice of the skyrmion crystal at the same temperature of 28.5\,K we find experimentally $k_{\rm SkL} \approx 0.037$/\AA. Using $g \approx 2$ for the $g$-factor, this yields the energy scale $E_{c2} \approx 0.04$ meV.

\subsection{Linear spin wave theory for the skyrmion lattice}
\label{subsec:SkX1}

\subsubsection{Variational Ansatz and spin wave equation}

In order to obtain the equilibrium magnetization for the skyrmion lattice, we employ the variational Ansatz
\begin{align} \label{SkXAnsatz}
\vec m_0({\bf r}) = \sum_{{\bf G}_\perp \in L_R} \vec m_0({\bf G}_\perp) e^{i {\bf G}_\perp {\bf r}}
\end{align}
with the Fourier components $\vec m_0({\bf G}_\perp)$. The vectors ${\bf G}_\perp$ belong to the two-dimensional hexagonal reciprocal lattice $L_R$ that is perpendicular to the applied magnetic field, where ${\bf G}_\perp \hat z = 0$. In practice, the reciprocal lattice is restricted to a finite number of primitive unit cells of the reciprocal lattice and the symmetries of the skyrmion crystal may be exploited in order to reduce the number of variational parameters $\vec m_0({\bf G}_\perp)$, for details see Ref.~\cite{PhDWaizner2016}. First, the free energy is minimized using the Ansatz \eqref{SkXAnsatz}. In a second step, the spin wave equation \eqref{SpinwaveEq2} is solved.

With the help of the Fourier transforms $\psi_\alpha({\bf r}, t) = \int \frac{d{\bf q}}{(2\pi)^3} \frac{d\omega}{2\pi} e^{- i \omega t + i {\bf q r}} \psi_\alpha({\bf q},\omega)$ and $\tilde \chi^{-1}({\bf r},{\bf r'}) = \int \frac{d{\bf q}}{(2\pi)^3}\frac{d{\bf q'}}{(2\pi)^3} e^{i {\bf q r}-i {\bf q' r'}} \tilde \chi^{-1}_{\alpha\beta}({\bf q},{\bf q'})$ this wave equation may be expressed as
\begin{align}
\alpha \hbar \omega \psi_\alpha({\bf q},\omega) =   \int \frac{d{\bf q'}}{(2\pi)^3} \tilde \chi^{-1}_{\alpha \beta}({\bf q},{\bf q'})   \psi_\beta({\bf q'},\omega).
\end{align}
It is convenient to decompose the wavevectors ${\bf q} = {\bf K}_\perp + {\bf k}$ and ${\bf q'} = {\bf K'}_\perp + {\bf k'}$ into reciprocal lattice vectors, ${\bf K}_\perp$ and ${\bf K'}_\perp$, and wavevectors ${\bf k}$ and ${\bf k'}$, where the components of ${\bf k}$ and ${\bf k'}$ that are perpendicular to the $z$-axis belong to the first Brillouin zone, ${\bf k}_\perp, {\bf k'}_\perp \in $ 1.BZ. Using Bloch's theorem, we exploit that the susceptibility is diagonal in wavevectors but only up to reciprocal lattice vectors, $\tilde \chi^{-1}_{\alpha \beta}({\bf K}_\perp + {\bf k},{\bf K'}_\perp + {\bf k'}) =
\tilde \chi^{-1}_{\alpha \beta;{\bf K}_\perp,{\bf K'}_\perp}({\bf k})(2\pi)^3 \delta ({\bf k} - {\bf k'})$. In turn, the wave equation simplifies to
\begin{align} \label{MatrixEq}
\alpha\, \hbar\omega\, \psi_{\alpha;{\bf K}_\perp}({\bf k},\omega) =   \sum_{{\bf K'}_\perp \in L_R}  \tilde \chi^{-1}_{\alpha \beta;{\bf K}_\perp,{\bf K'}_\perp}({\bf k}) \psi_{\beta;{\bf K'}_\perp}({\bf k},\omega),
\end{align}
where we abbreviated $\psi_{\alpha;{\bf K}_\perp}({\bf k},\omega) = \psi_{\alpha}({\bf K}_\perp+{\bf k},\omega)$. 

The solution of this matrix equation for a given frequency $\omega$ and wavevector ${\bf k}$ yields the magnon dispersion $E({\bf k}) = \hbar \omega({\bf k})$ and the eigenvectors
$\psi_{\alpha;{\bf K}_\perp}({\bf k},\omega)$. The matrix $\tilde \chi^{-1}_{\alpha \beta;{\bf K}_\perp,{\bf K'}_\perp}({\bf k})$ has double indices with
$\alpha, \beta = \pm 1$ and reciprocal lattice vectors ${\bf K}_\perp$ and ${\bf K'}_\perp$. In practice, we restrict the reciprocal lattice vectors to belong only to a subset of $L_R$ that maintains the symmetry of the problem such that the matrix assumes a finite dimensionality where we refer to Ref.~\cite{PhDWaizner2016} for details. Whereas the matrix $\tilde \chi^{-1}$ is Hermitian, it is important to note that the prefactor $\alpha = \pm 1$ on the left-hand side in general requires the equation to be diagonalized with a Bogoliubov transformation.

For what follows, it proves to be convenient to introduce a  retarded Green function obeying the equation
\begin{align} \label{GreenFunction}
\Big[\tau_{\alpha \beta}^z \delta_{{\bf K}_\perp,{\bf K'}_\perp} (\hbar \omega +i 0) - \tilde \chi^{-1}_{\alpha \beta;{\bf K}_\perp,{\bf K'}_\perp}({\bf k})
\Big] g^R_{\beta \gamma;{\bf K'}_\perp,{\bf K''}_\perp}(\vec k,\omega) = \delta_{\alpha,\gamma} \delta_{{\bf K}_\perp,{\bf K''}_\perp},
\end{align}
where $\tau^z_{\alpha\beta} = \alpha \delta_{\alpha,\beta}$ and a summation over repeated indices is implied. The Green function may be  obtained by inverting the matrix within the square brackets.

\subsubsection{Limit of high-energy spin waves in the skyrmion lattice}
\label{subsub:HighEnergy}

It is expected that spin wave excitations at high energies will be governed by  contributions to the susceptibility $\chi^{-1}$ in Eq.~\eqref{SpinwaveEq} that grow with wavevectors, i.e., that involve the gradients \cite{schroeter_scattering_2015}
\begin{align}
\chi^{-1}_{ij}({\bf r},{\bf r'}) \approx \delta({\bf r} - {\bf r'}) (J \delta_{ij} \nabla^2_{\bf r'} + 2 D \epsilon_{ikj} \nabla^k_{\bf r'} ).
\end{align}
If this reduced susceptibility is inserted into Eq.~\eqref{SpinwaveEq2} we obtain for the component $\psi_+$
\begin{align} \label{SpinwaveEq - Landau1}
i \hbar \partial_t \psi_+({\bf r},t) = \frac{\gamma \hbar }{M_s} \Big(J (\nabla + i \frac{e}{\hbar} \vec A)^2 - 2 D i (\vec m_0 \nabla) \Big) \psi_+({\bf r},t),
\end{align}
where we have neglected terms that correspond to a potential for this effective Schr\"odinger equation. 

Here we have introduced the vector potential $\vec A_i = \frac{\hbar}{e} \hat e_1 \partial_i \hat e_2$, that is proportional to the spin connection $\hat e_1 \partial_i \hat e_2$. In order to emphasize the analogy with the Schr\"odinger equation of a charged particle in an orbital magnetic field, we have introduced the factor $e/\hbar$ in front of the vector potential in Eq.~\eqref{SpinwaveEq - Landau1} that cancels the factor $\hbar/e$ in the definition of $\vec A$. According to the Mermin-Ho relation \cite{mermin_circulation_1976}, the flux attributed to this vector potential is related to the topological charge density $\rho_{\rm top} = \frac{1}{4\pi} \hat e_3 (\partial_x \hat e_3 \times \partial_y \hat e_3)$ as follows
\begin{equation}
(\nabla \times \vec A)_z = \frac{4\pi \hbar}{e} \rho_{\rm top}.
\end{equation}
The vector potential may be decomposed into a part that generates a constant homogeneous flux and a part that generates the variations around the constant average, $\vec A = \vec A_{\rm hom} + \vec A_{\rm inhom}$, with $(\nabla \times \vec A_{\rm hom})_z = -\frac{4\pi \hbar}{e \mathcal{A}_{\rm UC}}$ where $\mathcal{A}_{\rm UC}$ is the area of the magnetic unit cell of the skyrmion lattice with $\int_{\rm UC} d^2 \vec r \rho_{\rm top} = -1$.

The energy spectra for wavevectors within the two-dimensional plane of the skyrmion lattice are governed by the homogeneous component $\vec A_{\rm hom}$. Neglecting the variation of the emergent magnetic field attributed to $\vec A_{\rm inhom}$ as well as to $\vec m_0$ with $\int_{\rm UC} d^2 \vec r (\nabla \times \vec m_0)_z = 0$, the Schr\"odinger equation reduces to that of a particle 
with the magnon mass $m = \hbar M_s/(2 J\gamma)$ in the presence of a homogeneous orbital magnetic field with magnitude $\langle B_{\rm em}\rangle = - \frac{4\pi \hbar}{e \mathcal{A}_{\rm UC}}$ corresponding to two flux quanta per skyrmion area. The spectrum is then given by Landau levels with cyclotron energy $E_{\rm cyclo} = \frac{\hbar e}{m} \frac{4\pi \hbar}{e \mathcal{A}_{\rm UC}}=\frac{4\pi \hbar^2}{m \mathcal{A}_{\rm UC}}$. 
Using that the area of the primitive unit cell of the hexagonal lattice is given by $\mathcal{A}_{\rm UC} = \frac{\sqrt{3} a^2}{2}$ with the lattice constant $a = 4\pi/(\sqrt{3} k_{\rm SkL})$ where $k_{\rm SkL}$ is the magnitude of the reciprocal primitive lattice vectors, the cyclotron energy reduces to $E_{\rm cyclo} = \frac{\sqrt{3}}{\pi} \hbar^2k_{\rm SkL}^2/(2m)$. This energy may be compared to the energy $E_{c2} = \hbar^2 k^2_{h}/(2m)$, where $k_h$ is the wavevector of the conical helix so we get $E_{\rm cyclo}/E_{c2} = \frac{\sqrt{3}}{\pi} (k_{\rm SkL}/k_h)^2$. 

For the estimate of the density of states reported in the main text, the ratio $k_{\rm SkL}/k_h$ was approximated  to be one. Experimentally, we find $k_{\rm SkL}/k_h \approx 0.95$ in excellent agreement with our theoretical result $k_{\rm SkL}/k_h \approx 0.97$ (strictly speaking, this value also depends weakly on the applied magnetic field). Taking into account that each skyrmion contributes two flux quanta, the Landau levels within the first Brillouin zone are doubly degenerate for each wavevector $\vec q_\perp$ corresponding to an average density of approximately $2 \pi/\sqrt{3} (k_h/k_{\rm SkL})^2 \approx 3.9$ states per energy unit $E_{c2}$ is expected. 

The non-homogeneous contributions $\vec A_{\rm inhom}$ and $\vec m_0$ in Eq.~\eqref{SpinwaveEq - Landau1}, as well as the neglected potentials, split the double degeneracy and also lead to modulations of the Landau levels within the first Brillouin zone, see Fig.~1 of the main text. It is straight-forward to confirm that the component $\psi_-({\bf r},t)$ of the magnon wavefunction fulfils the complex conjugate of the wave equation \eqref{SpinwaveEq - Landau1}. 

As widely reported in the literature, the averaged emergent field $|\langle B_{\rm em}\rangle| = \frac{2\pi \hbar}{e \mathcal{A}_{\rm UC}}$ of the emergent electrodynamics of electrons may be expressed in Tesla, where $e$ is the elementary electric charge. This permits to compare the strength of the effects of the non-trivial topology to orbital degrees of freedom of charged particles coupled to an applied magnetic field \cite{neubauer_topological_2009}. In contrast, since magnons do not carry an electrical charge they do not couple to the vector potential of the applied magnetic field and a numerical value of the averaged emergent field $|\langle B_{\rm em}\rangle| = \frac{4\pi \hbar}{e\mathcal{A}_{\rm UC}}$, reflecting the effects of the non-trivial topology, is not meaningful for the emergent electrodynamics of magnons.

\subsection{Inelastic scattering cross section for polarized neutron scattering}
\label{subsec:InScattering}

\subsubsection{Dynamic structure factor of spin waves of the skyrmion lattice
\label{subsec:spinwaves}}

The dynamic structure factor associated with excitations causing changes of the magnetization, $\delta \vec m$, is defined by
\begin{align}
\tilde{\mathcal{S}}_{ij}(\vec q, \omega) = \frac{M_s^2}{V} \int d\vec r d \vec r' \int d t
e^{-i \vec q \vec r + i \vec q \vec r' + i \omega t} \langle \delta m_i(\vec r, t) \delta m_j(\vec r',0)\rangle,
\end{align}
where $V$ is the volume of the system and we average over the center of mass coordinate as the system is in general not translationally invariant. The dynamic structure factor is related to the dissipative part of the susceptibility by the fluctuation-dissipation theorem
\begin{align}
\tilde{\mathcal{S}}_{ij}(\vec q, \omega) =
\frac{2\hbar}{1-e^{-\hbar \omega/(k_B T)}} \chi''_{ij}(\vec q, \omega) \approx
\frac{2 k_B T}{\omega} \chi''_{ij}(\vec q, \omega),
\label{eq:structfact}
\end{align}
where the last approximation is valid in the limit $\hbar \omega \ll k_B T$ that applies to the present experiment. The dissipative part of the susceptibility is given by $ \chi''_{ij}(\vec q, \omega) = \frac{1}{2 i} (\chi^R_{ij}(\vec q, \omega) - \chi^A_{ij}(\vec q, \omega))$ where the retarded susceptibility is defined by
\begin{align} \label{RetSusc}
\chi^R_{ij}(\vec q, \omega) &= \frac{M_s^2}{V} \int d\vec r d \vec r' \int d t
e^{-i \vec q \vec r + i \vec q \vec r' + i (\omega + i 0) t}
\frac{i}{\hbar} \Theta(t) \langle [\delta m_i(\vec r, t), \delta m_j(\vec r',0)]\rangle\\\nonumber
&= - \frac{g \mu_B M_s}{V} \int d\vec r d \vec r' \int d t
e^{-i \vec q \vec r + i \vec q \vec r' + i \omega t}  \hat e_{\alpha,i}(\vec r) g^R_{\alpha\beta}(\vec r,t; \vec r',0) \hat e^\dagger_{\beta,j}(\vec r'),
\end{align}
and $\chi^A_{ij}(\vec q, \omega) = (\chi^R_{ji}(\vec q, \omega))^\dagger$ and the Theta function $\Theta(t) = 1$ for $t>0$ and zero otherwise. In the last equation we introduced the retarded Green function
\begin{align}
g^R_{\alpha\beta}(\vec r,t; \vec r',0) = - \frac{i}{\hbar} \Theta(t) \langle [\psi_\alpha(\vec r, t),\psi_\beta^\dagger(\vec r', 0)]\rangle,
\end{align}
using the decomposition $\delta \vec m(\vec r, t) = \sqrt{\frac{\gamma \hbar}{M_s}}(\hat e_+(\vec r) \psi_+(\vec r, t) + \hat e_-(\vec r) \psi_-(\vec r, t))$ with $\delta \vec m(\vec r, t) = \delta \vec m^*(\vec r, t)$. Performing the Fourier transformation we express the retarded susceptibility in terms of the Green function defined in Eq.~\eqref{GreenFunction},
\begin{align}
\chi^R_{ij}(\vec q, \omega) &= -  g \mu_B M_s \sum_{{\bf K}_\perp,{\bf K'}_\perp}
 \hat e_{\alpha,i}(\vec q_{\rm RL} - {\bf K}_\perp) g^R_{\alpha\beta;{\bf K}_\perp,{\bf K'}_\perp}(\vec q_{\rm BZ}, \omega) \hat e^\dagger_{\beta,j}(\vec q_{\rm RL} - {\bf K'}_\perp),
\end{align}
where we decomposed the external wavevector $\vec q = \vec q_{\rm RL} + \vec q_{\rm BZ}$ into a part $\vec q_{\rm RL}$ associated with the reciprocal lattice of the skyrmion lattice and a part $\vec q_{\rm BZ}$ whose components perpendicular to the $z$-axis belong to the 1. BZ. The Fourier transform of the vectors $\hat e_{\alpha}(\vec r) = \sum_{\vec G} e^{i \vec G \vec r} \hat e_{\alpha}(\vec G)$ only contain components associated with the reciprocal lattice, $\vec G \in L_R$. The notation $\hat e_{\alpha,i}(\vec G)$ denotes the $i^{\rm th}$ component with $i = x,y,z$ of the vector with $\alpha = \pm 1$.

These equations allow to evaluate the spectral weight associated with the scattering processes in terms of the weight of the poles of the susceptibility $\chi''$.

\subsubsection{Dynamic structure factor of longitudinal fluctuations of the skyrmion lattice
\label{subsec:longfluc}}

Our study and the theory presented so far focuses on transverse spin excitations. In this section we comment on the contribution to the dynamic structure factor arising from longitudinal fluctuations. A proper theory for longitudinal fluctuations is demanding and goes well beyond the scope of the present work. For this reason, we limit ourselves here to a phenomenological discussion. The longitudinal fluctuations yield an additive contribution to the retarded susceptibility that possesses a form similar to Eq.~\eqref{RetSusc},
\begin{align}
\chi^R_{\parallel ij}(\vec q, \omega) &= - \frac{g \mu_B M_s}{V} \int d\vec r d \vec r' \int d t
e^{-i \vec q \vec r + i \vec q \vec r' + i \omega t} \hat e_{3,i}(\vec r) g^R_\parallel(\vec r,t; \vec r',0) \hat e^\dagger_{3,j}(\vec r'),
\end{align}
with the difference that the vectors $\hat e_{\pm}$, which are locally transverse, get replaced with the unit vector $\hat e_3 \parallel \vec m_0$, i.e., $\hat e_3$ is longitudinal to the local magnetization $\vec m_0$. In the following, we will limit ourselves to the behaviour in the ordered phase sufficiently far away from the transition to the paramagnetic phase. In this regime, the magnitude of $\vec m$  will vary only weakly as a function of position such that backscattering of longitudinal fluctuations by the periodicity of the skyrmion lattice is expected to be weak. As a consequence, we focus on the dependence on the relative position $\vec r - \vec r'$ of the Green function $g^R_\parallel(\vec r,t; \vec r',0) = g^R_\parallel(\vec r - \vec r',t)$. In turn, the susceptibility may be expressed in terms of the Fourier transforms as 
\begin{align}
\chi^R_{\parallel ij}(\vec q, \omega) &= - g \mu_B M_s \sum_{\vec K_\perp  \in L_R}
\hat e_{3,i}(\vec K_\perp) g^R_\parallel(\vec q -\vec K_\perp,\omega) \hat e^\dagger_{3,j}(\vec K_\perp),
\end{align}
where $\vec K_\perp$ belongs to the reciprocal lattice $L_R$. 

Even for conventional itinerant ferromagnets without Dzyaloshinskii-Moriya interactions, the dynamics of longitudinal spin fluctuations is complex as it is  influenced both by the damping of the particle-hole continuum as well as the scattering of virtual transverse fluctuations due to non-linear magnon-magnon interactions \cite{Boeni_1985,Boeni_2002,Vaks_1968,Solontsov_2005}. Here, we  assume instead a phenomenological form of the Green function given by
\begin{align}
g^R_\parallel(\vec q ,\omega) =  \frac{\chi_{\parallel \vec q}}{g\mu_B M_s} \, \frac{\Gamma_{\vec q} }{i \omega - \Gamma_{\vec q}},
\end{align}
where $\chi_{\parallel \vec q}$ is a static suceptibility and $\Gamma_{\vec q}$ is a phenomenological relaxation rate, that depends in general on the wavevector $\vec q$. This will give rise to a contribution to the dissipative part of the susceptibility,
\begin{align}
\frac{\chi''_{\parallel ij}(\vec q, \omega)}{\omega} &=\sum_{\vec K_\perp\in L_R}
\hat e_{3,i}(\vec K_\perp) \hat e^\dagger_{3,j}(\vec K_\perp) 
 \chi_{\parallel \vec q - \vec K_\perp}   \frac{\Gamma_{\vec q-\vec K_\perp} }{\omega^2 + \Gamma_{\vec q-\vec K_\perp}^2}.
\end{align}
This amounts to a quasi-elastic contribution in the excitation spectrum of the skyrmion lattice in the form of a sum of Lorentzians, each of which is attributed to a wavevector vector $\vec K_\perp$ of the skyrmion lattice, i.e., to the positions of magnetic Bragg peaks in reciprocal space. In analogy to ferromagnets , the relaxation rate $\Gamma_{\vec q} \propto |\vec q|^\alpha$ is expected to vanish with wavevector $\vec q$. The exponent depends on the relaxation mechanism, and depends in general on the region of phase space as well as on temperature \cite{Boeni_1985,Solontsov_2005}. If Landau damping by particle-hole pairs prevails, the exponent is  $\alpha = 1$, whereas $\alpha = 3$ in case magnon-magnon interactions dominate. The quasi-elastic contribution is thus expected to become sharp for wavevectors $\vec q$ close to a magnetic Bragg peak $\vec K_\perp$. In any case, the total energy-integrated weight is independent of $\Gamma_{\vec q}$,
\begin{align}
\int \frac{d\omega}{\pi} \frac{\chi''_{\parallel ij}(\vec q, \omega)}{\omega} &=  \sum_{\vec K_\perp\in L_R}
\hat e_{3,i}(\vec K_\perp) \hat e^\dagger_{3,j}(\vec K_\perp) 
\chi_{\parallel \vec q - \vec K_\perp} . 
\end{align}
It is determined by the static susceptibility $\chi_\parallel$ that is expected to decreases with increasing distance from the  transition to the paramagnetic phase. 

\subsubsection{Polarization-resolved inelastic neutron scattering}

As the neutrons are scattered by the magnetic flux associated with the fluctuating magnetization, the dynamic structure factor needs to be projected onto a subspace perpendicular to the wavevector transfer $\vec Q$. In our experiment, the wavevector transfer $\vec Q = \vec G_{\rm nucl} + \vec q$ is measured with respect to a nuclear Bragg peak with wavevector $\vec G_{\rm nucl}$ that is large compared to $\vec q$. As a result we can approximate the projection operator $P_{ij} = \delta_{ij} - \hat Q_i \hat Q_j \approx \delta_{ij} - \hat G_{{\rm nucl},i} \hat G_{{\rm nucl},j}$ where $\hat G_{\rm nucl} = \vec G_{\rm nucl}/|\vec  G_{\rm nucl}|$. The inelastic neutron scattering intensity is thus proportional to the correlation function
\begin{align}
\mathcal{S}_{ij}(\vec q, \omega) = P_{i k}\tilde{\mathcal{S}}_{k\ell}(\vec q, \omega)P_{\ell j} = \frac{2 k_B T}{\omega} P_{i k} \chi''_{k\ell}(\vec q, \omega)P_{\ell j}.
\end{align}
where $\vec q$ is measured with respect to the nuclear Bragg peak. $\tilde{\mathcal{S}}_{ij}$ as well as $\chi''_{k\ell}$ were discussed in the previous section. In a polarized scattering experiment, we may address the components of the matrix $\mathcal{S}_{ij}$ individually. We assume that the polarizer and analyzer selects the neutron spin to be aligned parallel or antiparallel with respect to the applied magnetic field. It is convenient to introduce a frame of reference given by the unit vectors $\hat s^i$ with $\hat s^1 \times \hat s^2 = \hat s^3$ where $\hat s^3$ is parallel to the applied field.

The non-spin flip (NSF) scattering contribution is then defined by the projection onto the direction of the magnetic field,
\begin{align}
\mathcal{S}_{\rm NSF}(\vec q, \omega) =  \hat s^3_i \mathcal{S}_{ij}(\vec q, \omega) s^3_j .
\end{align}
The two spin-flip (SF) contributions are obtained with the help of the vectors $\hat s^\pm = 
\frac{1}{\sqrt{2}} (\hat s^1 \pm i \hat s^2)$,
\begin{align}
\mathcal{S}_{\pm\mp}(\vec q, \omega) = \hat s^\pm_i \mathcal{S}_{ij}(\vec q, \omega) \hat s^\mp_j.
\end{align}
For example, the contribution $\mathcal{S}_{-+}(\vec q, \omega)$ describes the process of an incoming neutron with a magnetic moment parallel to the field, and an out-going neutron with a magnetic moment antiparallel to the field. In addition, one has to account for the efficiencies of the polarizer and analyzer in terms of their specific probabilities.

The neutron spin-echo spectroscopy is performed in the time-domain and probes the Fourier transform of the dynamic structure factor $\mathcal{S}_{ij}(\vec q,t) = \int \frac{d\omega}{2\pi} e^{-i \omega t} \mathcal{S}_{ij}(\vec q,\omega)$, which is also known as the intermediate scattering function. Moreover, the MIEZE spectroscopy represents essentially a half-polarized small-angle scattering experiment, where the incoming polarization of the neutron is specified but not the outgoing configuration. Summing over the outgoing spin polarization of the neutron using $$\sum_{\sigma_{\rm out} = \pm 1} \langle \sigma_{\rm in} | \sigma^i | \sigma_{\rm out} \rangle \langle \sigma_{\rm out} | \sigma^j | \sigma_{\rm in} \rangle = \delta_{ij} + i \sigma \epsilon_{ij k} \hat s^3_k,$$ where $\sigma^i$ is the $i^{\rm th}$ Pauli matrix, we obtain the intermediate scattering function relevant for our specific experiment 
\begin{align}
\mathcal{S}_{\sigma}(\vec q, t) &= (\delta_{ij} + i \sigma \epsilon_{ij k} \hat s^3_k) \mathcal{S}_{ij}(\vec q, t)\\\nonumber
&= \mathcal{S}_{\rm NSF}(\vec q, t) + (1+\sigma) \mathcal{S}_{-+}(\vec q, t) + (1-\sigma) \mathcal{S}_{+-}(\vec q, t) 
\end{align}
where $\sigma \hat s^3$ with $\sigma = \pm 1$ specifies the direction of the magnetic moment of the incoming neutron. 


\clearpage
\newpage



\end{document}